\documentclass[11pt,aps,prd,preprint,preprintnumbers,showpacs,groupedaddress,floatfix]{revtex4}

\usepackage{graphicx}
\usepackage{epsfig}
\usepackage{dcolumn}
\usepackage{bm}
\usepackage{subfigure}
\usepackage{lscape}
\usepackage{amsmath}
\usepackage{amssymb}
\usepackage{bbm}
\usepackage{multirow}
\usepackage{setspace}
\usepackage{xcolor}
\usepackage[hypertex,
            dvips,ps2pdf,
            colorlinks,
            linkbordercolor={0 0 0},
            pdfborder={0 0 0},
            linkcolor=blue,
            citecolor=blue,
            urlcolor=blue,
            breaklinks=true]{hyperref}

\allowdisplaybreaks

\def\rm #1{\mbox{\scriptsize #1}}
\def\eq#1{\begin{equation} #1 \end{equation}}
\def\eqarray#1{\begin{eqnarray} #1 \end{eqnarray}}
\newlength{\x}
\newlength{\y}
\newlength{\z}
\begin{document}

\phantomsection
\addcontentsline{toc}{chapter}{Main}


\begin{flushright}{\bf {UK/11-08}}\end{flushright}

\title{Moments of Nucleon's Parton Distribution for the Sea and Valence Quarks from Lattice QCD}

\author{M.\ Deka$^{1,a}$,\ T.\ Streuer$^{2,b}$\, T.\ Doi$^a$,\ S.\ J.\ Dong$^a$,\
T.\ Draper$^a$,\ K.\ F.\ Liu$^a$,\ N.\ Mathur$^c$,\ A.\ W.\
Thomas$^d$}

\affiliation{%
\centerline{$^{a}$Department of Physics and Astronomy, University of
Kentucky, Lexington, KY 40506}
\centerline{$^{b}$John von Neumann Institute NIC/DESY Zeuthen, 15738 Zeuthen, Germany}
\centerline{$^{c}$Department of Theoretical Physics, Tata Institute of
Fundamental Research, Mumbai 40005, India}
\centerline{$^{d}$Thomas Jefferson National Accelerator Facility, Newport News, VA 23606,
USA}}

\bigskip

\begin{abstract}
\bigskip
We extend the study of lowest moments,\ $\langle x\rangle$ and  $\langle x^2\rangle$,\
of the parton distribution function of the nucleon to include those of the sea quarks; this entails
a disconnected insertion calculation in lattice QCD. This is carried out on a
$16^3 \times 24$ quenched lattice with Wilson fermion. The quark loops are calculated with
$Z_2$ noise vectors and unbiased subtractions, and multiple nucleon
sources are employed to reduce the statistical errors. We obtain 5$\sigma$ signals for
$\langle x\rangle$ for the $u,d,$ and $s$ quarks, but $\langle x^2\rangle$ is consistent
with zero within errors. We provide results for both the connected and disconnected insertions.
The perturbatively renormalized $\langle x\rangle$ for the strange quark at $\mu = 2$ GeV is
$\langle x\rangle_{s+\bar{s}} = 0.027 \pm 0.006$ which is consistent with the experimental result.
The ratio of $\langle x\rangle$ for $s$ vs. $u/d$ in the disconnected insertion with quark loops is
calculated to be $0.88 \pm 0.07$. This is about twice as large as the phenomenologically fitted
$\displaystyle\frac{\langle x\rangle_{s+\bar{s}}}{\langle x\rangle_{\bar{u}}+\langle x\rangle_{\bar{d}}}$
from experiments where $\bar{u}$ and $\bar{d}$ include both the connected and disconnected insertion parts.
We discuss the source and implication of this difference.

\vspace{1pc}
\end{abstract}


\maketitle
\thispagestyle{empty}

\pagestyle{plain}
\setcounter{page}{1}
\pagenumbering{arabic}
\singlespacing
\parskip 5pt


\section{Introduction}

Recently, there has been a good deal of interest in the study of sea quarks,\ both
in theory and experiment.\ In such studies,\ strange quarks play an important role
in observables involving sea quarks.\ For more than a decade,\ intensive studies
have been made in measuring and understanding the strangeness contribution to the
nucleon spin~\cite{hughes}, the electromagnetic form factors~\cite{beck-mckeown,musolf,G0,Young:2006jc},
the strangeness condensate~\cite{hohler}, and the parton distribution function in the
nucleon~\cite{adams,CTEQ_heavy,MRST}.\ Another important aspect of studying strangeness
content is to address the issue of the {\it NUTEV anomaly}.\ The NUTEV experiment
($\nu + N \rightarrow \mu + X$)~\cite{zeller}, which measures the Weinberg angle or
weak mixing angle,\ an important parameter in the Standard Model of particle physics,\
finds a value which is three standard deviations away from the world average value.\ One
suggestion to explain this discrepancy is the asymmetry in strange and anti-strange parton
distribution~\cite{athomas1,brod_ma},\ a non-perturbative effect.\ Attempts by various
theoretical models to calculate the asymmetry give inconsistent results (sometimes with
completely opposite signs)~\cite{brod_ma,cao_signal,alwall_ingelman,ding,ma}.\ Also,
phenomenological extractions by CTEQ and NUTEV~\cite{olness,mason} for leading order
and next-to-leading order give different results.\ Lattice QCD can assess this asymmetry
from first principles in terms of $\langle x^2 \rangle_{s-\bar{s}}$ to address whether it
is strange or anti-strange which is leading in large $x$.\ This information will be helpful
to constrain and analyze the experimental data.\ Similarly,\ the first moment of the
strange parton distribution,\ $\langle x \rangle_{s+\bar{s}}$ is not well known. It ranges
between 0.018 and 0.04 from the fitting of parton distribution functions to
experiments~\cite{CTEQ_heavy}.

In the present work,\ we will study the first and second moments of quark distribution
for up,\ down and strange quarks.\ The first moment provides the measure of the symmetric
contribution from parton and anti-parton distributions ($q+\bar{q}$) and the second moment
provides the measure of the asymmetry in parton and anti-parton distributions ($q-\bar{q}$).\
These moments have contributions both from connected and disconnected insertions
for up and down quarks and only disconnected insertion for strange quarks.\ Since lattice
calculations on connected insertions,\ for the first and second moments,\ have
been done before,\ this work is going to focus mainly on disconnected insertion contributions
(particularly for strange quarks) to the first and second moments,\ which has not been
attempted in lattice QCD.

This paper is organized as follows.\ We give the formalism and lattice operators in Sec.~II.
The disconnected insertion calculation is presented in Sec.~III. The perturbative
renormalization is given in Sec.~IV. Sec.~V presents numerical parameters and error studies
of the noise estimate. The results for both the disconnected insertions and connected insertions
are given in Sec.~VI. Finally, we offer a conclusion and some discussion in Sec.~VII. Some details
of the three-point correlation functions are given in the Appendices.


\section{Formalism}
\label{sec:formulation}

In deep inelastic scattering~\cite{peskin}, it is useful to consider and analyze the moments of
the structure function via the operator product expansion (OPE) where,
in the limit of distance $x \rightarrow 0$ or equivalently
$q  \rightarrow \infty$, the product of two operators can be expanded in terms
of local operators
\eqarray{
\lim_{x \rightarrow 0} {\cal O}_i (x)\, {\cal O}_j (0)
     &=& \displaystyle\sum_k c_{ijk}(x, \mu) \,{\cal O}_k (\mu),
\label{ope1}
}
where $c$'s are the Wilson coefficients.

The leading term for such an expansion has the lowest
{\em twist},\ $t=2$.\ For unpolarized structure functions with vector currents $J^{\mu}$,
\ the twist-two operators for quarks have the bilinear form
\eqarray{
{\cal O}^{(n)\mu_1\cdots\mu_n}_f = \overline {\psi}_f\gamma^{\{\mu_1}
  (i \stackrel{\leftrightarrow}{\cal D})^{\mu_2}\cdots
  (i\stackrel{\leftrightarrow}{\cal D})^{\mu_n\}}\psi_f - \mbox{traces},
\label{eq:operator_min}
}
where $\psi_f$ denotes the quark field operator for the flavor $f$,\
${\stackrel{\leftrightarrow}{\cal D}}=
\frac{1}{2}({\stackrel{\rightarrow}{\cal D}}-{\stackrel{\leftarrow}{\cal D}})$,\
and $\{\cdots\}$ stands for symmetrization of the indices,\ $\mu$'s.\ The
subtracted trace terms are proportional to $g^{\mu_i \mu_j}$,\ so that
the operator is traceless on all pairs of indices.

In the leading twist,\ the moments of structure functions $F_1$ and $F_2$ can be written as
\eqarray{
2 \int^1_0 dx \,x^{n-1} \,F_1 (x, Q^2) &=& \displaystyle\sum_f c_{f}^{1,n} (\mu^2/Q^2, g(\mu))\,
             A^n_f (\mu), \nonumber\\
          & & \, \nonumber\\
 \int^1_0 dx \,x^{n-2} \,F_2 (x, Q^2) &=& \displaystyle\sum_f c_{f}^{2,n} (\mu^2/Q^2, g(\mu))\,
             A^n_f (\mu), \hspace{10mm}\left(\mbox{for }n\geq 2 \right)
}
where $A^n_f$ is defined through the forward matrix elements
\eqarray{
\langle \, P \, | \, {\cal O}^{(n)\mu_1\cdots\mu_n}_f \, | \, P \,\rangle &=& 2 \, A^n_f \,
           P^{\mu_1} \cdots  P^{\mu_n} - \mbox{traces}.
\label{ch:matelements}
}
In the parton model,  $A^n_f$ has the interpretation as the $(n-1)$th moment of the momentum
fraction carried by the quarks with flavor $f$ at some scale $\mu$, i.e.
\eqarray{
A^n_f(\mu)&=&\int^1_0 dx \,x^{n-1} \left[ f(x) + (-1)^n \, \bar{f}(x) \right],
}
where $f(x)$ is the quark distribution function and $\bar{f}(x)$ is the anti-quark
distribution function for the flavor $f$.\ We see that the first
moment ($n=2$) has the symmetric combination of the quark and anti-quark
distribution and the second moment ($n=3$), due to the interference between
the vector and axial-vector part of the weak interaction current, has the asymmetric
combination of the quark and anti-quark distribution.\ Our goal is to compute the first
and second moments for up,\ down, and strange quarks.

\subsection{Lattice Operators}
\label{subsec:Latticeoperators}

Lattice calculations are carried out in Euclidean path-integral. Thus, we need to transform the
twist-two operators from Minkowski space to Euclidean space.\ Following the convention~\cite{montvay,best1}
\eqarray{
\gamma^{(M)0} &\longrightarrow& \gamma^{(E)}_4  \ , \ \gamma^{(M)j} \longrightarrow  i\gamma^{(E)}_j,
\nonumber \\
 iD^{(M)0}  &\longrightarrow&  -D^{(E)}_4   \  , \  iD^{(M)j} \longrightarrow -iD^{(E)}_j,
\label{app:covdme}
}
with the $\gamma$ matrices defined as
\eqarray{
\{\gamma_\mu, \gamma_\nu \} = 2 \delta_{\mu\nu} \ ,
\ \gamma_5 &=&\gamma_1\gamma_2\gamma_3\gamma_4 \ ,
\ \sigma_{\mu\nu} = \frac{1}{2i}\,\big{[}\gamma_{\mu},\gamma_{\nu}\big{]},
}
we can transform the twist-two operators in Eq.~(\ref{eq:operator_min}) to its
Euclidean counterpart by using the notation~\cite{hagler1}
\eq{
{\cal O}^{(n)(E)}_{(f) \mu_1\cdots\mu_n} \longleftarrow h_{\mu_1 \nu_1}d_{\mu_2 \nu_2}\cdots
  d_{\mu_n \nu_n}{\cal O}^{(M)(n)\nu_1\cdots\nu_n}_f,
}
where\ $h_{\mu\nu}=\mbox{diag}\,(i,i,i,1)$ and  $d_{\mu\nu}=\mbox{diag}\,(-1,-1,-1,i)$.\
We have also set $a^{(M)0}=a^{(M)4}$ for any four vector,\ $a$. \ From now on we will consider
Euclidean operators only and drop the superscript,\ $E$.\ Since the Euclidean
signature is (+ + + +),\  the subtracted trace terms are proportional to
$\delta_{\mu_i \mu_j}$. To be specific, we use the Pauli-Sakurai $\gamma$ matrix convention
in our calculation.

We discretize our current operators by using the following relations for right and
left derivatives in lattice~\cite{kronfeld}
\eqarray{
 \stackrel{\rightarrow} D_\mu \psi^L(x) &=& \frac{1}{2a}\,
         \left[ U_\mu (x)\, \psi^L (x + a_\mu) - U^\dag_\mu(x-a_\mu)\, \psi^L(x-a_\mu) \right],\\
        & & \nonumber\\
\overline{\psi}^L(x)\stackrel{\leftarrow} D_\mu &=&
         \frac{1}{2a}\, \left[ \overline\psi^L (x + a_\mu)\, U^\dag_\mu (x) -
         \overline\psi^L(x - a_\mu)\, U_\mu(x-a_\mu) \right],
\label{discretn}
}
where $a$ is the lattice spacing.\ For example,\ the two-index operator,\ ${\cal O}_{\mu\nu}$,\ can
be written as
\eqarray{
{\cal O}_{\mu\nu} (x) &=& \frac{\lambda}{8a}\,\left[ \overline\psi^{(f)}(x)\,\gamma_\mu\,
           U_\nu (x) \, \psi^{(f)}(x + a_\nu)\, - \, \overline\psi^{(f)}(x)\,\gamma_\mu\,
           U^\dag_\nu (x - a_\nu) \, \psi^{(f)}(x - a_\nu)\right. \nonumber\\
          & &\nonumber\\
          &+& \overline\psi^{(f)}(x - a_\nu)\,\gamma_\mu\, U_\nu (x - a_\nu) \, \psi^{(f)}(x) -
          \overline\psi^{(f)}(x + a_\nu)\,\gamma_\mu\, U^\dag_\nu (x) \, \psi^{(f)}(x)\nonumber\\
          & &\nonumber\\
          &+& \overline\psi^{(f)}(x)\,\gamma_\nu\, U_\mu (x) \, \psi^{(f)}(x + a_\mu) -
          \overline\psi^{(f)}(x)\,\gamma_\nu\, U^\dag_\mu (x - a_\mu) \, \psi^{(f)}(x - a_\mu)\nonumber\\
          & &\nonumber\\
          &+& \left. \overline\psi^{(f)}(x - a_\mu)\,\gamma_\nu\, U_\mu (x - a_\mu) \, \psi^{(f)}(x) -
          \overline\psi^{(f)}(x + a_\mu)\,\gamma_\nu\, U^\dag_\mu (x) \, \psi^{(f)}(x) \right],
\label{ch:4idiscrete}
}
where $\lambda = -i$ for $\mu = 4,\ \nu = 1,2,3$;\ $\lambda = +1$ for $\mu = \nu = 1,2,3$ and
$\lambda = -1$ for $\mu = \nu = 4$. Similar expressions can be obtained for the
three-index operators.

Since in lattice QCD the continuous space-time space is described on a four-dimensional cubic
lattice, the $O(4)$ group in the continuum reduces to the hyper-cubic group
$H(4)$~\cite{baake, mandula}.\ This implies that operators belonging to irreducible
representations of $O(4)$ may transform in a reducible way under $H(4)$.\ This
will allow them to mix with lower dimensional operators under renormalization.
In order to avoid such mixing, it is suggested to adopt  the following combination of
operators which have minimal mixing (or no mixing)~\cite{martinelli, martinelli1,capitani1, beccarini}.\
For two-index operators (for $\langle x\rangle$), we choose ${\cal O}_{4i}\, (i=1,2,3)$ and
$\tilde{{\cal O}}_{44}={\cal O}_{44} - \displaystyle\frac{1}{3} ({\cal O}_{11}+ {\cal O}_{22}+{\cal O}_{33})$
which does not suffer from any mixing~\cite{martinelli,capitani1}. The best choice for three-index
operator (for $\langle x^2\rangle$) is
$\tilde{{\cal O}}_{4ii}={\cal O}_{4ii}-\displaystyle\frac{1}{2}({\cal O}_{4jj}+{\cal O}_{4kk})$,
where $i, j, k = 1, 2, 3 (\,i\neq j\neq k$), which still suffers from some mixing~\cite{martinelli,beccarini}.\
The matrix elements for these operators are
\eqarray{
\langle \,P \,|\,{\cal O}^f_{4i}\,|\, P \,\rangle
      &=&-\frac{2}{2m}\, \langle x\rangle_{f + \bar f}\,E \, P_i,\nonumber\\
\langle \, P \, | \, {\tilde{\cal O}}^f_{44}\, | \, P \,\rangle
      &=& \frac{2}{2m}\, \langle x\rangle_{f + \bar f}\, E^2,\nonumber\\
\langle \, P \, | \,{\tilde{\cal O}}^f_{4ii}\, | \, P \,\rangle
      &=& \frac{2}{2m}\, \langle x^2\rangle_{f - \bar f}\, E \, p^2_i,
\label{matrixelements_euc_all_op}
}
where the $2m$ factor, with $m$ being the nucleon mass, is due to the normalization of the
spinors with $\bar{u}(p,s)u(p,s') = \delta_{ss'}$.

\subsection{Two-Point and Three-Point Correlation Functions}
\label{subsec:twopoint}

The proton two-point function we use (with the color indices suppressed) is
\eqarray{
G^{\alpha\beta}_{NN}(t, \vec{p} ) &=&\displaystyle\sum_{\vec{x}}\,
      e^{-i\vec{p}.(\vec{x}-\vec{x}_0)}\,
      \langle \,0\,|\,\mbox{T}\,[\,\chi^\alpha(\vec{x}, t)\, \bar{\chi}^\beta(\vec{x}_0, t_0)\,]\,|\,0 \,\rangle,
}
where $t$ is the nucleon sink time,\ and $\vec{p}$  is the momentum of the nucleon.\
\ The interpolating fields~\cite{ioffe, ychung, draper1, sharpe, gupta3} we use are
\eqarray{
 \chi_\gamma(x)&=&\epsilon_{abc}\,\psi\,^{\mbox{\scriptsize T} (u)a}_\alpha(x)\,
             (C \gamma_5)_{\alpha\beta} \, \psi^{(d)b}_\beta(x)\, \psi^{(u)c}_\gamma(x), \\
             & &\nonumber\\
\bar{\chi}_{\gamma^\prime} (x)&=& -\epsilon_{def}\,\overline{\psi}^{(u)f}_{\gamma^\prime}(x)\,
             \overline{\psi}^{(d)e}_{\rho}(x)\,(\gamma_5 C)_{\rho\sigma}\,
             \overline{\psi}\,^{\mbox{T}(u)d}_{\sigma}(x),
}
where $u$ and $d$ stand for up and down quarks,\ respectively.\ $C=\gamma_2\gamma_4$,\ is
the charge conjugation operator with the Pauli-Sakurai $\gamma$ matrices. \ The letters,\
$a$,\ $b$,\ $\cdots$,\ stand for the color indices.\ The Greek letters,\ $\alpha$,\ $ \beta$,\
 ...,\ are the spin indices.

Since we are interested only in nucleon with $J^P={\frac{1}{2}}^+$, we use the projection
operator $\displaystyle\Gamma=\frac{1}{2}(1+\frac{m^-}{E^{0-}_p}\gamma_4)$\cite{lhp} to
eliminate the contamination from negative parity $S_{11}$ state. Here $m^{-}$ and $E_p^{0-}$ are
the mass and energy of the $S_{11}$ state. After applying the projection operator, we
get the two-point function as
\eqarray{
\mbox{Tr}\, \left[ \Gamma\, G_{NN}(t, \vec{p} ) \right]&=& \frac{a^6}{(2\kappa)^3}\,
    \,|\phi^+|^2 \frac{m^+}{E^{0+}_p}\, \left(1 + \frac{m^-}{E^{0-}_p}\, \frac{E^{0+}_p}{m^+}\right)\,
    e^{-E^{0+}_p (t-t_0)}\nonumber\\
    & &\nonumber\\
    &+&  \displaystyle\sum_{\theta=+,-}\, \displaystyle\sum^\infty_{n^{(\theta)}=1} e^{-E^{n^{(\theta)}}_p(t-t_0)}\,
    \tilde{f}(n^{(\theta)}, \vec{p}),
\label{ch:nucleon}
}
where the superscript $\theta = +, -$ represents positive (negative) parity states
and
\eqarray{
\tilde f (n^{+,-},\vec{p}) &=&
          N\, \Gamma^{\alpha\beta}\displaystyle\sum_s \langle\,0\,|\,\chi^\alpha(x_0)\,|\,n^{+,-}, \vec{p},s\,
          \rangle\langle\,n^{+,-}, \vec{p},s\,|\,\bar{\chi}^\beta (x_0)\,|\,0\,\rangle,
}
$N$ being the number of lattice points.

As a result, the projected two-point function with momentum $\vec{p}$ at large time
separation, i.e. $t \gg t_0$,\ will filter out the excited states, leaving only the
nucleon state remaining asymptotically
\eq{
\mbox{Tr}\,\left[\Gamma\,G_{NN}(t,\vec{p})\right]\hspace{2mm}
{\overrightarrow{\hspace{\x}}}\hspace{-\x}\raisebox{2ex}{$(t-t_0)\gg 1$}\hspace{2mm}
\left[\frac{a^6}{(2\kappa)^3}\,|\phi^+|^2\, \frac{m^+}{E^0_p}\left(1+\frac{m^-}{E^{0-}_p}\,
  \frac{E^{0+}_p}{m^+}\right)\right] \,e^{-E^{0+}_p (t - t_0)},
}
where $m^+$ and $E^{0+}_p$ are the nucleon mass and energy,\ respectively. $\kappa$ is the
hopping parameter in the Wilson fermion action. From now on, we will drop the superscript $+$.

The three-point function for any general operator ${\cal O}$ (with color indices suppressed) is defined as
\eqarray{
G^{\alpha\beta}_{N {\cal O} N}(t_2, t_1, \vec{p}_f, \vec{p}_i)
          &=&\displaystyle\sum_{\vec{x}_2,\vec{x}_1}
          e^{-i\vec{p}_f.(\vec{x}_2 -\vec{x}_1)} e^{-i\vec{p}_i.(\vec{x}_1-\vec{x}_0)}
          \langle 0|\mbox{T}(\chi^\alpha (\vec{x}_2,t_2){\cal O}(\vec{x}_1,t_1)
          \bar{\chi}^\beta (\vec{x}_0,t_0))|0\rangle,
}
where  $t=t_2$\ is the nucleon sink time, $t=t_1$\ is the current insertion
time,\ $t=t_0$ \ is the nucleon source time, and $\vec{p}_i$  and $\vec{p}_f$
are the initial and final momenta of the nucleon,\ respectively.\
For forward matrix element,\ $\vec{p}_f= \vec{p}_i=\vec{p}$.\ In this case,
\eqarray{
 G^{\alpha\beta}_{N {\cal O} N}(t_2, t_1, \vec{p})
          &=&\displaystyle\sum_{\vec{x}_2, \vec{x}_1}\,
          e^{-i\vec{p}.(\vec{x}_2 -\vec{x}_0)}\, \langle\, 0\, |\, \mbox{T}\,
          ( \chi^\alpha (\vec{x}_2, t_2)\, {\cal O}(\vec{x}_1, t_1)\,\bar{\chi}^\beta (\vec{x}_0, t_0))\,|\,0 \,\rangle.
}

The three-point functions can be classified according to two
different topologies of the quark paths~\cite{liu1, liu2, liu3, wilcox1}
between the source and the sink of the proton\---- one is
{\em quark line connected} and the other is {\em quark line disconnected}.

\begin{figure}[h]
\centering
\subfigure[]
{\rotatebox{0}{\includegraphics[width=0.4\hsize]{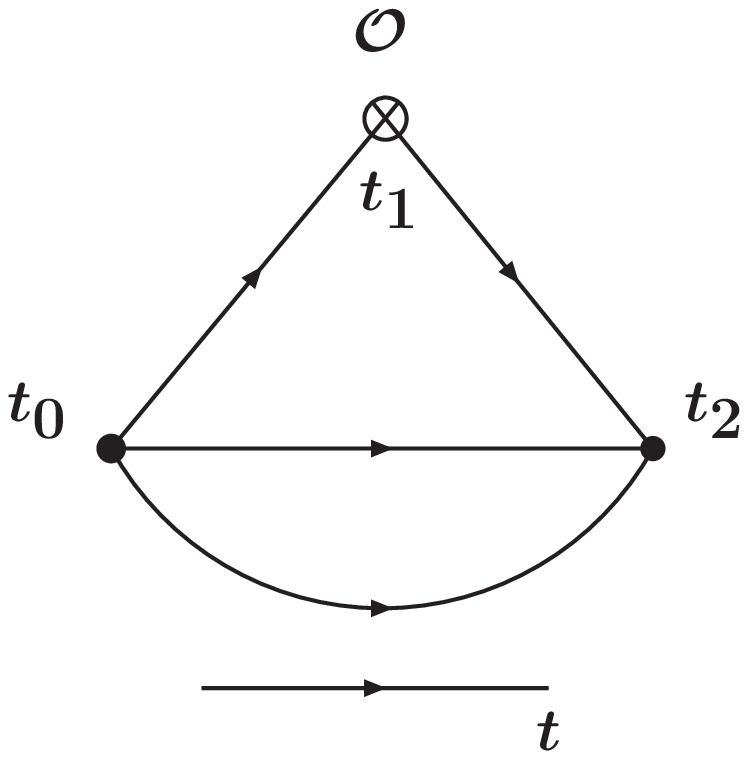}}
\label{connected_fig}}
\hspace{0.6cm}
\subfigure[]
{\rotatebox{0}{\includegraphics[width=0.4\hsize]{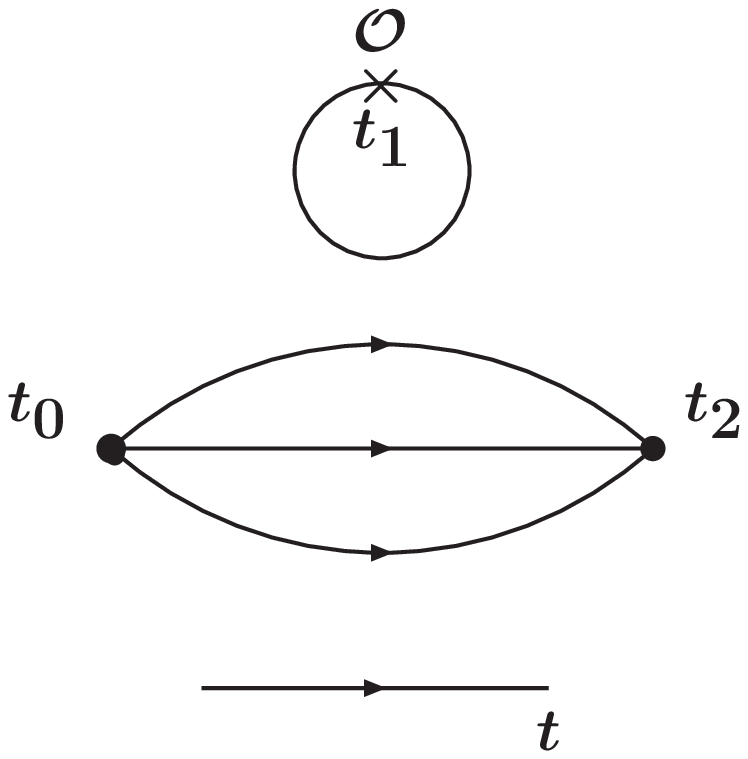}}
\label{disconnected_fig}}
\caption{Quark line diagrams of the three-point function in the Euclidean
         path integral formalism.\ (a) Connected insertion  and
         (b) disconnected insertion.}
\label{conanddisconfig}
\end{figure}

The {\em quark line connected} part of the three-point function in the path integral is
represented diagrammatically in Fig.~\ref{connected_fig}. It needs to be
stressed that it is not a Feynman diagram in perturbation theory.\ We see that the nucleon
interpolating quark fields contract with the quark fields of the
current so that the quark lines flow continuously from $t=t_0$ at the
nucleon source to $t=t_2$ at the nucleon sink.\ This is termed  the {\em connected insertion}
(C.I.).\ The  {\em quark line disconnected} part is represented
diagrammatically in Fig.~\ref{disconnected_fig}.\ In this case,\ we see that
the quark fields in the current self contract to form a loop,
which is disconnected from the nucleon interpolating quark
fields with regard to their quark lines.\ This is termed  the
{\em disconnected insertion} (D.I.). Since the quarks are propagating in the
gauge background, if we were to consider a similar situation in terms of
Feynman diagrams in the perturbative approach, it would involve gluon lines between the quark loop
and the nucleon propagator so that the corresponding Feynman diagrams are {\it connected} in this
sense. Indeed the corresponding disconnected Feynman diagrams are subtracted as the uncorrelated
part in the definition of the {\em disconnected insertion} (D.I.) i.e.
\eq{
G_{N{\cal O}N} (\mbox{D.I.}) = \langle  \chi {\cal O}\overline{\chi} \rangle
    - \langle {\cal O} \rangle\langle \chi \overline{\chi}\rangle,
}
where $\chi$ is the nucleon interpolation field. In the literature, this
disconnected insertion is sometimes referred to as the ``disconnected diagram'' which
can cause some confusion.

The computation of the C.I. is relatively straight forward. We shall use the
sequential source technique~\cite{bernard3, draper2, bernard4, martinelli1} to
calculate it.\ This fixes the source point $t_0$ and the sink time slice $t_2$.\
However, the computation of the D.I.\ poses a major numerical challenge.\
The D.I. contains not only the usual propagators from the source,\ $x_0$,\ to
any point,\ $x$,\ but also the propagators from any insertion position ($x_1$)
to any other lattice points.\ This amounts to inverting the fermion matrix at
each point of the lattice to construct the {\em all-to-all} propagators.\ This
entails inversion of a million by million ($\sim16^3\times24\times3\times4$)
sparse matrix on a $16^3\times24$ lattice (3 and 4 are number of color
and spin indices) for each gauge configuration.\ This is unattainable even by
using the computing powers of today's supercomputers.\ Instead,\ we shall calculate with the
stochastic method. Specifically, we adopt the complex $Z_2$ noise for the estimation with
unbiased subtraction. The detailed description of the method and the usefulness of
discrete symmetries will be presented in Sec.~\ref{subsec:discretesymmetries}.

\subsection{Ratios of Correlation functions}
\label{subsec:ratioscorrelationfunction}

In order to extract $\langle x \rangle$ and $\langle x^2 \rangle $,\
we take the suitable ratios of the three-point to two-point correlation functions.\
Since,\ for the case of C.I.,\ the nucleon sink time is fixed,\
and for the case of D.I.,\ the sink time is not fixed,\ the procedures
for extracting the matrix elements,\ after taking the ratios,\ are
different for the two cases.

\subsubsection{Connected Insertions}
\label{subsubsec:rati_connected_insertions}

In view of the fact that the matrix elements  of ${\cal O}_{4i},$
between the equal momentum nucleon states, are proportional to $p_i$, and that of
$\tilde{{\cal O}}_{4ii}$ are proportional to $p^2_i$, and that the momentum projection
is folded in the sequential source at the sink time $t_2$ in the connected insertion
calculation, we have chosen only one momentum for the nucleon to be $p_i= 2\pi/La$
(the lowest available non-zero momentum) along the $x$ direction in order to reduce the
computational cost.

After inserting complete sets of physical states between the interpolation fields and
the operators ${\cal O}$, we arrive at the asymptotic relations for the following ratios of
three- to two-point functions at $(t_1-t_0)\gg 1$ \mbox{and} $(t_2-t_1)\gg 1$:
\eq{
\frac{2 \kappa}{p_i}\,\frac{\mbox{Tr}\,\left[ \Gamma \,G_{N {\cal O}_{4i} N}(t_2, t_1,p_i)\right]}
 {\mbox{Tr}\,[\Gamma\,G_{NN}(t_2,p_i)]}\hspace{4mm}
 {\overrightarrow{\hspace{\z}}}\hspace{-\z}\raisebox{2ex}{$(t_1-t_0)\gg 1,(t_2-t_1)\gg 1$}
\hspace{4mm} \langle x\rangle.
\label{ch:ciratio4i}
}
In practice, one takes a plateau in the insertion time,\ $t_1$, to define the asymptotic region.

Similarly,\ we get for the other two operators
\eq{
\frac{2 \kappa E^0_p}{(E^0_p)^2 -\frac{1}{3}p^2_i}\,\frac{\mbox{Tr}\,
\left[ \Gamma \,G_{N \tilde{{\cal O}}_{44} N}(t_2, t_1,p_i)\right]}{\mbox{Tr}\,[\,\Gamma\,G_{NN}(t_2,p_i)]}
\hspace{4mm}{\overrightarrow{\hspace{\z}}}\hspace{-\z}\raisebox{2ex}{$(t_1-t_0)\gg 1,(t_2-t_1)\gg 1$}
\hspace{4mm}\langle x\rangle,
\label{ch:ciratio44}
}
and
\eq{
\frac{2 \kappa}{p^2_i}\,\frac{\mbox{Tr}\,\left[ \Gamma \,G_{N \tilde{{\cal O}}_{4ii} N}(t_2, t_1,p_i)\right]}
 {\mbox{Tr}\, [\,\Gamma\,G_{NN}(t_2,p_i)]}\hspace{4mm}
 {\overrightarrow{\hspace{\z}}}\hspace{-\z}\raisebox{2ex}{$(t_1-t_0)\gg 1,(t_2-t_1)\gg 1$}
\hspace{4mm}\langle x^2\rangle.
\label{ch:ciratio4ii}
}

\subsubsection{Disconnected Insertions}
\label{subsubsec:ratio_disconnected_insertions}

As seen from Fig.~\ref{disconnected_fig}, the calculation of the valence quark propagators in the
disconnected insertion is separate from the loop calculation in each configuration; this means that
the momentum in the three-point function can be chosen in the nucleon two-point functions independent
of the expensive loop calculation. For the present calculation, we choose $p_i = \pm 2\pi/La$
for the three-point function for the cases of ${\cal O}_{4i}$ and
$\tilde{{\cal O}}_{4ii}$, and zero momentum for $\tilde{{\cal O}}_{44}$.

In contrast to C.I.,\ the sink time need not be fixed in D.I. We can sum over the insertion time
to gain more statistics~\cite{maiani, gupta4, liu6, viehoff1}. There are various methods by which
the summation is performed~\cite{liu8, mathur00, wilcox3, viehoff1}. All these methods, except
in~\cite{viehoff1}, considered summation over insertion time up to or beyond sink time. From
Appendix~\ref{app:threepoint}, we see that the matrix elements for the twist-two operators
are analytically zero if $t_1 < t_0$ and $t_1 > t_2$.\ (It may not be zero for other operators. Even
then it will not contribute to the physical quantity intended to be measured). So the summation outside
the nucleon source and sink times will contribute to unnecessary noise and possible contribution from
higher states which are unrelated to the target matrix element. Also, at the source and
sink time it can have contributions from the contact term. In view of the above, we shall take the sum to
be from (source time $+ 1$) to (sink time $- 1$)~\cite{viehoff1}.\ According to the derivations in
Appendix~\ref{appsubsec:threept_ratio_di} (see Eq.~(\ref{ratio_disconnected})),\ under the
condition $t_2-t_0\gg 1$,\ we get for the operator ${\cal O}_{4i}$
\eq{
\displaystyle\sum^{t_2-1}_{t_1=t_0+1} \frac{2 \kappa}{p_i}\, \frac{\mbox{Tr}\,
\left[G_{N{\cal O}_{4i} N}(t_2, t_1, p_i)\right]}{\mbox{Tr}\,[\,\Gamma\,G_{NN}(t_2, p_i)]}
\hspace{4mm}{\overrightarrow{\hspace{\y}}}\hspace{-\y}\raisebox{2ex}{$(t_2-t_0)\gg 1$}
\hspace{4mm} \langle x\rangle_{4i}\, t_2 \, + \, \mbox{const.}
\label{ch:diratio4i}
}

Similarly,\ for the other two operators,\ we get
\eq{
\displaystyle\sum^{t_2-1}_{t_1=t_0+1} \frac{2 \kappa}{E^0_p}\, \frac{\mbox{Tr}\,
\left[G_{N\tilde{{\cal O}}_{44} N}(t_2, t_1, 0)\right]}{\mbox{Tr}\,[\,\Gamma\,G_{NN}(t_2, 0)]}
\hspace{4mm}{\overrightarrow{\hspace{\y}}}\hspace{-\y}\raisebox{2ex}{$(t_2-t_0)\gg 1$}
\hspace{4mm} \langle x\rangle_{44}\, t_2 \, + \, \mbox{const.},
\label{ch:diratio44}
}

\noindent
and
\eq{
\displaystyle\sum^{t_2-1}_{t_1=t_0+1} \frac{2 \kappa}{p^2_i}\,
\frac{\mbox{Tr}\,\left[G_{N \tilde{{\cal O}}_{4ii} N}(t_2, t_1, p_i)\right]}{\mbox{Tr}\,[\,\Gamma\,G_{NN}(t_2, p_i)]}
\hspace{4mm}{\overrightarrow{\hspace{\y}}}\hspace{-\y}\raisebox{2ex}{$(t_2-t_0)\gg 1$}
\hspace{4mm} \langle x^2\rangle_{4ii}\, t_2 \, + \, \mbox{const.}
\label{ch:diratio4ii}
}

\section{Disconnected Insertion Calculation}

The D.I. calculation is the most numerically intensive part. We shall discuss
the various aspects of the calculation in more detail.

\subsection{Discrete Symmetries and Transformations}
\label{subsec:discretesymmetries}

Since the D.I. calculations are performed by using a stochastic noise estimator,\ the
signals for the current loop are always noisy.\ We can reduce the errors by making
good use of some of the discrete symmetries, specifically parity, $\gamma_5$ hermiticity,
and charge-$\gamma_5$ hermiticity ($CH$ transformation)~\cite{ber89,draper3,mathur00,nilmani1}.

By applying these symmetries and transformations,\ one can then work out the effects of gauge
averaging in order to find out the correct part (e.g.\ even or odd parity,\ real or
imaginary part, etc.) of the two-point and three-point correlation
functions,\ current operators etc., and discard the irrelevant part.

\subsubsection{Two-point Functions}

Since in the case of D.I. the three-point functions are constructed by multiplying
the nucleon propagator with the current loop in each gauge configuration (Fig.~\ref{disconnected_fig}),\
we have the advantage of taking into account nucleons propagators with equal and opposite momenta in
order to increase statistics.\ While doing so,\ we have to consider the fact that
such combinations have appropriate parity and have appropriate real or imaginary part
w.r.t. the loop. In Table~\ref{tab:discrete_trans_nuc},\ we show the effect of parity and
$CH$ transformations on various such combinations.\ We denote
$\mbox{Tr}\,\left[ \Gamma \, G_{NN} (t,\vec{p}; U) \right]$ to be the nucleon propagator
on each gauge-configuration with momentum $\vec{p}$.

\begin{table}[h]
\centering
\begin{tabular}{|c|c|c| }
\hline\hline
\multirow{2}{*}{\bf Nucleon Propagators}& $\bm CH$ & \multirow{2}{*}{\bf Parity} \\
& {\bf Transformations} & \\
\hline\hline
&&\\
$\mbox{Tr}\,[\Gamma \, G_{NN} (t,\vec{p}; U)]$ & $\mbox{Tr}\,[\Gamma \, G_{NN} (t,-\vec{p}; U^*)]^*$ & $\mbox{Tr}\,[\Gamma \, G_{NN} (t,-\vec{p}; U^p)]$   \\
\hline
&&\\
$\mbox{Tr}\,\left[ \Gamma \, G_{NN} (t,\vec{p}; U) \right]$ & $ - \Big\{\mbox{Tr}\,\left[\Gamma\, G_{NN} (t,\vec{p}; U^*)\right]$ & \multirow{3}{*}{Odd} \\
&&\\
$ - \mbox{Tr}\,\left[ \Gamma \, G_{NN} (t,- \vec{p}; U) \right]$ & $ - \mbox{Tr}\,\left[ \Gamma \, G_{NN} (t,- \vec{p}; U^*)\right] \Big\}^*$ & \\
\hline
&&\\
$\mbox{Tr}\,\left[ \Gamma \, G_{NN} (t,\vec{p}; U) \right]$ & $\Big\{\mbox{Tr}\,\left[ \Gamma \, G_{NN} (t,\vec{p};U^*)\right] $
&  \multirow{3}{*}{Even}\\
&&\\
$+\mbox{Tr}\,\left[ \Gamma \, G_{NN} (t,- \vec{p}; U) \right] $ & $+ \mbox{Tr}\,\left[ \Gamma \, G_{NN} (t,- \vec{p};U^*)\right]\Big\}^* $ & \\
\hline\hline
\end{tabular}
\caption{Table showing the outcome of the parity and $CH$ and parity transformations on the combinations of
nucleon propagators with equal and opposite momenta. $U^p$ is the parity transformed gauge link.}
\label{tab:discrete_trans_nuc}
\end{table}

\subsubsection{Current Loop}

The outcome of the parity,\ $\gamma_5$ hermiticity, and $CH$ transformations for
the operators  ${\cal O}_{4i}$, $\tilde{{\cal O}_{44}}$, and  $\tilde{{\cal O}}_{4ii}$
on each gauge configuration are shown in Table~\ref{tab:discrete_trans_loop}. The notation
``Im'' includes the factor $i$ of the imaginary part of each operator.

\begin{table}[h]
\centering
\begin{tabular}{|c|c|c|c|c|}
\hline\hline
\multirow{3}{*}{\bf Loops} & \multirow{3}{*}{\bf Parity} & \multirow{2}{*}{$\bf \gamma_5$} & \multirow{2}{*}{$\bm CH$}& {$\bf \gamma_5$ \bf Hermiticity} \\
& &\multirow{2}{*}{\bf Hermiticity } & \multirow{2}{*}{\bf Transformations} & {\bf \& $\bm CH$ Transformations} \\
&&&& {\bf Combined} \\
\hline\hline
&&&&\\
${\cal O}_{4i}$ & Odd & Imaginary & $[\mbox{L}]_U = - [\mbox{L}_{U^*}]^*$ & $\mbox{Im\,[L]}_U = \mbox{Im\,[L]}_{U^*}$\\
\hline
&&&&\\
${\cal O}_{44}$ & \multirow{2}{*}{Even} & \multirow{2}{*}{Real} & \multirow{2}{*}{$[\mbox{L}]_U = [\mbox{L}_{U^*}]^*$} & \multirow{2}{*}{$\mbox{Re\,[L]}_U = \mbox{Re\,[L]}_{U^*}$} \\
$-\frac{1}{3}\left({\cal O}_{11}+{\cal O}_{22}+{\cal O}_{33}\right)$ &&&&\\
\hline
&&&&\\
${\cal O}_{4ii}-\frac{1}{2}\left({\cal O}_{4jj}+{\cal O}_{4kk}\right)$ & Even & Imaginary& $[\mbox{L}]_U = [\mbox{L}_{U^*}]^*$ & $\mbox{Im\,[L]}_U = - \mbox{Im\,[L]}_{U^*}$\\
\hline\hline
\end{tabular}
\caption{Table showing the outcome of the parity transformations,\ $\gamma_5$
Hermiticity and $CH$ Transformations on current loops.\ L stands for current loop and ``Im'' includes the factor $i$
of the imaginary part.}
\label{tab:discrete_trans_loop}
\end{table}

\subsubsection{Correlations between the Nucleon Propagator and the Loop}

When determining the correlation between the appropriate parts of the nucleon propagator and the loop,
we have to consider the three-point function as a whole. As an example, if we consider
two-point functions with both the momenta $\pm p_i$ for the operator ${\cal O}_{4i}$, then
from Eq.~(\ref{ch:diratio4i}), we get
\eqarray{
\displaystyle\sum^{t_2-1}_{t_1=t_0+1} \frac{\kappa}{p_i}\, \frac{\mbox{Tr}\,
        \left[G_{N{\cal O}_{4i}  N}(t_2, t_1, p_i)\right]-\mbox{Tr}\,\left[G_{N{\cal O}_{4i}  N}(t_2, t_1, -p_i)\right]}
        {\mbox{Tr}\,[\,\Gamma\,G_{NN}(t_2, p_i)]}&=& \langle x\rangle\, t_2 + \mbox{const.}
\label{bothmom4i}
}
So, for ${\cal O}_{4i}$ which is odd in parity, one can explicitly consider the odd combination of
the nucleon propagators in Table~\ref{tab:discrete_trans_nuc}, and similarly,
the even combinations for $\tilde{{\cal O}}_{44}$ and $\tilde{{\cal O}}_{4ii}$, to reduce noise.
According to the $CH$ theorem~\cite{ber89,draper3,mathur00}, the path integral for $\langle {\cal O}\rangle$
in QCD is either real or imaginary (except in the case with chemical potential). It can be shown from
Tables~\ref{tab:discrete_trans_nuc} and \ref{tab:discrete_trans_loop} that if we combine the $CH$
transformation of the nucleon propagators (with the appropriate combination or parity) and the loop,
the three-point  functions transform to the positive complex conjugate of themselves on the link $U$
for all the three D.I. cases considered here. This, according to the
$CH$ theorem, means that the three-point functions are real. Since, by using $\gamma_5$ hermiticity we see
that the loops are either real or imaginary (Table~\ref{tab:discrete_trans_loop}); one needs to multiply
them with only the real or imaginary part of the nucleon propagators to make the three-point functions real.
For example, the L.H.S. of Eq.~(\ref{bothmom4i}) can be written as
\eqarray{
& &\frac{1}{2}\,\left[\mbox{Tr}\,\left[G_{N{\cal O}_{4i}  N}(t_2, t_1, p_i)\right]
        - \mbox{Tr}\,\left[G_{N{\cal O}_{4i}  N}(t_2, t_1, -p_i)\right]\right]\nonumber\\
        & & \nonumber\\
&=&\frac{1}{2N_g}\displaystyle\sum_j\bigg[\frac{1}{2}\,
         \left[\mbox{Tr}\,\left[\Gamma \, G_{NN}(t,p_i; U^j) \right] -
         \mbox{Tr}\,\left[\Gamma \, G_{NN}(t,-p_i; U^j) \right]\right]
         \times\mbox{Im} (\mbox{Loop})_{U^j} , \nonumber\\
&+& \frac{1}{2}\,\left[\mbox{Tr}\,\left[\Gamma \, G_{NN} (t,p_i; U^{*j}) \right]
         - \mbox{Tr}\,\left[\Gamma \, G_{NN} (t,-p_i; U^{*j})\right]\right]
        \times\mbox{Im} (\mbox{Loop})_{U^{*j}}\bigg]\nonumber\\
        & & \nonumber\\
&=&\frac{1}{2N_g}\displaystyle\sum_j\,\mbox{Im}\,
        \left\{\mbox{Tr}\,\left[ \Gamma \, G_{NN} (t,p_i; U^j) \right]
        - \mbox{Tr}\,\left[ \Gamma \, G_{NN} (t,-p_i; U^j) \right]\right\}\times\mbox{Im}(\mbox{Loop})_{U^j}.
}
where $N_g$ is the number of gauge configurations. As illustrated above, one can exclude the real part of the
nucleon propagator and the real part of the loop for ${\cal O}_{4i}$ operator which contributes to noise with
finite number of noise vectors. The utilization of $\gamma_5$ hermiticity and $CH$ theorem has been shown to
reduce noise effectively in the loop calculation of the quark angular momentum~\cite{mathur00}.
Similar procedures can be applied to the other two operators. We show in
Table~\ref{tab:loop_2pt_part} the relevant parts of the nucleon propagator(s) and the corresponding
loop in each row which are to be correlated on each gauge configuration.

\begin{table}[h]
\centering
\begin{tabular}{|c|c|}
\hline\hline
{\bf Nucleon Propagators} & {\bf Loops} \\
\hline\hline
&\\
$\mbox{Im} \left[\mbox{Tr}\,\left[ \Gamma \, G_{NN} (t,\vec{p}; U) \right] -
\mbox{Tr}\,\left[ \Gamma \, G_{NN} (t,- \vec{p}; U) \right]\right]$ & $\mbox{Im}[{\cal O}_{4i}]$ \\
\hline
&\\
$\mbox{Re}\left[\mbox{Tr}\,\left[ \Gamma \, G_{NN} (t,0; U)\right]\right]$ &
$\mbox{Re}\left[{\cal O}_{44}-\frac{1}{3}\left({\cal O}_{11}+{\cal O}_{22}+{\cal O}_{33}\right)\right]$\\
\hline
&\\
$\mbox{Im}\left[\frac{1}{2}\,\left[\mbox{Tr}\,\left[ \Gamma \, G_{NN} (t,p_i; U) \right]
+ \mbox{Tr}\,\left[ \Gamma \, G_{NN} (t,-p_i; U) \right]\right]\right]$ &
$\mbox{Im}\left[{\cal O}_{4ii}-\frac{1}{2}\left({\cal O}_{4jj}+{\cal O}_{4kk}\right)\right]$\\
\hline\hline
\end{tabular}
\caption{The relevant parts considered for D.I. calculations for the nucleon propagator
and the corresponding loops based on parity, $\gamma_5$ hermiticity, and CH theorem.}
\label{tab:loop_2pt_part}
\end{table}

\subsection{Complex $Z_2$ Noise and Unbiased Subtraction Method}
\label{sub:complexz2noise}

As we mentioned in Sec.~\ref{subsec:twopoint}, it is a numerical challenge
to evaluate the quark loop. We shall adopt the complex $Z_2$ noise with unbiased
subtraction to calculate it.

The basic idea of the complex $Z_2$ noise method~\cite{liu5}
is to construct $L$ noise vectors, $\eta^1, \eta^2......\eta ^L$
(each of dimension $N\times 1$),\ where $\eta^j={\{\eta_1^j,\eta_2^j,....\eta_N^j}\}$,\
in order to stochastically estimate the inversion of an $N\times N$ matrix.\ Each element $\eta^j_n$ takes
one of the four values,\ $\displaystyle\frac{1}{\sqrt{2}}\{\pm 1 \pm i\}$,\ chosen independently
with equal probability.\ They have the properties of a white noise
\eqarray{
\langle \eta_i\rangle=\lim_{L \rightarrow \infty}
      \frac{1}{L}\displaystyle\sum_{n=1}^L \eta_i^n=0, \hspace{5mm}
      \langle \eta_i^\dag\eta_j\rangle = \lim_{L \rightarrow \infty }
      \frac{1}{L}\displaystyle\sum_{n=1}^L\eta_i^{\dag n}\eta_j^n =
       \delta_{ij},\ \hspace{5mm} \eta_i^\dag\eta_i=1.
\label{z_2prop}
}

\noindent
Then,\ the expectation value of the matrix element $M^{-1}_{ij}$ is
obtained by solving for $X_i$ in the matrix equations $MX=\eta$, so that
\eqarray{
 E[M^{-1}_{ij}]=\langle \eta_j^\dag X_i\rangle = \displaystyle\sum_k M^{-1}_{ik}
      \langle \eta_i^\dag\eta_k\rangle = M^{-1}_{ij}.
}

It has been shown~\cite{bernardson,liu5,thron} that the variance
corresponding to the estimator is given by
\eqarray{
\sigma^2_M = \frac{1}{L} \displaystyle\sum^N_{m\neq n}\bigg{|} M^{-1}_{m, n} \bigg{|}^2,
\label{eq:variance}
}
which is minimal, since it does not involve the positive contribution from the diagonal matrix
elements as do in other type of noises, such as the Gaussian noise. It is in this sense,
$Z_2$ noise is considered optimal~\cite{bernardson,liu5,thron}. Similarly, one can show that
$Z(N)$ and $U(1)$ noises are also optimal.

It has been further shown that the off-diagonal matrix element contributions to the variance in
Eq.~(\ref{eq:variance}) can be reduced by subtracting a judiciously
chosen set of traceless $N\times N$ matrices $Q^{(p)}$~\cite{thron, mathur00,nilmani1},\ which satisfy
$\displaystyle\sum^N_{n=1}Q^{(p)}_{n, n} =0,\ p=1,\cdots, P$.\ Then the expectation value is unchanged
when $M^{-1}$ is substituted with  $M^{-1} - \displaystyle\sum^P_{p=1}\lambda_p\, Q^{(p)}$
($\lambda_p$ is a constant)
\eqarray{
E [\langle\, \eta^\dag\, (M^{-1} - \displaystyle\sum^P_{p=1}\lambda_p\, Q^{(p)})\, \eta\, \rangle ]
      &=& \mbox{Tr}\,  M^{-1},
}
while the variance becomes
\eqarray{
\sigma^2_M = \mbox{Var} [\langle\, \eta^\dag\, (M^{-1} - \displaystyle\sum^P_{p=1}\lambda_p\, Q^{(p)})\, \eta\, \rangle]
      &=& \frac{1}{L} \displaystyle\sum^N_{m\neq n}\left| M^{-1}_{m, n} -
      \displaystyle\sum^P_{p=1}\lambda_p\, Q^{(p)}_{m, n}\right|^2.
}
So,\ with a judicious choice of traceless matrix,\ the variance may be reduced when the
off-diagonal matrix elements of $Q^{(p)}$ are correlated with those of $M^{-1}$.\
This subtraction is unbiased,\ because it doesn't change the expectation value
of the trace.\ The natural choice for the set of traceless matrices is the hopping
parameter expansion of the inverse of the Wilson fermion matrix,\ $M$~\cite{thron}, given by
\eqarray{
 M^{-1}& =& I + \kappa D + \kappa^2 D^2 + \kappa^3 D^3 + \cdots,
\label{hopping_expansion}
}
where
\eqarray{
D_{x, y} &=& \displaystyle\sum^4_{\mu=1}\, \bigg{[} (1-\gamma_\mu)_{\alpha\beta}\,
         U^{ab}_\mu (x)\, \delta_{x, y - a_\mu}\, + \,   (1+\gamma_\mu)_{\alpha\beta}\,
         U^{\dag ab}_\mu (x - a_\mu)\, \delta_{x, y + a_\mu} \bigg{]},
\label{hopping}
}
which is off-diagonal in space-time.

In evaluating the quark loop, we shall consider the following subtraction
\eq{
\langle\, \eta^\dag\, Q'\, [ M^{-1} - (I + \kappa D + \kappa^2 D^2 + \kappa^3 D^3 + ...) ]\,
\eta\, \rangle  + \mbox{traces},
\label{subtraction}
}
where $Q'$ represent operators between the fermion fields in
${\cal O}_{4i}, \tilde{\cal{O}}_{44}$, and $\tilde{\cal{O}}_{4ii}$.
Traces may arise while combining the operators $Q'$ with the hopping expansion in
Eq.~(\ref{hopping_expansion}), leading to  non-zero traces at certain power of $D$. In this case,
one needs to evaluate these traces and add them back to cancel out those which are subtracted out
in the noise estimation. The details of the trace calculations for these operators are given in
Appendix~\ref{app:subtraction}.\ Since the calculations of traces are cumbersome for large loops,
we will restrict ourselves only up to $\kappa^4 D^4$ for two-index operators and
$\kappa^3 D^3$ for three-index operators. We shall now discuss these trace terms here.

\subsubsection{Traces for Two-Index Operators $O_{4i}$ and $O_{\mu\mu}$}
\label{subsubsec:sub_traces_two_index}

Since $Q'$ for ${\cal O}_{4i}$ and ${\cal O}_{\mu\mu}$ are point-split, the trace for the first subtraction term,
i.e.\ $\mbox{Tr}(Q'\,{\bf I})$, is zero. So we can use it for subtraction.\ However,\ from
Appendix~\ref{app:subtraction},\ we see that the noise estimate is real for ${\cal O}_{4i}$ while
the exact loop  $\mbox{Tr}(Q' M^{-1})$ is imaginary (see Sec.~\ref{subsec:discretesymmetries}).
Similarly, the noise estimate is imaginary for ${\cal O}_{\mu\mu}$ while the exact loop
$\mbox{Tr}(Q' M^{-1})$ is real.\ Therefore, there
is no use to include this subtraction term. The subtraction with $\kappa^2 D^2$ and $\kappa^4 D^4$ terms
in the hopping expansion are also traceless, since the multiplication of $Q'$ to these terms does not lead to a
plaquette which is the lowest order in $\kappa$ that contributes to a trace.\ On the other hand, the $\kappa D$
and  $\kappa^3 D^3$ terms can lead to a plaquette, thus can have non-zero traces. Since the operators,\
${\cal O}_{4i}$ and ${\cal O}_{\mu\mu}$,\ lead to different traces,\ we shall consider them separately.

In the case of ${\cal O}_{4i}$, the contribution from the hopping expansion terms $\kappa D$, $\kappa^2 D^2$,\ $\kappa^3 D^3$ and
$\kappa^4 D^4$ are traceless (Appendix~\ref{app:subtraction}).\ One does not have to worry about
the complication of having to add back the trace contributions.

\begin{figure}[h]
   \centering
  \subfigure[]
         {\label{subtract_two_index}
           \rotatebox{0} {\includegraphics[width=0.46\hsize]{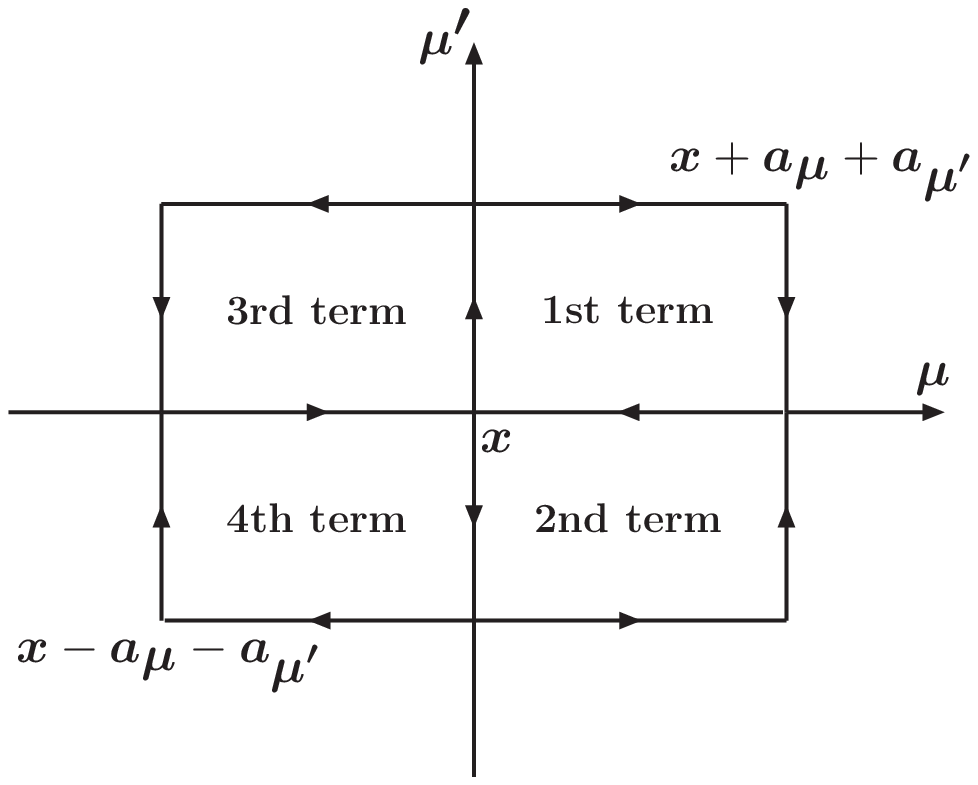}}}
        \hspace{0.4cm}
   \subfigure []
          {\label{subtract_three_index}
        \rotatebox{0}{\includegraphics[width=0.46\hsize]{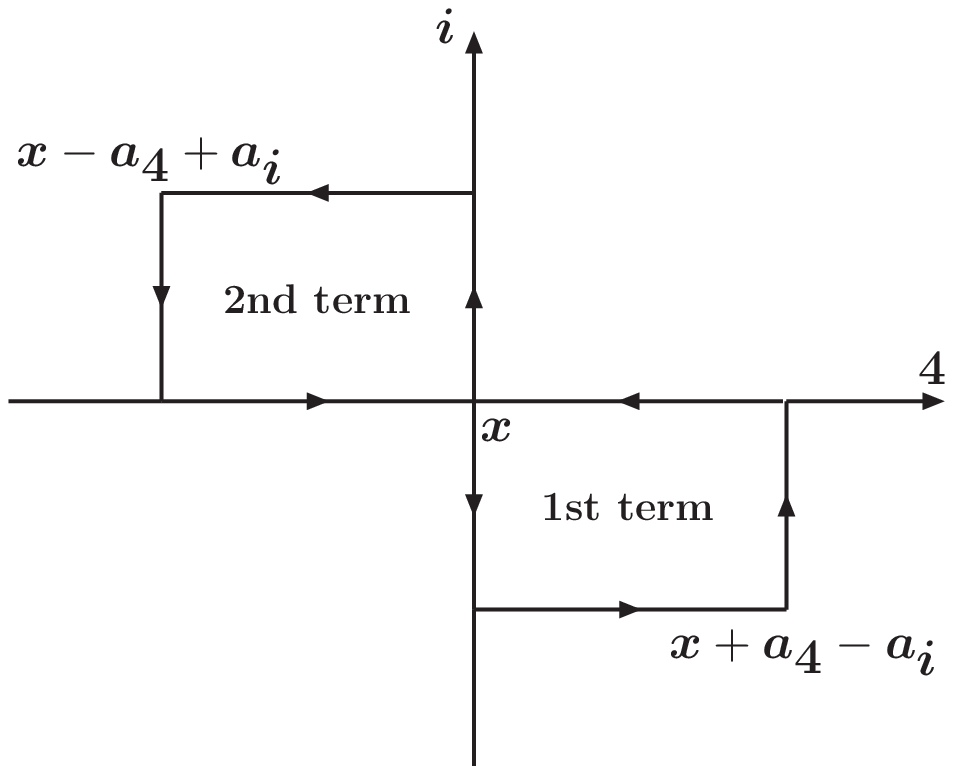}}}
\caption{Plaquette terms (a) for the operator ${\cal O}_{\mu\mu}$ when $\kappa^3 D^3$
term is considered and (b) for ${\cal O}_{4ii}$ when $\kappa^2 D^2$ term is considered.}
\label{ch:ubs_two_three}
\end{figure}

But,\ ${\cal O}_{\mu\mu}$ operator has non-zero traces when combined with $\kappa D$
and $\kappa^3 D^3$ (see Appendix~\ref{app:subtraction} for a full derivation).\ The
trace for the  $\kappa D$ term is $\displaystyle\frac{\lambda}{8a} 96\kappa V_3$,\ where $V_3$ is
the three-volume of the lattice used, and $\lambda = +1$ for $\mu = \nu = 1,2,3$ and
$\lambda = -1$ for $\mu = \nu = 4$.\ And,\ for $\kappa^3 D^3$
[see Fig.~\ref{subtract_two_index}],\ it is
\eqarray{
          & & \frac{\lambda}{8a} 32\kappa^3 \displaystyle\sum_{\vec{x}_1} \, \displaystyle\sum_{m}\,
          \displaystyle\sum_{\mu^\prime\neq\mu}\mbox{Re}\bigg{[}\{U_{\mu^\prime} (x_1+a_\mu)
          U^\dag_\mu (x_1+a_{\mu^\prime})U^\dag_{\mu^\prime}(x_1)U_ \mu(x_1)  \nonumber\\
          & &\nonumber\\
          &+&U_\mu (x_1) U^\dag_{\mu^\prime} (x_1+a_\mu-a_{\mu^\prime})U^\dag_{\mu}(x_1-a_{\mu^\prime})
          U_{\mu^\prime} (x_1- a_{\mu^\prime})  \nonumber\\
          & &\nonumber\\
          &+&U_\mu(x_1-a_{\mu}+a_{\mu^\prime})U^\dag_{\mu^\prime}(x_1)U^{\dag}_\mu (x_1-a_\mu)
          U_{\mu^\prime} (x_1-a_\mu)\nonumber\\
          & &\nonumber\\
          &+& U_{\mu^\prime} (x_1 - a_{\mu^\prime})U^{\dag}_\mu (x_1-a_\mu)
          U^\dag_{\mu^\prime}(x_1-a_\mu-a_{\mu^\prime})U_{\mu} (x_1-a_\mu-a_{\mu^\prime})\}^{mm}\bigg{]}.
\label{ch:3pref44}
}
These traces need to be added as shown in Eq.~(\ref{subtraction}).

\subsubsection{Traces for Three-Index Operators}
\label{subsubsec:sub_traces_three_index}

For the three-index operator,\ the noise estimator is real for the first term,\ ${\bf I}$.
But from $\gamma_5$ hermiticity we see that the loop for ${\tilde{\cal O}}_{4ii}$ is imaginary.
Thus, it is not useful to consider this term in subtraction. Up to $\kappa^3 D^3$, the only trace
contribution comes from the $\kappa^2 D^2$ term, which is (see
Appendix~\ref{app:subtraction} for a full derivation)
\eqarray{
       & &\frac{1}{24 a^2} 16 \kappa^2 \displaystyle\sum_{\vec{x}_1}
       \bigg{[}\displaystyle\sum_{m} \mbox {Im} \,\{U^\dag_4(x_1-a_i)
       U_i (x_1-a_i)U_4(x_1)U^{\dag}_i(x_1+a_4-a_i)\}^{mm} \nonumber\\
       & &\nonumber\\
       &+&\displaystyle\sum_{m} \mbox {Im} \,\{U_4(x_1-a_4+a_i)
       U^\dag_i (x_1)U^{\dag}_4(x_1-a_4)U_i(x_1-a_4)\}^{mm}\,\bigg{]}.
}
The graphical representation for this trace is illustrated in Fig.~\ref{subtract_three_index}.

\subsubsection{Numerical Test}
\label{subsubsec:numerical_test}

It is necessary to check that the analytical calculations for trace for each subtraction term
is correct.\ In order to do that we have to calculate the subtraction terms exactly.\ We did it on a smaller
($4^4$) lattice as described below.

From the matrix equation $MX = \eta$, we have
\begin{eqnarray}
X\,\eta^\dag &=& M^{-1}\,\eta\, \eta^\dag\, \nonumber\\
                  &=& (I + \kappa D + \kappa^2 D^2 + \kappa^3 D^3 + \kappa^4 D^4 + \cdots )
                  \eta \,\eta^\dag.
\label{numtest1}
\end{eqnarray}
From Eq.~(\ref{numtest1}) we see that we can exactly calculate $M^{-1}$\ {\it i.e.} each subtraction
term, only if $\eta \,\eta^\dag$ is a unit matrix. By choosing the following orthogonal set of vectors
$\{\eta_i\}$
\eqarray{
\{\eta_i\} = \left\{
\bordermatrix{
& \cr
& 1 \cr
& 0 \cr
& 0 \cr
& \vdots\cr
& \vdots\cr} \hspace{1mm}, \hspace{2mm}
\bordermatrix{
& \cr
& 0 \cr
& 1 \cr
& 0 \cr
& \vdots\cr
& \vdots\cr}\hspace{1mm},\hspace{2mm}
\bordermatrix{
& \cr
& 0 \cr
& 0 \cr
& 1 \cr
& \vdots\cr
& \vdots\cr}\hspace{1mm},\hspace{2mm}
\cdots\hspace{1mm},\hspace{2mm}
\bordermatrix{
&\cr
& \vdots\cr
 & \vdots\cr
& 0 \cr
& 0 \cr
& 1 \cr}
\right\},
}
we obtain $\displaystyle \eta \,\eta^\dag = \frac{1}{L} \sum^L_i \eta_i \,\eta^\dag_i = {\mathbbm 1}$.\
Using these vectors, we compute ${\bf I}, \kappa D,  \kappa^2 D^2, \cdots$ for each operator. We have
found that our numerical results matched with those of analytical expressions e.g. for ${\cal O}_{4i}$ operator,
and $\kappa D$, $\kappa^2 D^2$,\ $\kappa^3 D^3$ and $\kappa^4 D^4$ terms have no traces.

\section{Renormalization}
\label{sec:renormalization}

The physical matrix elements,\ which are determined from a linear extrapolation to the chiral limit,\
are extracted in lattice units from the Monte Carlo calculation.\ In order to relate to the experimental
values,\ they have to be expressed in physical units and renormalized at a certain scale. \ The
renormalized operators,\ at a finite energy scale $\mu$,\ are related to the bare lattice operators
through the renormalization constant
\eqarray{
{\cal O}(\mu)&=& Z_{{\cal O}}(a\mu, g(a))\, {\cal O}(a),
\label{renorm}
}
where $g$ is the bare coupling constant,\ which is equal to one in our case.\ The
renormalization constants,\ $Z$'s,\ are computed by using perturbation theory.\ We will use the
values of $Z$ factors,\ which are computed in~\cite{gockeler1} by using tadpole improved
perturbation theory~\cite{lapage1}.\ Since the experimental results are often renormalized in the
$\overline{\mbox{MS}}$ scheme,\ we will also use the calculated $Z$ factors matched to this scheme.\ In
the quenched approximation,\ the renormalization constants are
\eqarray{
Z_{{\cal O}}(a\mu, g^*)&=& \frac{u_0}{u^{n_D}_0}\, \left(1  - \frac{g^{*2}}{16\pi^2}\,
           C_F \left(\gamma_{{\cal O}}\, \ln(a\mu)  +  B^{\overline{\mbox{MS}}}_{{\cal O}}
         + (n_D - 1)\, 8 \pi^2 Z_0\right)  +  O\left(g^{*4}\right)\right),
\label{zfac1}
}
where
\eqarray{
u_0 = \langle \frac{1}{3} \, \mbox{Tr}\, U_{\mbox{plaq}}\rangle^{\frac{1}{4}}\,,\, \,
g^{*2} = \frac{g^2}{u^4_0},\ \ C_F = \frac{4}{3},\ \ Z_0=0.155,
}
and $\gamma_{\cal O}$ is the anomalous dimension of the operator,\
$\overline{B}_{{\cal O}}$ is the finite part of $Z_{{\cal O}}$,\ and  $n_D$ is the number
of covariant derivative(s) in the operator. We list $\gamma_{\cal O}$ and
$\overline{B}_{\cal O}^{\overline{MS}}$ in Table~\ref{renormbo} for the three operators
we consider.

\begin{table}[!hbtp]
\centering
{\begin{tabular}{|c|c|c|}
\hline\hline
{\bf Operators} & $\gamma_{{\cal O}} $ & {\bf $B^{\overline{\mbox{MS}}}_{{\cal O}}$}  \\
\hline
&&\\
${\cal O}_{4i}$ & $\frac{16}{3}$   & 1.279 \\
\hline
&& \\
${\tilde{\cal O}}_{44}$ & $\frac{16}{3}$  &  2.561  \\
\hline
&& \\
${\tilde{\cal O}}_{4ii}$ & $\frac{25}{3}$ &  -12.128 \\
\hline\hline
\end{tabular}
}
\caption{The values of  $\gamma_{{\cal O}} $ and $B^{\overline{\mbox{MS}}}_{{\cal O}}$ for all the
three operators under consideration.}
\label{renormbo}
\end{table}

Our inverse lattice spacing is determined to be $1/a=1.74$ GeV~\cite{mathur00} by using nucleon mass.\
The values of $B^{\overline{\mbox{MS}}}_{{\cal O}}$'s are given in Table~\ref{renormbo}.\ The value
of $u_0 = \langle \frac{1}{3} \, \mbox{Tr}\, U_{\mbox{plaq}}\rangle^{\frac{1}{4}}=0.88$ is obtained
from Ref.~\cite{liu7}.\ By using all the relevant factors for a particular operator,\ we get the
renormalization factors at $\mu = 2$ GeV ($\mu^2 = 4 \mbox{GeV}^2)$ scale for the following
operators as
\eqarray{
{\cal O}_{4i}: \hspace{2mm}Z_{{\cal O}_{4i}}&=&0.972, \nonumber\\
{\tilde{\cal O}}_{44}: \hspace{2mm}Z_{{\cal O}_{44}}&=& 0.953,\nonumber\\
{\tilde{\cal O}}_{4ii}: \hspace{2mm}  Z_{{\cal O}_{4ii}} &=& 1.116.
\label{renorm1.74}
}

Also,\ in the tadpole-improved mean-field approach,\ there is a finite $ma$ correction factor $f$ for the
Wilson fermion in the case of fermion bilinear operators~\cite{gusken1, gupta4, liu7}
\eqarray{
f = \frac{e^{m_q a}}{8\kappa_c}\frac{1}{\langle \frac{1}{3} \,
\mbox{Tr}\, U_{\mbox{plaq}}\rangle^{1/4}}\,,\hspace{5mm}m_q a=\ln (\frac{4\kappa_c}{\kappa}-3),
\label{tadpole_fac}
}
with the critical $\kappa_c=0.1568$~\cite{liu7}.\ The values of $f$ for $\kappa = 0.154,0.155$, and 0.1555
(which are used for our calculation) are 0.972, 0.948, and 0.936, respectively.


\section{Numerical Parameters and Error Studies}

We use 500 gauge configurations on a $16^3 \times 24$ lattice
generated with Wilson action at $\beta=6.0$ in the quenched approximation.
They are produced by the pseudo-heatbath algorithm with 10,000 sweeps between
consecutive configurations. The values of the hopping parameter we have used are
$\kappa = 0.154,\ 0.155$ and 0.1555.\ The critical hopping parameter,\
$\kappa_c = 0.1568$ is obtained by a linear extrapolation to the zero pion
mass~\cite{liu7}. Using the nucleon mass to set the lattice
spacing at $a = 0.11$ fm, the corresponding pion
masses are 650(3),\ 538(4), and 478(4) MeV,\ and the nucleon
masses are 1291(9),\ 1159(11), and 1093(13) MeV,\ respectively.\ We used
Dirichlet boundary condition in the present work.
We should note that there is a large uncertainty in determining the
lattice spacing in the quenched approximation, as much as
$\sim 20\%$. For example, using $r_0 = 0.5$ fm to set the scale, the
lattice spacing would be $a= 0.09$ fm. Thus, using this scale, the
dimensionful quantities will be shifted by $\sim 20\%$. Since we
are calculating the moments $\langle x \rangle$ and
$\langle x^2 \rangle$ which are dimensionless, the results we report
here will not depend on the scale of the lattice spacing except for
the renormalization constant which has an negligible difference between
these two lattice spacings.

\begin{itemize}

\item For the connected insertion,\ we have chosen the number of independently
generated gauge configurations to be 200 and for disconnected
insertion it is 500.\ The maximum number of noise used is 400 for each gauge
configuration.

\item  The error analysis has been performed by using the jackknife
procedure~\cite{efron, berg}.\ Since the computations for all the
time slices and quark masses has been performed using the same set of gauge
configurations,\ the data we obtain are correlated in Euclidean time.\ The correlation
among different quantities are taken into account by constructing the corresponding
covariance matrices~\cite{degrand1, detar1, gottileb1}. The error bars
are obtained by using this method. In order to extract various physical
quantities,\ we have used correlated least-$\chi^2$ fits.

\end{itemize}

\subsection{Error Studies in DI}
\label{error-study}

Since noise estimate plays an essential role for the case of the
disconnected insertion,\ we shall study the effect of number
of noise and gauge configurations on the error for the signal we are
extracting.\ We will present a few results of such studies before
discussing the results for the moments of the quark distribution.

The standard error due to the $Z_2$ noise estimation of the loop
averaged over the gauge configurations is given
by~\cite{wilcox2,wilcox4,wol02}
\eqarray{
\sigma=\sqrt{\frac{\sigma^2_g}{N_g} + \frac{\sigma^2_n}{N_n N_g}}\,\, .
\label{z2errorbarresults}
}
where\ $\sigma^2_g$ is the variance of the ensemble of
gauge configurations,\ $\sigma^2_n$ is the variance coming from the
noise estimator,\ and $N_g$ and $N_n$ are number of gauge
configurations and noise estimator respectively.

To make sure that our results are generated correctly, we verify that
their errors are in conformity with Eq.~(\ref{z2errorbarresults})
and extract the standard deviations $\sigma_g$ and $\sigma_n$ from the
existing data with and without unbiased subtractions..

\begin{figure}[h]
   \centering
  \subfigure[]
         {\label{confvserrornosub}
           \rotatebox{270} {\includegraphics[width=5.5cm,height=0.46\hsize]{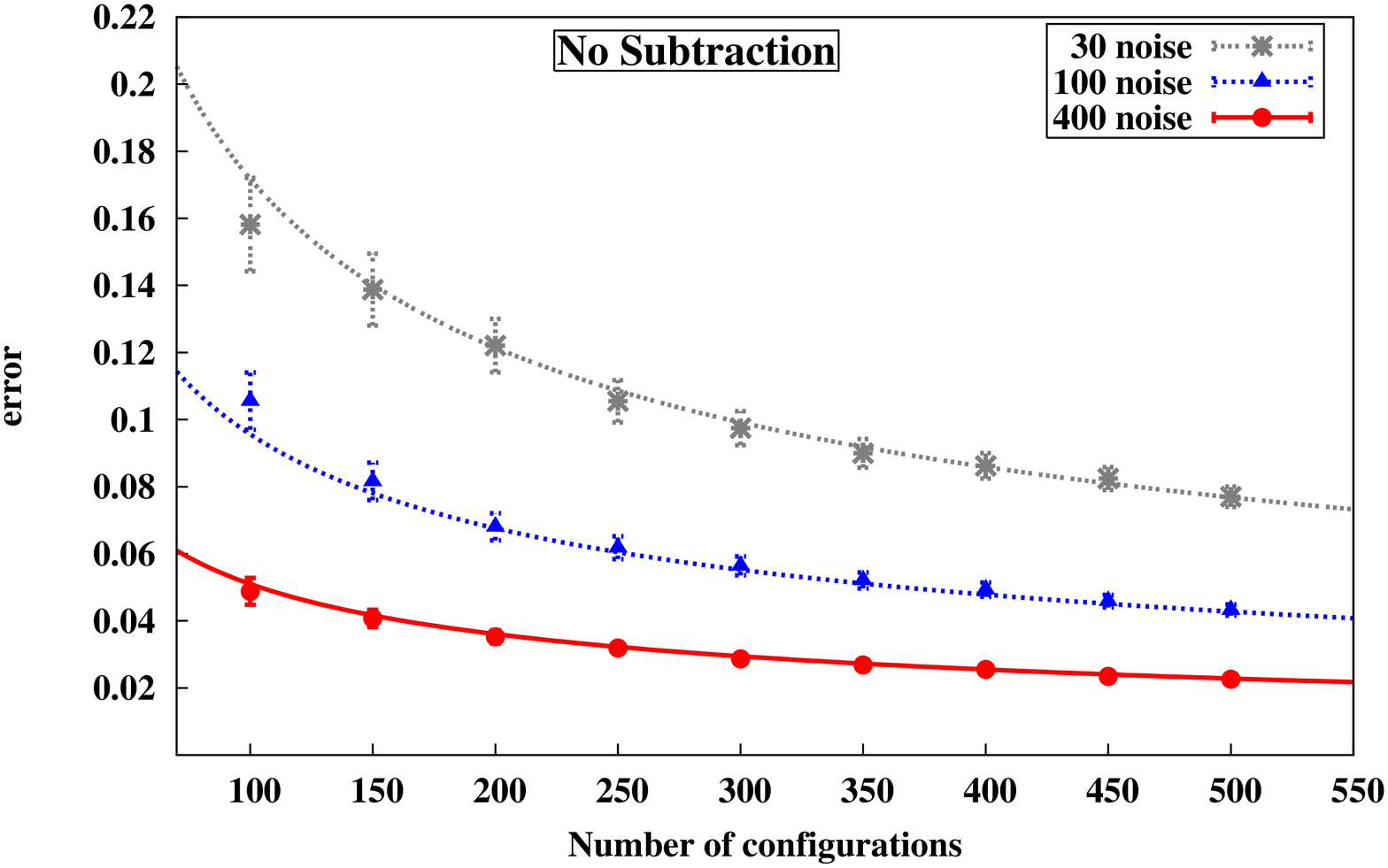}}}
       \hspace{0.5cm}
   \subfigure []
          {\label{confvserrorwsub}
        \rotatebox{270}{\includegraphics[width=5.5cm,height=0.46\hsize]{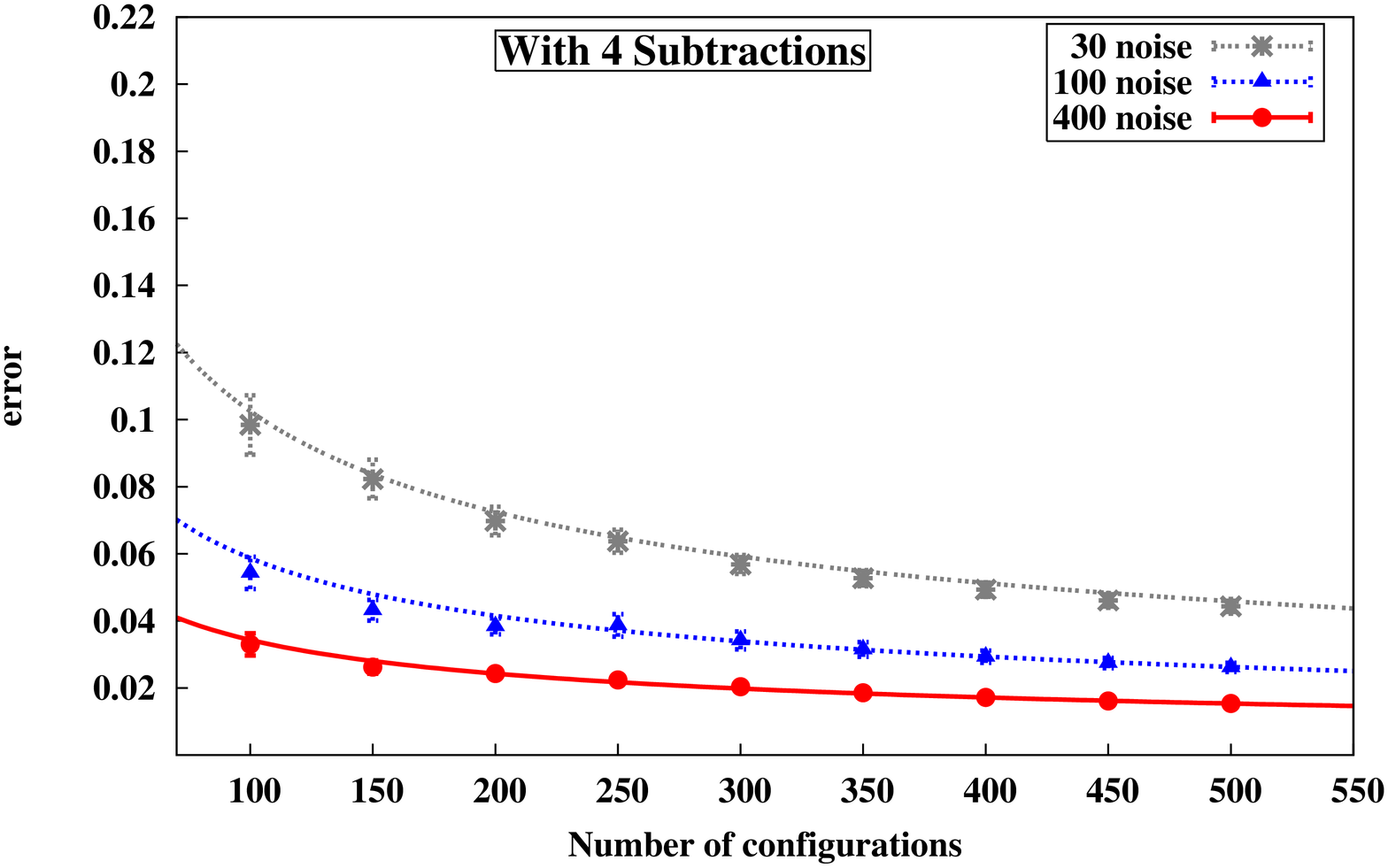}}}
          \caption{Errors of the noise estimation plotted against the number of configurations for different sets of noise
vectors for the loop part of the current,\ ${\cal O}_{4i}$ at $\kappa_s = 0.154$ nd insertion time,\ $t_1 = 14$
(a) without subtraction and (b) with four subtraction terms.}
          \label{fig:sub_1}
\end{figure}

\begin{figure}[h]
   \centering
  \subfigure[] 
         {\label{noisevserrornosub}
           \rotatebox{270} {\includegraphics[width=5.5cm, height=0.46\hsize]{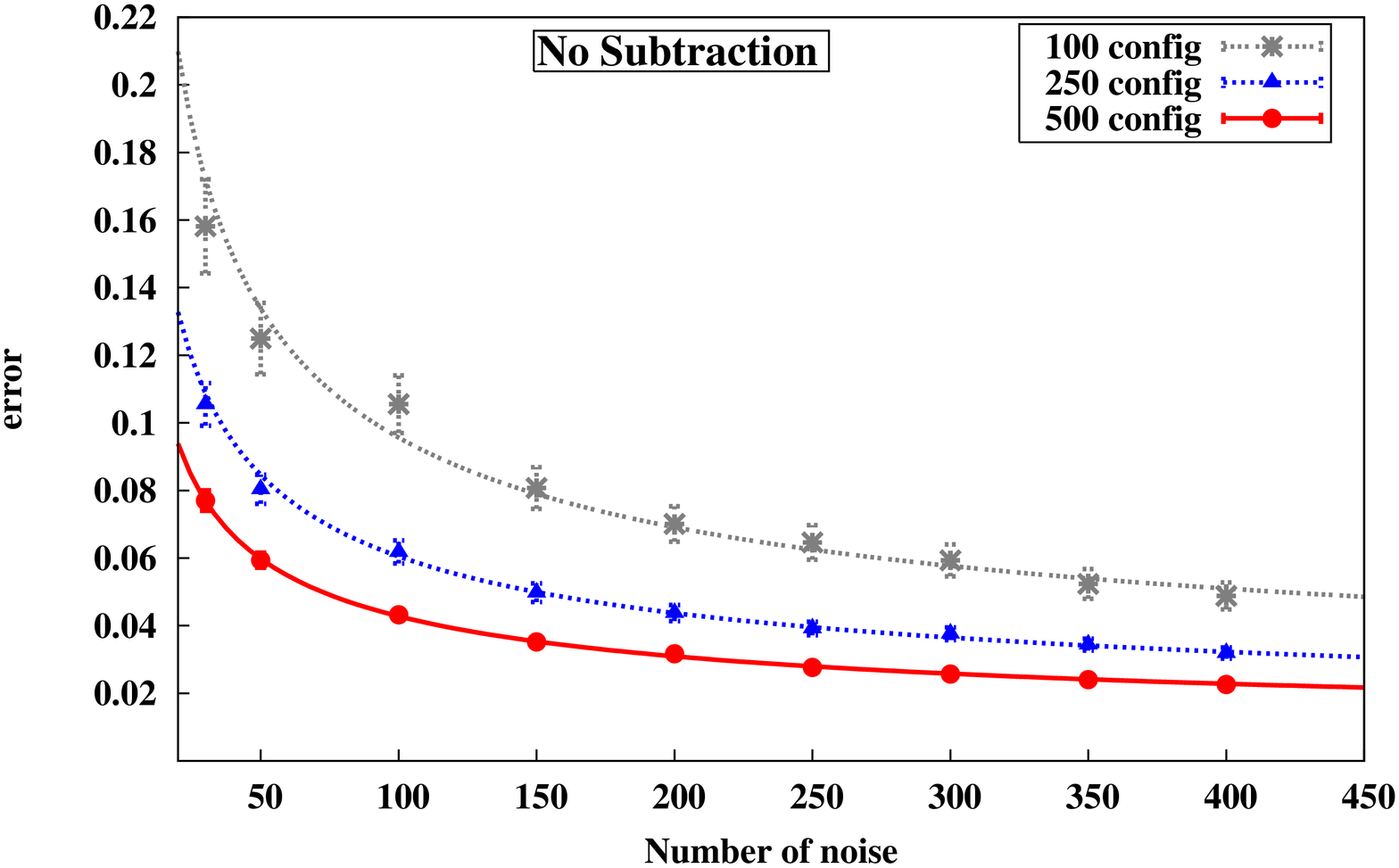}}}
        \hspace{0.5cm}
   \subfigure []
          {\label{noisevserrorwsub}
        \rotatebox{270}{\includegraphics[width=5.5cm, height=0.46\hsize]{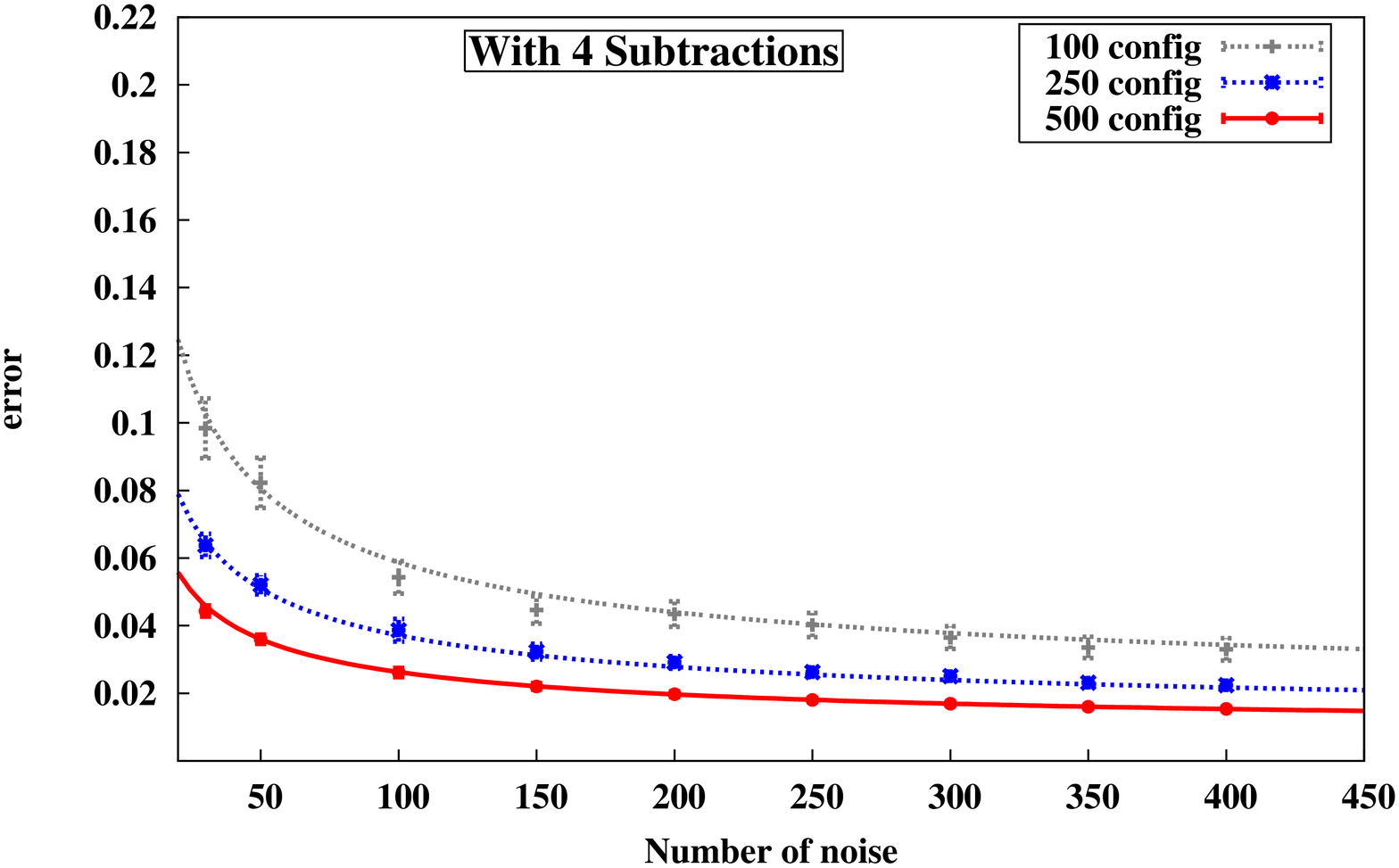}}}
          \caption{Errors of the noise estimation plotted against the number of noise vectors for different sets of configurations for the
loop part of the current ${\cal O}_{4i}$ at $\kappa_s = 0.154$ and insertion time,\ $t_1 = 14$
(a) without subtraction and (b) with four subtraction terms.}
          \label{fig:sub_2}
\end{figure}

In Figs.~\ref{confvserrornosub} and~\ref{confvserrorwsub},\ we
plot the errors of the noise estimation for the loop part of
the current ${\cal O}_{4i}$ in Eq.~(\ref{z2errorbarresults}) at the time slice 14
against the number of gauge configurations for 30,\ 100 and 400 noise
without unbiased subtraction and with four subtraction terms ($\kappa D$,\ $\kappa^2
D^2$,\ $\kappa^3 D^3$ and $\kappa^4 D^4$) respectively. The errors on the
errors are obtained by using the double jackknife method on the data.\ After
fitting for $\sigma^2_g$ and $\sigma^2_n$ from Eq.~(\ref{z2errorbarresults}),\ we see that
all the curves can be well described by Eq.~(\ref{z2errorbarresults}).\ Similarly,\ we
plot the errors against the number of noises for 100,\ 250 and 500
gauge configurations without subtraction and with four
subtraction terms in Figs.~\ref{noisevserrornosub}
and~\ref{noisevserrorwsub} respectively.\ Again,\ we see that the
curves fit Eq.~(\ref{z2errorbarresults}) well.\ A similar
conclusion can be drawn for the other two operators.\ The central
values of the standard deviations,\ $\sigma_g$ and $\sigma_n$ along with their errors
are given in Table~\ref{sigmanoiseandgauge} for the cases with and without subtractions.\
These values show that the standard
deviations for the gauge configuration and noise ensemble are not of the
same order.\ The standard deviation for noise are,\ in fact,\ much
higher than that for the gauge configuration.
Another point to note is that the standard deviation $\sigma_n$ for the noise
is reduced by almost a half with 4-term unbiased subtraction whereas $\sigma_g$
remains the same.

\begin{table}
\centering
\begin{tabular}{|c|c|c|}
\hline\hline
& {\bf $\sigma_{\rm{gauge}}$} & {\bf $\sigma_{\rm{noise}}$} \\
\hline \hline
&&\\
No Subtraction & 0.204 $\pm$ 0.063 & 9.341 $\pm$ 0.301  \\
\hline
&&\\
With 4 Subtractions & 0.205 $\pm$ 0.028 & 5.500 $\pm$ 0.201 \\
\hline\hline
\end{tabular}
\caption{Table for the values of standard deviations of gauge
configurations and noise for the current ${\cal O}_{4i}$ without
subtraction and with four subtraction terms.}
\label{sigmanoiseandgauge}
\end{table}

\subsection{Analysis for DI}

\begin{figure}[h]
\centering \subfigure[]
{\rotatebox{270}{\includegraphics[width=5.5cm, height=0.46\hsize]{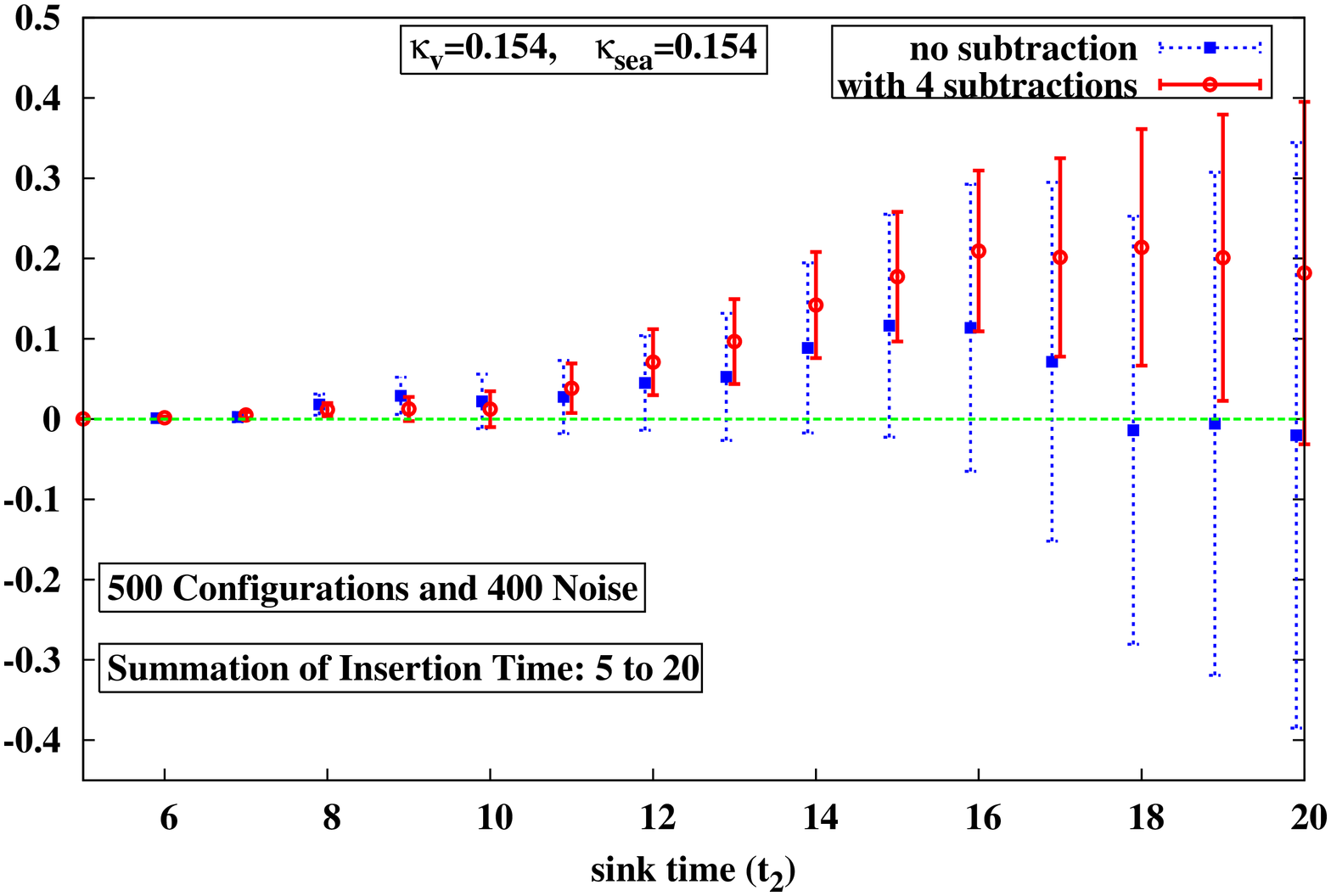}}
\label{fig:method_5_20}}
        \hspace{0.5cm}
\subfigure[]
{\rotatebox{270}{\includegraphics[width=5.5cm, height=0.46\hsize]{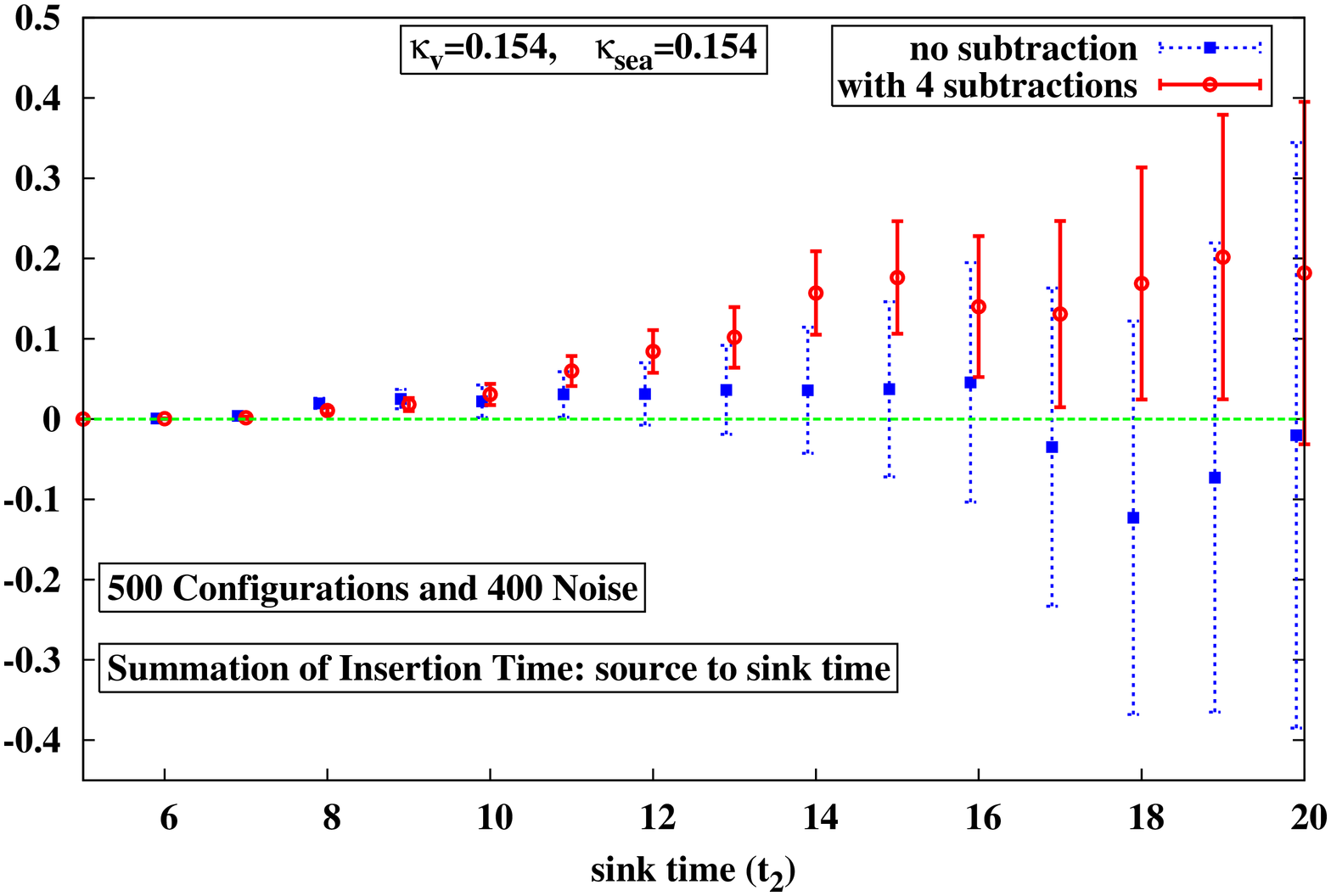}}
\label{fig:method_5_t}}
 \subfigure[]
{\rotatebox{270}{\includegraphics[width=5.5cm, height=0.46\hsize]{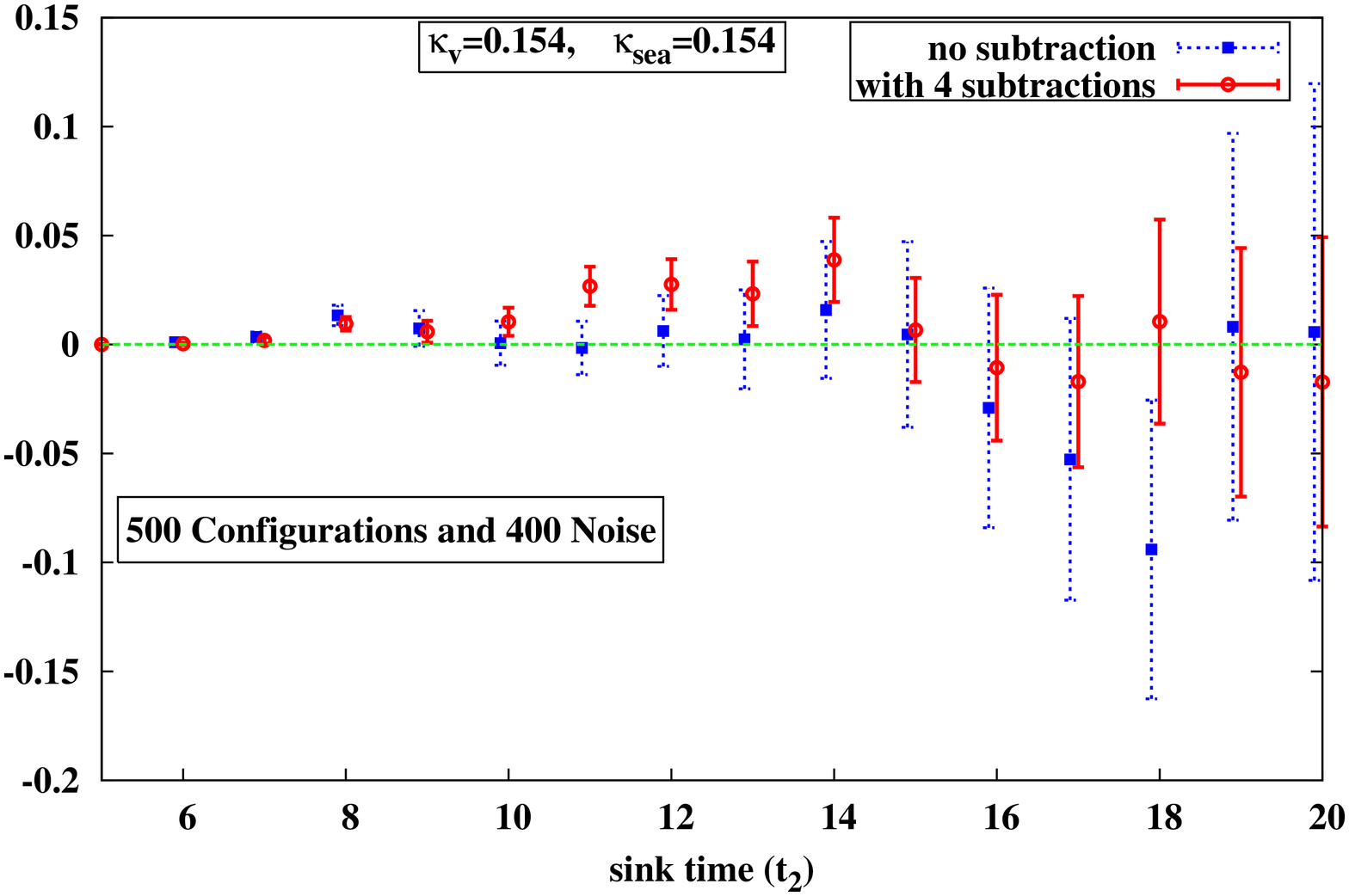}}
\label{fig:wilcox_method}}
      \hspace{0.5cm}
\subfigure[]
{\rotatebox{270}{\includegraphics[width=5.5cm, height=0.46\hsize]{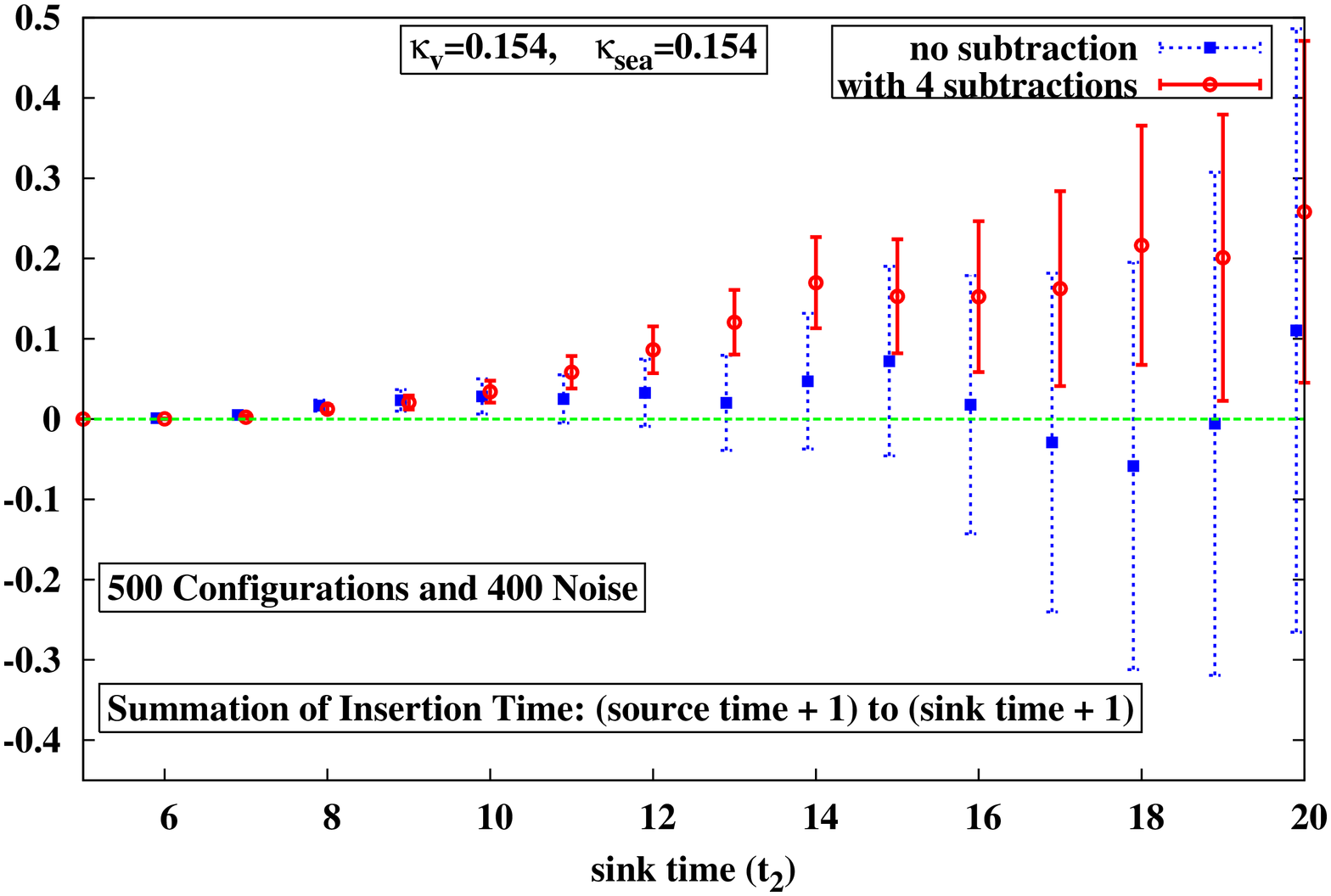}}
\label{fig:method_5_t+1}}
\caption{The ratio (D.I.) of the three-point to two-point functions at $\kappa_v=\kappa_s=0.154$ for the
${\cal O}_{4i}$ operator is plotted against the nucleon sink time ($t_2$) by using four
different methods:
(a) summation of insertion time from 5 to 20,
(b) summation of insertion time from source to sink time,
(c) method used in R. Lewis {\it et al.}~\cite{wilcox3}, and
(d) summation of insertion time from (source time $+$ 1) to (sink time + 1).
}
\label{fig:5_diff_method}
\end{figure}

As mentioned in Sec.~\ref{subsubsec:ratio_disconnected_insertions}, we have studied five
different methods of summation over insertion time by using the operator ${\cal O}_{4i}$
at $\kappa_v=\kappa_s=0.154$ shown in Fig.~\ref{fig:5_diff_method}. In the first
method~\cite{liu8}, we have performed the summation of the current insertion starting and
ending 4 time slices away from each of the boundary.\ In our case, it would be from 5 to 20
[Fig.~\ref{fig:method_5_20}].\ In the second method~\cite{mathur00}, the summation has been
performed from the source to the sink time of the nucleon propagator [Fig.~\ref{fig:method_5_t}].
The third method is described in~\cite{wilcox3} [Fig.~\ref{fig:wilcox_method}].\ The fourth
method is an additional study where the summation has been performed from (source time $+$ 1)
to (sink time $+$ 1) of the nucleon propagator [Fig.~\ref{fig:method_5_t+1}]. The fifth method
used in~\cite{viehoff1} and described in Sec.~\ref{subsubsec:ratio_disconnected_insertions}
[Fig.~\ref{resultfig6}]. In this method, the summation
has been performed from (source time $+$ 1) to (sink time $-$ 1) of the nucleon propagator.\ In
the first, second, fourth, and fifth methods, the slopes (given by Eq.~(\ref{ch:diratio4i}))
are fitted between the time slices 10 and 14 in order to extract the signal. And in the third
method, a constant is fitted between 11 and 14. The values are provided in Table~\ref{tab:5_diff_method}.
We see that the all these methods are consistent with each other. For our present work,
we adopt the fifth method.

\begin{table}
\centering
\begin{tabular}{|c|c|c|c|}
\hline\hline
{\bf Methods} & {\bf Fitting Range} & {\bf No Subtraction} & {\bf 4 Subtractions} \\
\hline\hline
Summation from 5 to 20 & 10 - 14 & 0.011 $\pm$ 0.019 & 0.030 $\pm$ 0.012 \\
\hline
Summation from source  to sink time & 10 - 14 & 0.007 $\pm$ 0.013 & 0.029 $\pm$ 0.009 \\
\hline
R. Lewis {\it et al.}~\cite{wilcox3} & 11 - 14 & 0.000 $\pm$ 0.012 & 0.027 $\pm$ 0.009 \\
\hline
Summation from (source time $+$ 1) & \multirow{2}{*}{10 - 14} & \multirow{2}{*}{-0.002 $\pm$ 0.014} & \multirow{2}{*}{0.026 $\pm$ 0.010}\\
to (sink time + 1) & & &\\
\hline
Current analysis & 10 - 14 & 0.004 $\pm$ 0.012  & 0.028 $\pm$ 0.008 \\
\hline\hline
\end{tabular}
\caption{Table for the vales of $\langle x \rangle$ (D.I.) at $\kappa_v = \kappa_s = 0.154$ for the ${\cal O}_{4i}$ operator by
using five different methods.}
\label{tab:5_diff_method}
\end{table}

\subsection{Multiple Sources for DI}

In the case of D.I., we have the liberty of choosing quark propagators at different source
locations on the lattice with much less overhead that the C.I. computation which involves
generating additional quark propagators using the sequential source method. We
correlate these nucleon propagators at different source locations with the already computed
loop to increase statistics. We have used 1, 4, and 16 different sources. This results in
significant reduction of error bars which is presented in Sec.~\ref{sec:results}.

\section{Results}
\label{sec:results}

In this section, we will present results for the first and second
moments of nucleon's parton distribution function for both the disconnected insertion and the
connected insertion.

\subsection{Disconnected Insertions}

The disconnected insertion [Fig.~\ref{disconnected_fig}] entails the correlation between the
quark loops with up, down, and strange quark currents and the nucleon two-point propagators as
discussed in Sec.~\ref{subsec:twopoint}.\ We consider the up and down to have the
same mass so that their moments are the same.

\subsubsection{First Moments}

First,\ we will discuss the results for the first moments from the
 ${\cal O}_{4i} (i=1, 2, 3)$ operator for the case where the valence quarks in the nucleon
propagators and the sea quark in the loop are the same. This is the case, when extrapolated
to the physical $u/d$ quark mass would give the disconnected insertion result for the
$u$ and $d$ quarks. We average over the three spatial directions on each configuration.\
From now on,\ ${\cal O}_{4i}$ stands for the average over 1, 2 and 3 directions.

\begin{figure}[h]
\centering \subfigure[]
{\rotatebox{270}{\includegraphics[width=5.5cm, height=0.46\hsize]{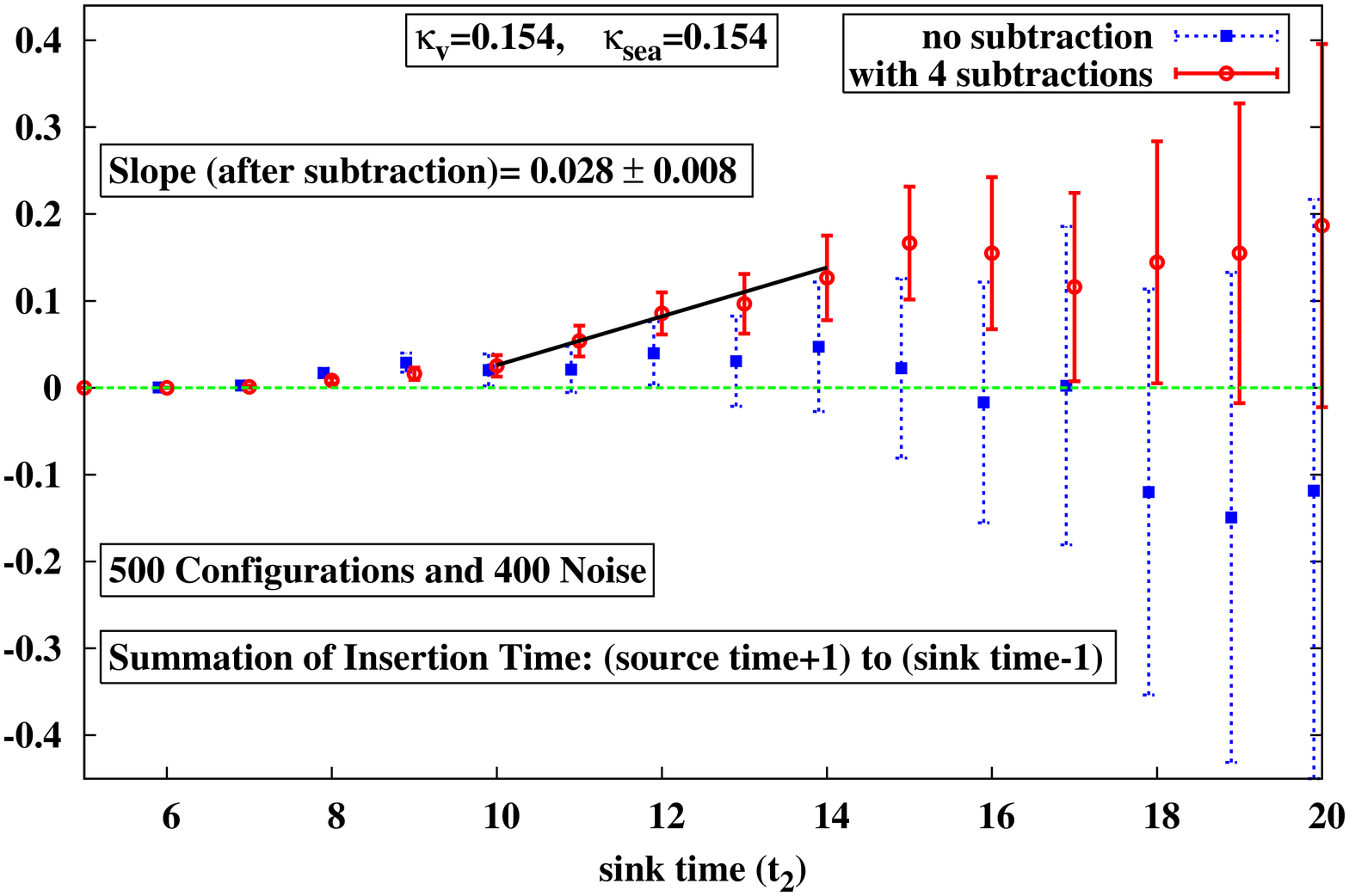}}
\label{resultfig6}}
        \hspace{0.5cm}
 \subfigure[]
{\rotatebox{270}{\includegraphics[width=5.5cm,height=0.46\hsize]{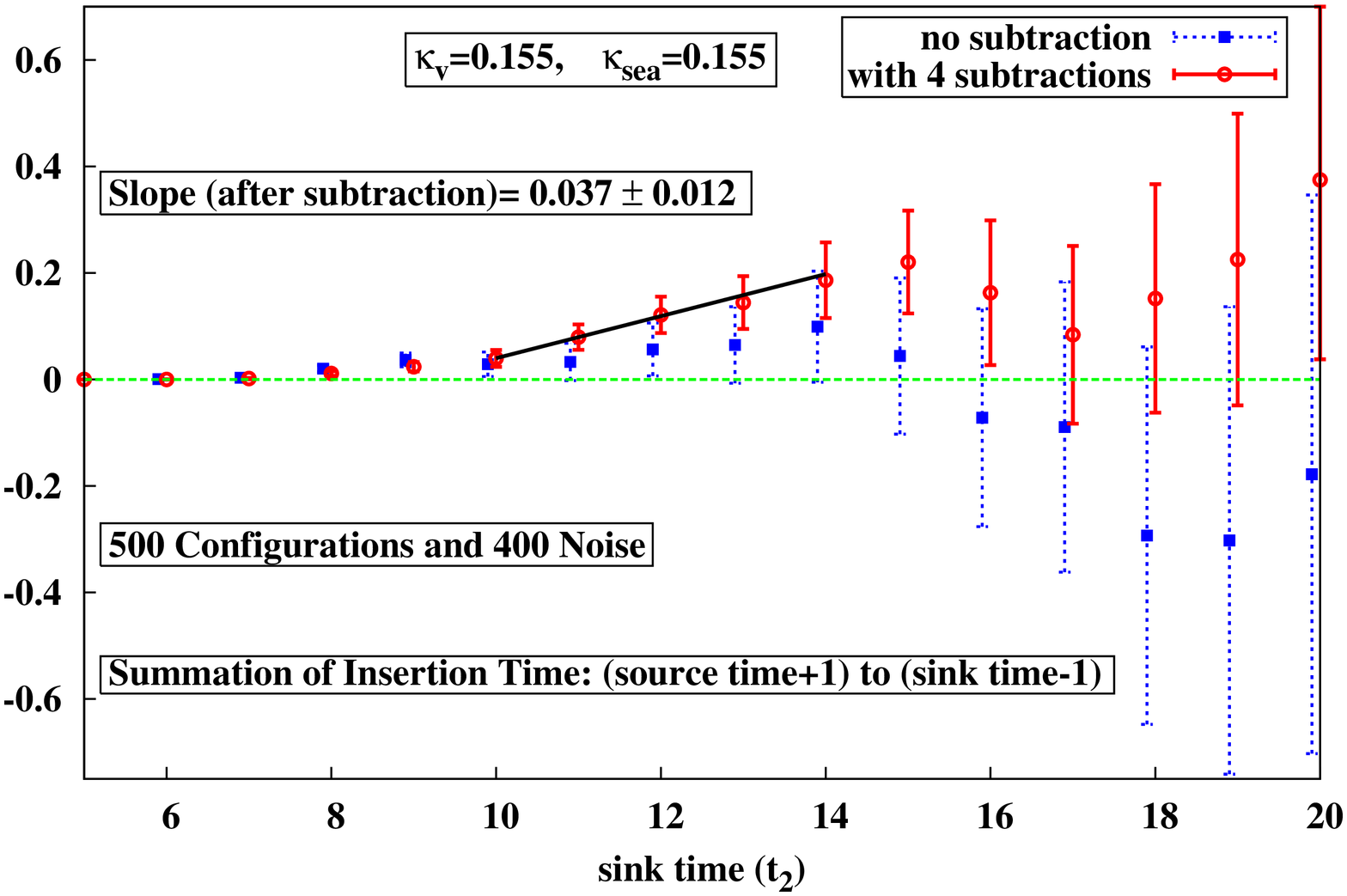}}
\label{resultfig7}}
\subfigure[]
{\rotatebox{270}{\includegraphics[width=5.5cm, height=0.46\hsize]{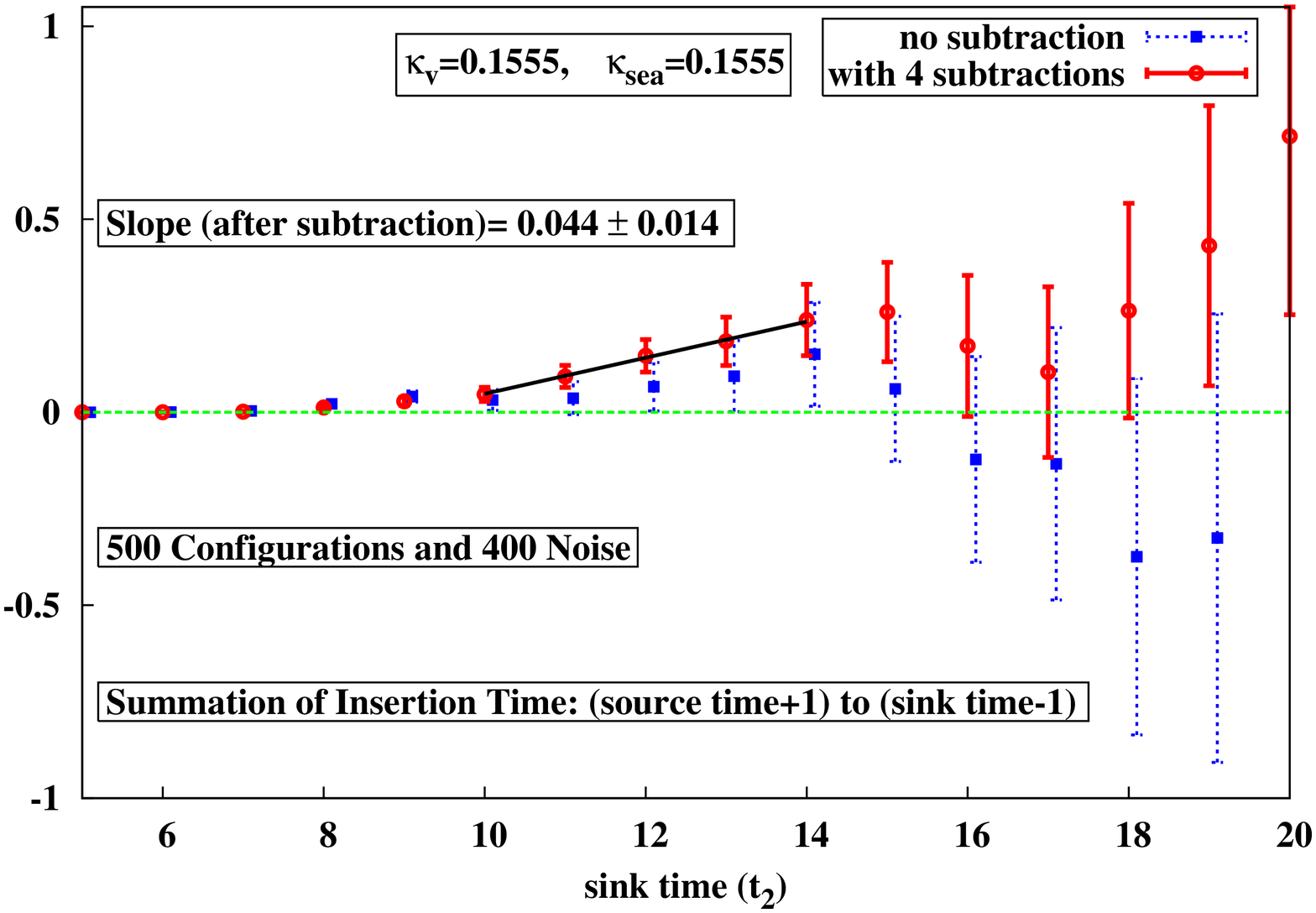}}
\label{resultfig8}}
        \hspace{0.5cm}
\subfigure[]
{\rotatebox{270}{\includegraphics[width=5.5cm, height=0.46\hsize]{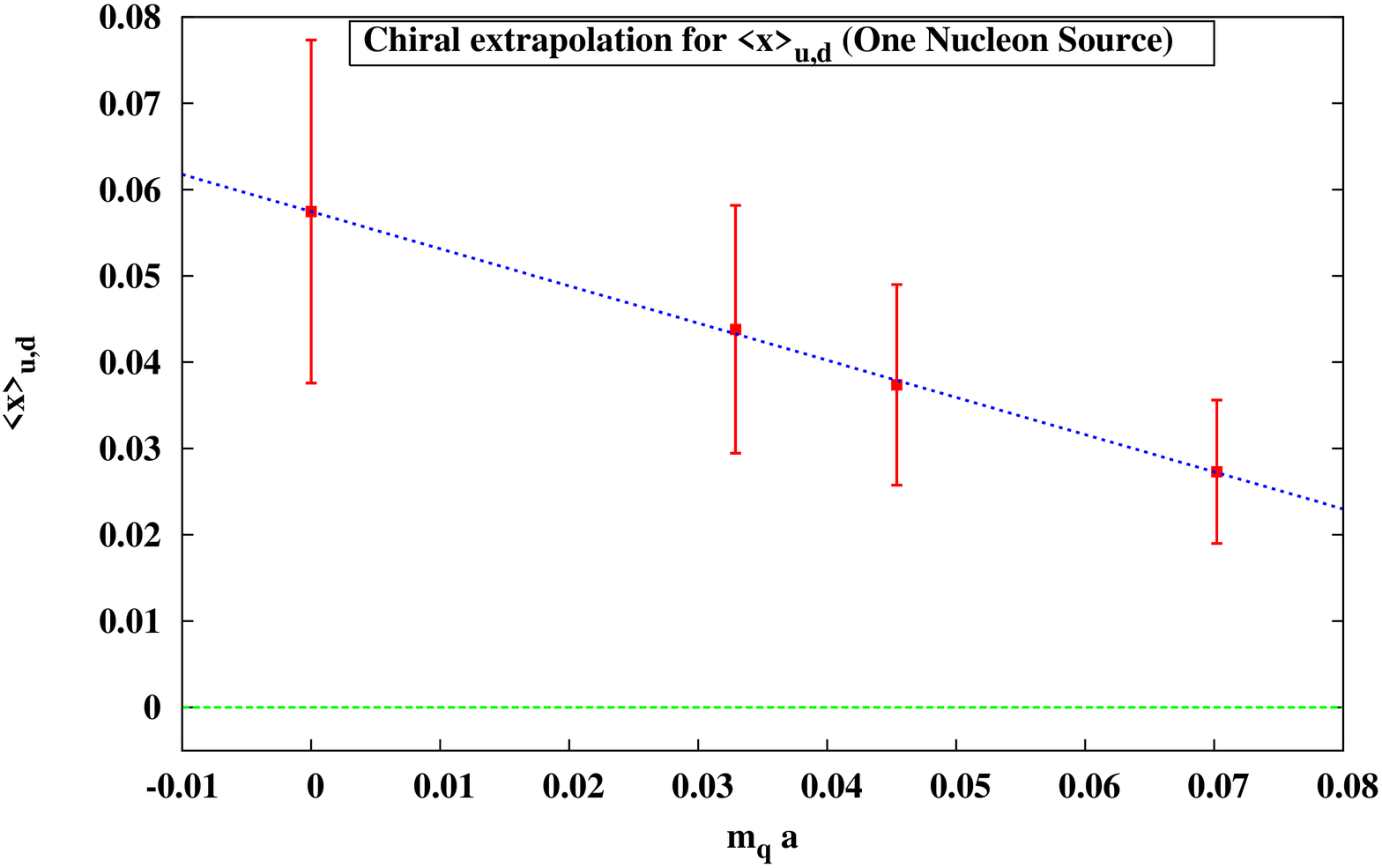}}
\label{resultfig10}}
\caption{The ratio of the three-point to
two-point functions (summed over insertion time) for the ${\cal O}_{4i}$
operator,\ for the case with equal valence and sea quark masses,\ is plotted against the nucleon
sink time ($t_2$) at
(a) $\kappa_v = \kappa_s = 0.154$,
(b) $\kappa_v = \kappa_s = 0.155$, and
(c) $\kappa_v = \kappa_s = 0.1555$.
(d) is a linear extrapolation to the chiral limit for the first moment,\ $\langle x
\rangle_{u,d}$,\ of the up (down) quark which is plotted against $m_q a$.}
\end{figure}

\begin{figure}[h]
\rotatebox{270}{\includegraphics[width=8cm,height=12cm]{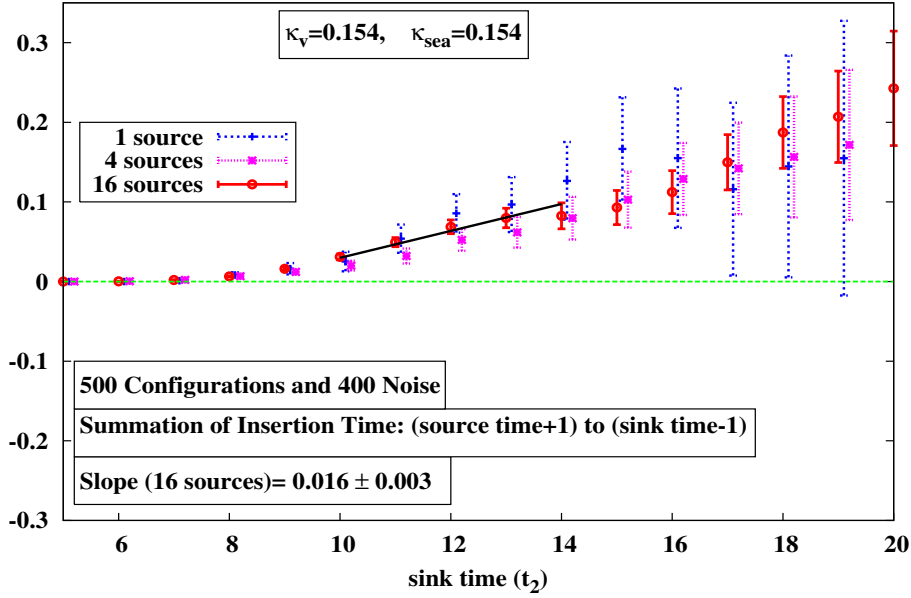}}
\caption{The ratio of the three-point to two-point functions for the
${\cal O}_{4i}$ operator is plotted against the nucleon sink
time ($t_2$)  at $\kappa_v = 0.154$ and $\kappa_s = 0.154$
for 1, 4, and 16 sources after 4 subtractions.}
\label{fig:firstmom_slope_all_source}
\end{figure}

\begin{figure}[!hbtp]
\subfigure[]
{\rotatebox{270}{\includegraphics[width=5.5cm, height=0.48\hsize]{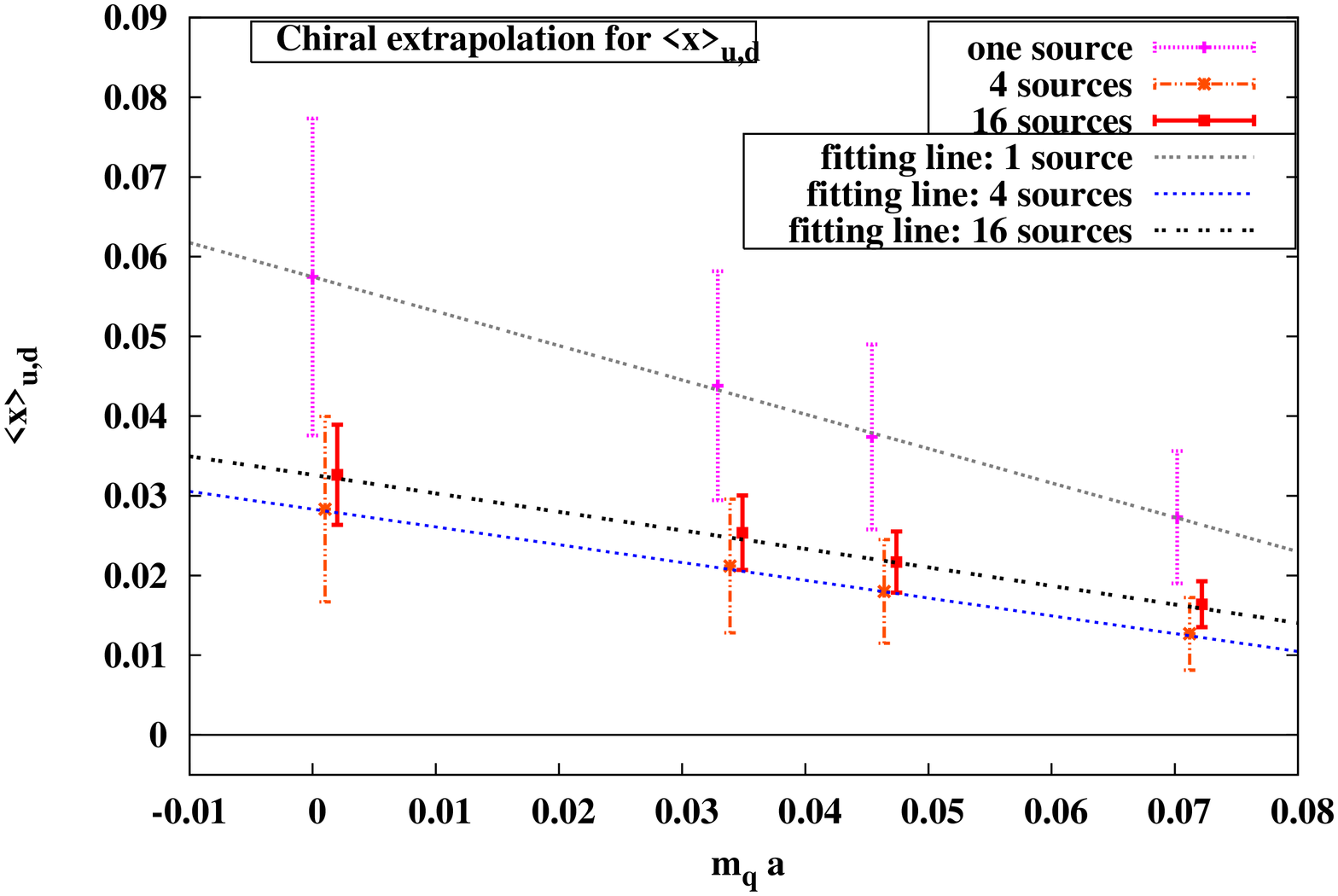}}
\label{fig:di_x_ud_154_all_source}}
        \hspace{0.1cm}
\subfigure[]
{\rotatebox{270}{\includegraphics[width=5.5cm, height=0.48\hsize]{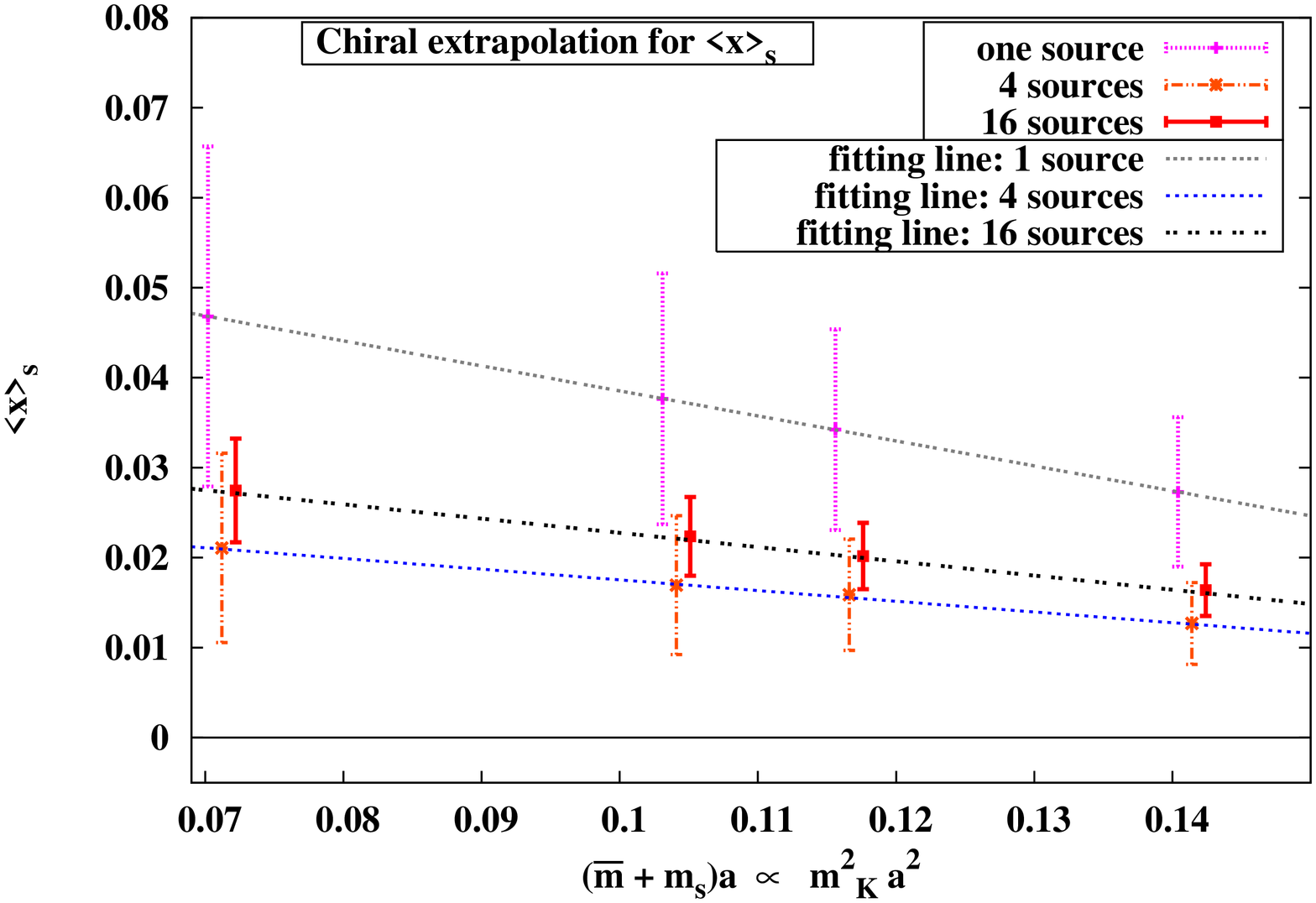}}
\label{fig:di_x_s_154_all_source}}
\label{chral}
\caption{Linear extrapolation to the chiral limit for the first moment with 1, 4, and 16 nucleon sources
(a) for $\langle x\rangle_{u,d}$ (D.I.) and (b) for $\langle x \rangle_{s}$ (D.I.).}
\end{figure}

\begin{figure}[h]
\centering \subfigure[]
{\rotatebox{270}{\includegraphics[width=5.5cm, height=0.46\hsize]{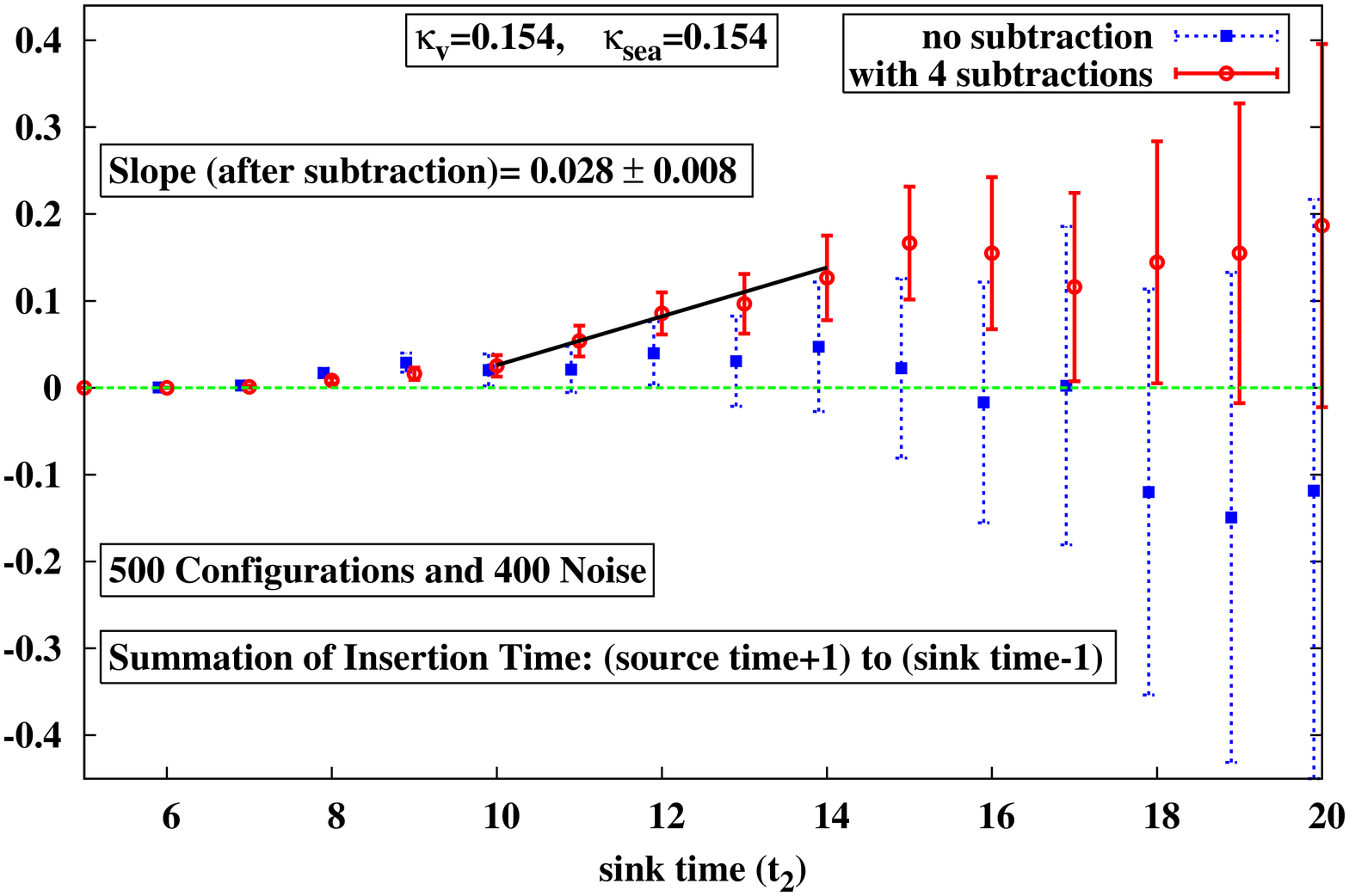}}
\label{resultfig1}}
        \hspace{0.5cm}
 \subfigure[]
{\rotatebox{270}{\includegraphics[width=5.5cm, height=0.46\hsize]{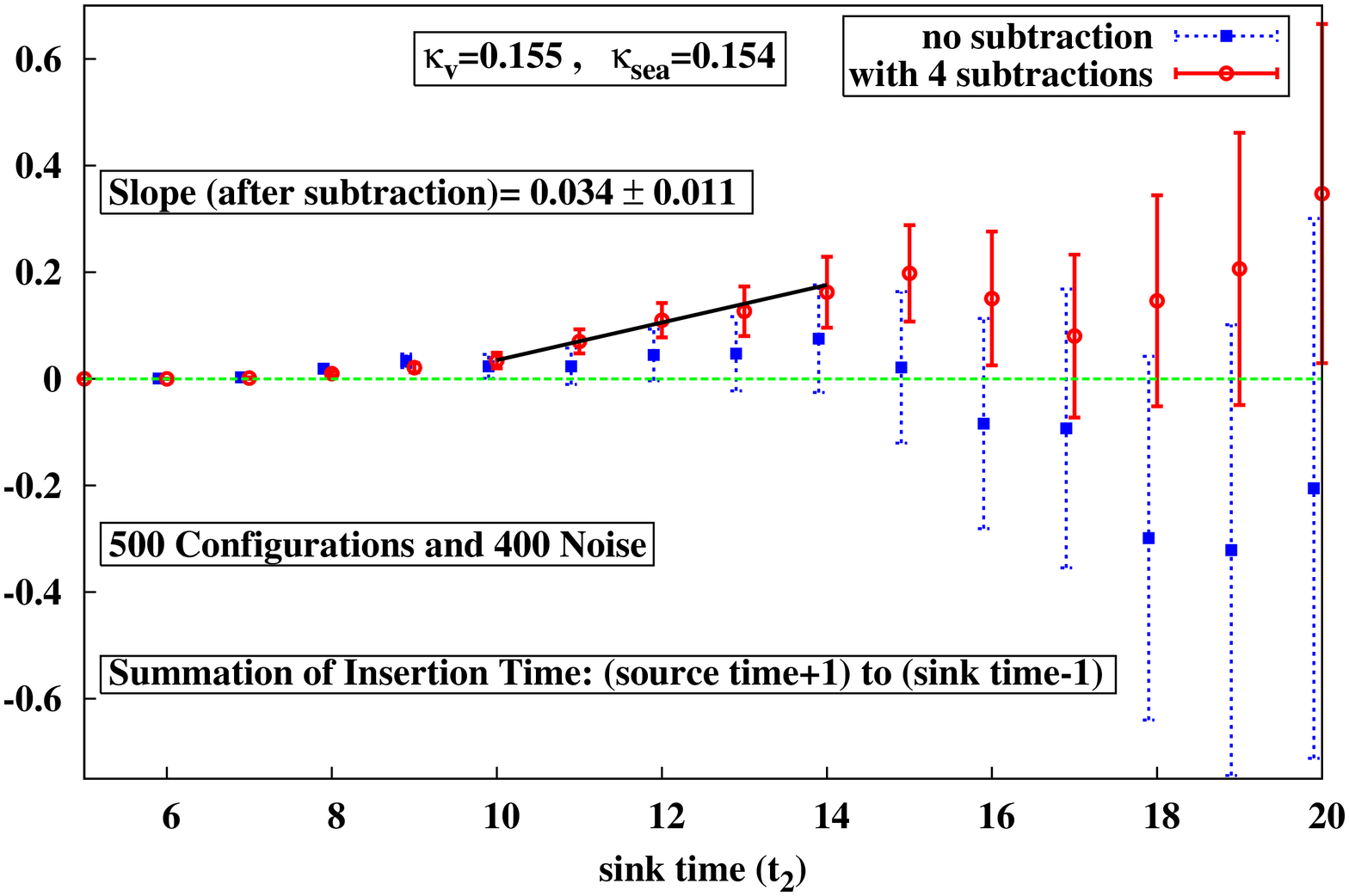}}
\label{resultfig2}}
\subfigure[]
{\rotatebox{270}{\includegraphics[width=5.5cm, height=0.46\hsize]{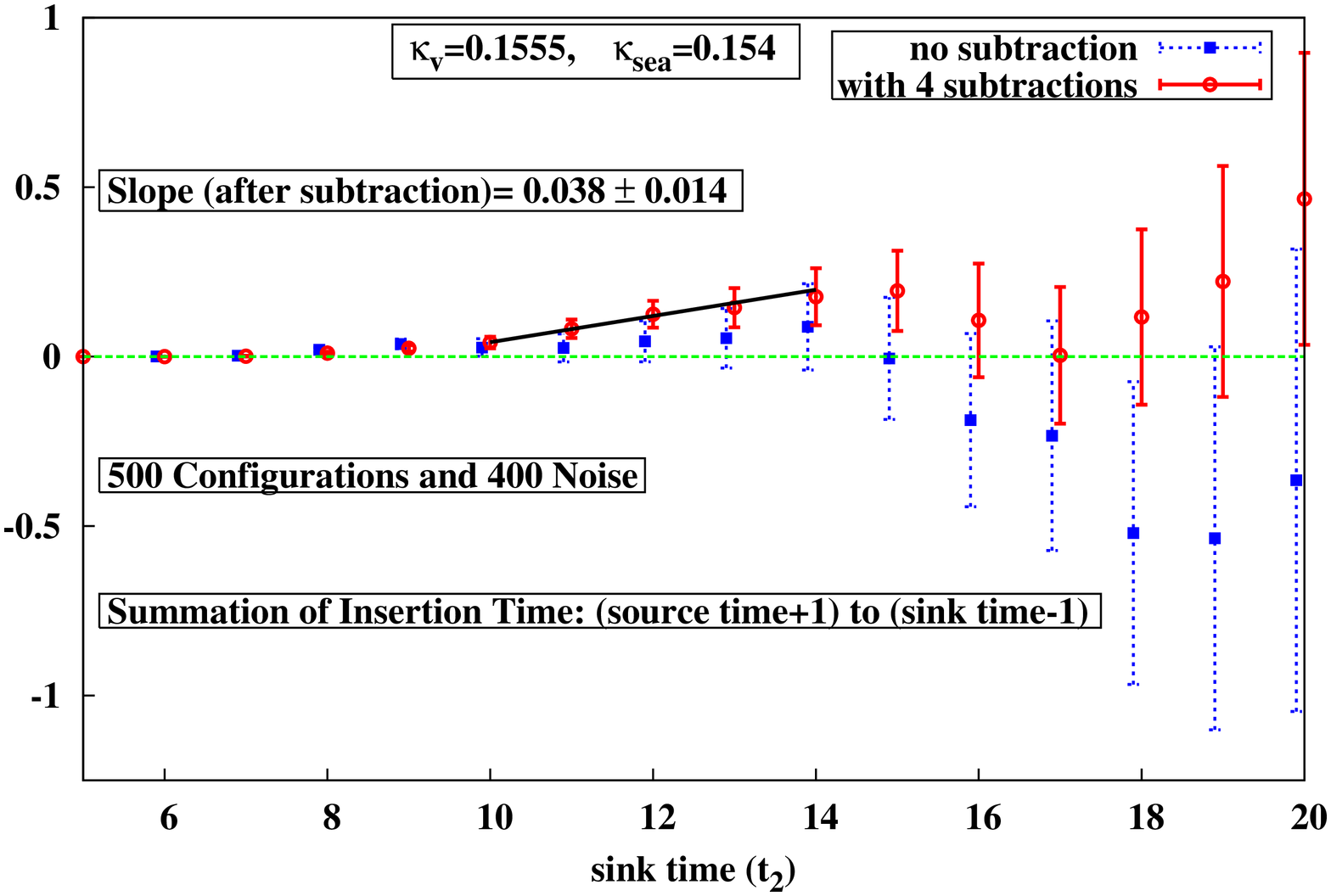}}
\label{resultfig3}}
        \hspace{0.5cm}
\subfigure[]
{\rotatebox{270}{\includegraphics[width=5.5cm, height=0.46\hsize]{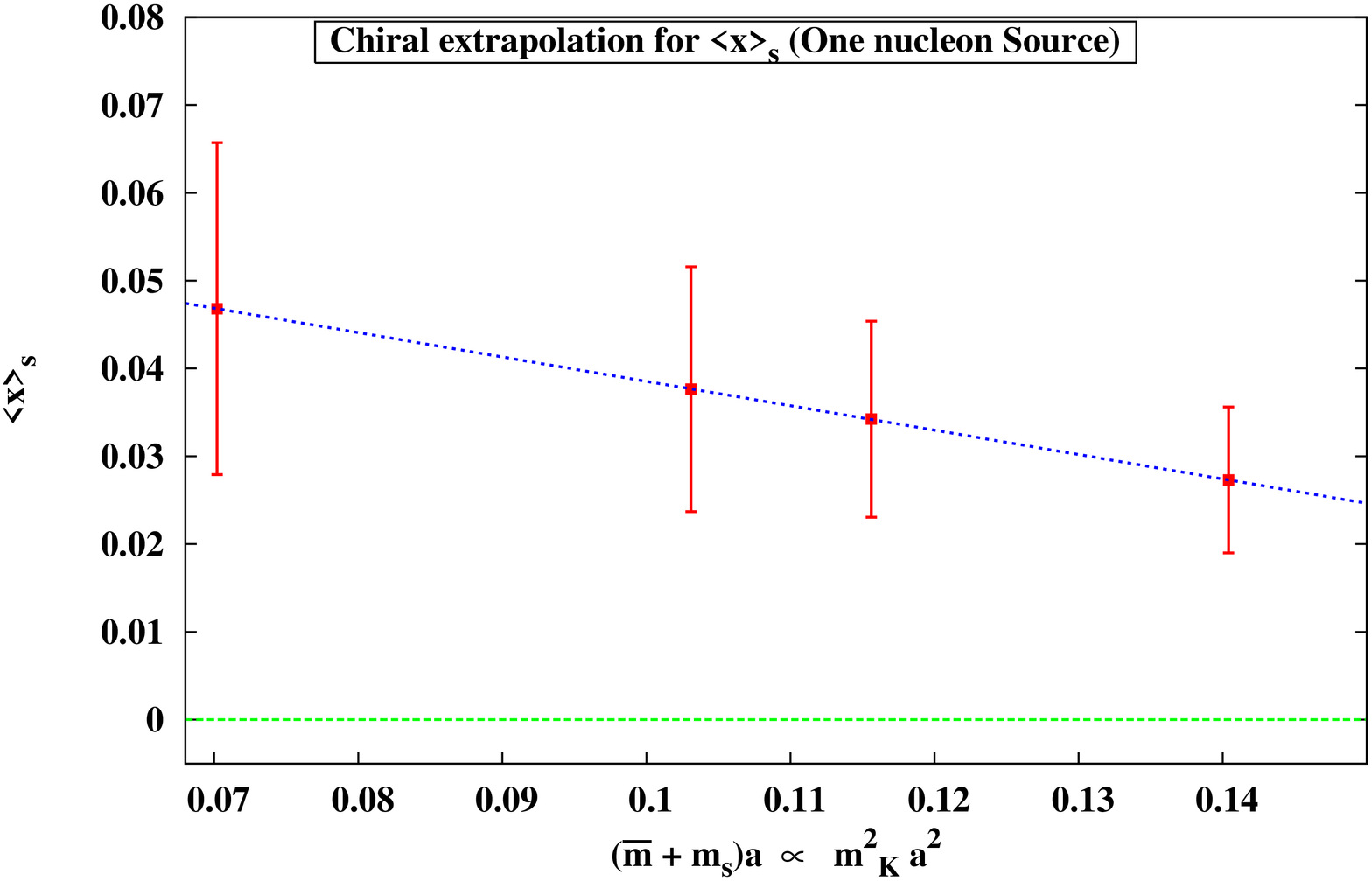}}
\label{resultfig5}}
\caption{The ratio of the three-point to
two-point functions (summed over insertion time) for the ${\cal O}_{4i}$
operator,\ for the strange quark which is fixed at $\kappa = 0.154$,\ is plotted against the nucleon sink
time ($t_2$) at
(a)  $\kappa_v = 0.154$ and $\kappa_s = 0.154$,
(b)  $\kappa_v = 0.155$ and $\kappa_s = 0.154$, and
(c)  $\kappa_v = 0.1555$ and $\kappa_s = 0.154$.
(d) is a linear extrapolation of the valence quarks to the chiral limit for the first
moment,\ $\langle x \rangle_s$,\ of the strange quark which is plotted
against $(\overline{m}+m_s) a$.}
\end{figure}

\begin{table}[h]	
\centering												
\begin{tabular}{|p{1.12cm}|c|c|c|c|c|}												
\hline\hline														
& & 1 source & 1 source & 4 sources & 16 sources\\	
& & (No sub) & (4 sub) & (4 sub) & (4 sub)\\
\hline \hline													
\multirow{3}{*}{$\langle x \rangle_{u,d}$} & $\kappa_v = \kappa_{\rm sea}=0.154$& 0.004 $\pm$ 0.012  & 0.028 $\pm$ 0.008 & 0.013 $\pm$  0.005 & 0.016 $\pm$  0.003 \\
\cline{2-6}														
\multirow{4}{*}{(D.I.)}& $\kappa_v = \kappa_{\rm sea}=0.155$& 0.009 $\pm$ 0.017 & 0.037 $\pm$ 0.012 & 0.018 $\pm$  0.007 & 0.022 $\pm$  0.004 \\
\cline{2-6}														
& $\kappa_v = \kappa_{\rm sea}=0.1555$& 0.011 $\pm$ 0.022 & 0.044 $\pm$ 0.014 & 0.021 $\pm$  0.008 & 0.025 $\pm$  0.005  \\	
\cline{2-6}														
& Linear Extrapolation & 0.033 $\pm$ 0.058 & 0.056 $\pm$ 0.019 & 0.028 $\pm$  0.011 & 0.032 $\pm$  0.006 \\
\hline\hline														
\multirow{3}{*}{$\langle x \rangle_{s + \bar s}$} & $\kappa_v = 0.154, \kappa_{\rm sea}= 0.154$& 0.004 $\pm$ 0.012 & 0.028 $\pm$ 0.008 & 0.013 $\pm$  0.005 & 0.016 $\pm$  0.003 \\	
\cline{2-6}														
\multirow{4}{*}{(D.I.)}& $\kappa_v = 0.155,  \kappa_{\rm sea} = 0.154$ & 0.005 $\pm$ 0.017 & 0.034 $\pm$  0.011 & 0.016 $\pm$ 0.006 & 0.020 $\pm$ 0.004 \\	
\cline{2-6}															
& $\kappa_v = 0.1555, \kappa_{\rm sea} = 0.154$& 0.004 $\pm$ 0.021 & 0.038 $\pm$  0.014 & 0.017  $\pm$ 0.008 & 0.023 $\pm$ 0.004\\	
\cline{2-6}														
& Linear Extrapolation & 0.005 $\pm$ 0.029  &  0.046 $\pm$ 0.018 & 0.021 $\pm$ 0.010 &  0.027 $\pm$ 0.006\\	
\hline
$\displaystyle\frac{\langle x \rangle_{s + \bar s}}{\langle x \rangle_{u + \bar u}}$& & 0.69 $\pm$ 0.64 & 0.85 $\pm$ 0.13 & 0.95 $\pm$ 0.18 & 0.88 $\pm$ 0.07\\
\hline\hline
\end{tabular}													
\caption{$\langle x \rangle$ (D.I.) for up (down) and strange quarks at various					
$\kappa$'s and the linearly extrapolated results to the chiral limit with different
number of nucleon sources for ${\cal O}_{4i}$ operator.}													
\label{table:dix4i}													
\end{table}

For the disconnected insertion,\ we shall define two
$\kappa$'s for the quark mass: $\kappa_v$ for valence quarks,\ and $\kappa_{\rm{sea}}$ for sea
quarks.\ For the strange quark currents we have fixed
$\kappa_{\rm{sea}} = 0.154$, which is close to the strange quark mass as determined from the
$\phi$ meson mass\ and $\kappa_v$ takes the values of 0.154, 0.155, and
0.1555.\ We consider the cases with equal valence and sea quark masses, i.e.
$\kappa_{\rm{sea}} = \kappa_v = 0.154, 0.155$, and 0.1555 in order to extrapolate to the chiral
limit to obtain $\langle x\rangle_{u + \bar u} = \langle x\rangle_{d + \bar d} = \langle x\rangle_{u,d}$ (D.I.).

In Figs.~\ref{resultfig6},~\ref{resultfig7}, and~\ref{resultfig8} we
plot the ratios in Eq.~(\ref{ch:diratio4i}) against the
nucleon sink time $t_2$.\
The insertion time is summed from [source time + 1] (i.e. $ t_0+1$) to
[sink time $-$ 1] (i.e. $t_2-1$).\ In these figures,\ we see that,
after the unbiased subtraction of four terms, there is a clear straight line behavior
starting from the time slice 10. On the other hand, without
the unbiased subtraction, there is not a clear signal. In fact, they are
consistent with zero slopes.\ Also,\ the plots show that the error bars get
reduced after subtraction by a factor of $\sim 1.5$.\ To extract the values of
$\langle x \rangle_{u,d}$ at each $\kappa_v$,\ we have performed a
correlated fit of the slope between the time slices 10 and 14.\ It
gives us the value of the $\langle x \rangle$ at the
corresponding $\kappa_v$.\ The values of  $\langle x \rangle_{u,d}$ (D.I.),\
along with their errors,\ are listed in Table~\ref{table:dix4i}.

In chiral perturbation theory, the first moment of $\langle x \rangle$ has a leading
non-analytic behavior $\propto~m_{\pi}^2 {\ln(m_{\pi}^2/\mu^2)}$ and leading analytic
behavior $\propto m_{\pi}^2$ ~\cite{Thomas:2000ny,arndt,chen,dgh07}. Since our pion
masses are relatively heavy, we do not expect to be in the region where the non-analytic
behavior is important. Furthermore, our present calculation is based on the quenched
approximation. In view of this and other systematic errors that we have not taken into
account, such as the large volume limit and continuum limit, we shall take the conservative
linear extrapolation of $\langle x \rangle_{u,d}$ to the chiral limit with the form
$A + B m_q a$ [Fig.~\ref{resultfig10}]. This linear extrapolation will inevitably introduce
systematic errors. This issue will be dealt with when the configurations with lighter quark
masses are available. We will certainly include the non-analytic behavior when the results
from the PAC-CS 2+1-flavor dynamical clover fermion configurations are available with lighter
sea quark masses~\cite{pacs-cs}.\ Before extrapolation,\ we have converted the values of
$\langle x \rangle_{u,d}$ (D.I.) to those of tadpole improved values by using the factors
in Eq.~(\ref{tadpole_fac}). We did the similar analysis for the cases with 4 and 16 nucleon
sources. Fig.~\ref{fig:firstmom_slope_all_source} shows the ratios between three-point to
two-point functions for all the sources with 4 subtractions, and
Fig.~\ref{fig:di_x_ud_154_all_source} shows the linear extrapolation to the chiral limit.\
In Table~\ref{table:dix4i},\ we list the renormalized and linearly extrapolated values
(to the chiral limit) of $\langle x \rangle_{u,d}$ along with their errors.\ As stated
in Eq.~(\ref{renorm1.74}),\ the renormalization factor for this operator is 0.972.\ We find
that the values for $\langle x \rangle_{u,d}$ are about $3\sigma$ away from zero for one and
four sources and more than $5\sigma$ for 16 sources. We consider this a solid affirmation
that we have been able to calculate the D.I. of $\langle x \rangle_{u,d}$ via the noise
method.


Next,\ we consider the strange quark loop.\ This time we have
fixed $\kappa_{\rm{sea}} = 0.154$ which corresponds to the strange quark mass.\
In Figs.~\ref{resultfig1},~\ref{resultfig2} and~\ref{resultfig3},\
we have plotted the ratio in Eq.~(\ref{ch:diratio4i}) against the
nucleon sink time by using the valence quark masses at $\kappa_v=0.154,
0.155$, and 0.1555, respectively.\ A similar procedure has been followed
as for the up (down) currents to obtain the slopes,\ which gives the
values of $\langle x \rangle_{s}$.\ Again,\ from these figures,\ it
is clear that if we do not use unbiased subtraction,\ we will not see
a signal.\ The error bars get reduced after
subtraction.\ The linear extrapolation to the chiral limit is performed with the form
$A + B(\overline{m}+m_s)$ where $(\overline{m}+m_s) \propto m^2_K$,\
with\ $\overline{m}$ being the average of the up and down quark masses and $m_s$
being the strange mass.\ In Table~\ref{table:dix4i},\ we have listed
values of $\langle x \rangle_{s+\bar{s}}$ (D.I.) for different $\kappa_v$'s and
fixed $\kappa_{\rm{sea}}$. Also listed are the linearly extrapolated values of
$\langle x \rangle_{s+\bar{s}}$ (D.I.) to the chiral limit. Again, we find that the value for
$\langle x \rangle_{s+\bar{s}}$ is 4.5$\sigma$ away from zero for 16 sources.
We have also listed the ratio of $\langle x \rangle_{s+\bar{s}}$ (D.I.) to $\langle x \rangle_{u+\bar{u}}$
(D.I.) in Table~\ref{table:dix4i}.\ We see that the ratio is close to 1 as expected~\cite{liu1}.
We should point out the caveat that these results are based on the linear extrapolation to the chiral limit
which are subjected to systematic corrections as mentioned above. The ratio is expected to be less susceptible
to the systematic errors except the chiral extrapolation due to the non-analytic terms. In view
of the fact that the lattice calculations of both the $\langle x\rangle_{u+d}$ and
$\langle x\rangle_{u-d}$ for the connected insertions at quark masses between the strange and the
physical $u/d$ mass are all larger than the respective experimental results
(see Table \ref{compresults} in Sec. \ref{CI}), we expect the ratio of $\langle x \rangle_{s+\bar{s}}$ (D.I.) to
$\langle x \rangle_{u+\bar{u}}$ (D.I.) for the physical $u/d$ mass to be larger than that obtained with
a linear extrapolation to the chiral limit.


We have performed the same analysis for the ${\cal
O}_{44}-\frac{1}{3}\left({\cal O}_{11}+{\cal O}_{22}+{\cal
O}_{33}\right)$ operator as in the case of the ${\cal O}_{4i}$
operator.\ The only difference is that we have used the nucleon with
zero momentum in this case.\ But,\ due to the subtraction of the spatial trace terms
in the current, it has significant numerical cancelations resulting in large
statistical errors~\cite{martinelli,capitani1}.\
The error bar of the slope is $\sim$ 3 - 7 times larger than that of
${\cal O}_{4i}$ operator.\ Due to this large error,\ the signal is
only $\sim 1\sigma$ to $1.5\sigma$ away from zero shown in Table~\ref{strange_varkappafirmom44nosub}.\
This happens for both $\langle x \rangle_{s + \bar s}$ and $\langle x \rangle_{u,d}$ (D.I.) for
this operator.

\begin{table} [!hbtp]
\begin{tabular}{|p{1.02cm}|c|c|c|c|c|}
\hline\hline
& & 1 source & 1 source & 4 sources & 16 sources\\	
& & (No sub) & (4 sub) & (4 sub) & (4 sub)\\
\hline
\multirow{3}{*}{$\langle x \rangle_{u,d}$} & $\kappa_v =\kappa_{\rm sea} = 0.154$  & 0.045  $\pm$  0.044   & 0.033  $\pm$ 0.040 &  0.025 $\pm$ 0.026 & 0.038  $\pm$ 0.020 \\
\cline{2-6}
\multirow{4}{*}{(D.I.)}& $\kappa_v =\kappa_{\rm sea} = 0.155$  & 0.075 $\pm$ 0.056    &  0.059  $\pm$ 0.053 &  0.032 $\pm$ 0.035 & 0.046 $\pm$ 0.027 \\
\cline{2-6}
& $\kappa_v =\kappa_{\rm sea} = 0.1555$  & 0.095 $\pm$ 0.065   &  0.073  $\pm$ 0.060  &  0.034 $\pm$ 0.042 & 0.049 $\pm$ 0.033 \\
\cline{2-6}
& Linear Extrapolation & 0.130 $\pm$ 0.081 & 0.095  $\pm$ 0.071 & 0.037 $\pm$ 0.050  & 0.058 $\pm$ 0.043 \\
\hline\hline
\multirow{3}{*}{$\langle x \rangle_{s + \bar s}$} & $\kappa_v = 0.154, \kappa_{\rm sea} = 0.154$  & 0.045  $\pm$  0.044   & 0.033  $\pm$ 0.040 &  0.025 $\pm$ 0.026 & 0.038  $\pm$ 0.020 \\
\cline{2-6}
\multirow{4}{*}{(D.I.)}& $\kappa_v = 0.155, \kappa_{\rm sea} = 0.154$  & 0.067 $\pm$  0.056   &  0.047  $\pm$ 0.050 &  0.026 $\pm$ 0.034 & 0.041 $\pm$ 0.027 \\
\cline{2-6}
& $\kappa_v = 0.1555, \kappa_{\rm sea} = 0.154$  & 0.087 $\pm$  0.065   &  0.062  $\pm$ 0.057  &  0.026 $\pm$ 0.039 & 0.041 $\pm$ 0.033 \\
\cline{2-6}
& Linear Extrapolation & 0.114 $\pm$ 0.080 & 0.077  $\pm$ 0.068 & 0.026  $\pm$ 0.048 & 0.043 $\pm$ 0.042 \\
\hline\hline
\end{tabular}
\caption{Table for the values of $\langle x \rangle$ (D.I.) for up (down) and
strange quarks at various kappa values and after linear
extrapolation to the chiral limit with different number of nucleon sources for the
${\cal O}_{44}-\frac{1}{3}\left({\cal O}_{11}+{\cal O}_{22}+{\cal O}_{33}\right)$
operator.}
\label{strange_varkappafirmom44nosub}
\end{table}


\subsubsection{Second Moments}

\begin{figure}[!hbtp]
\centering \subfigure[]
{\rotatebox{270}{\includegraphics[width=5.5cm, height=0.46\hsize]{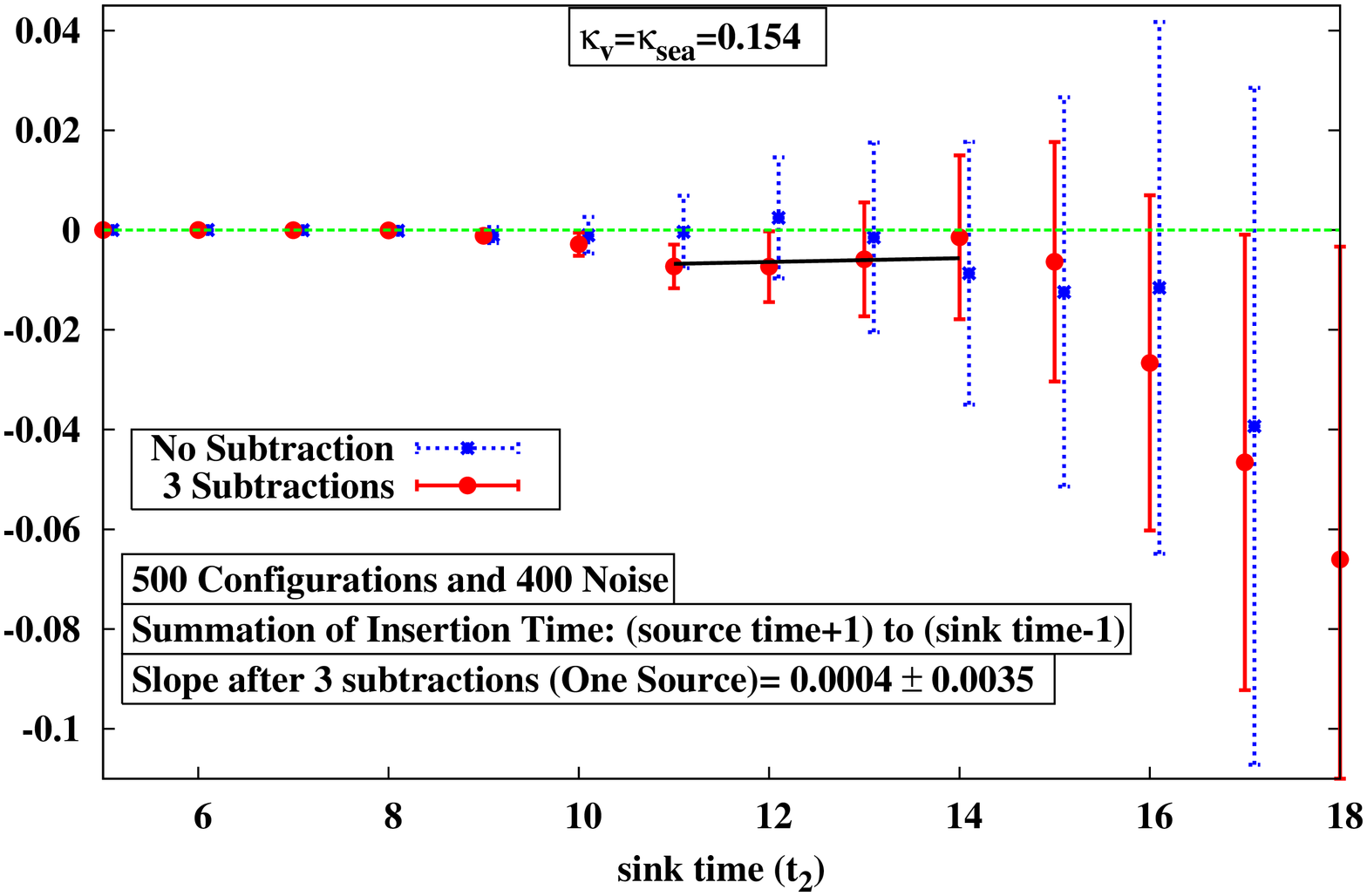}}
\label{dixsq_ud_4ii154}}
        \hspace{0.5cm}
 \subfigure[]
{\rotatebox{270}{\includegraphics[width=5.5cm, height=0.46\hsize]{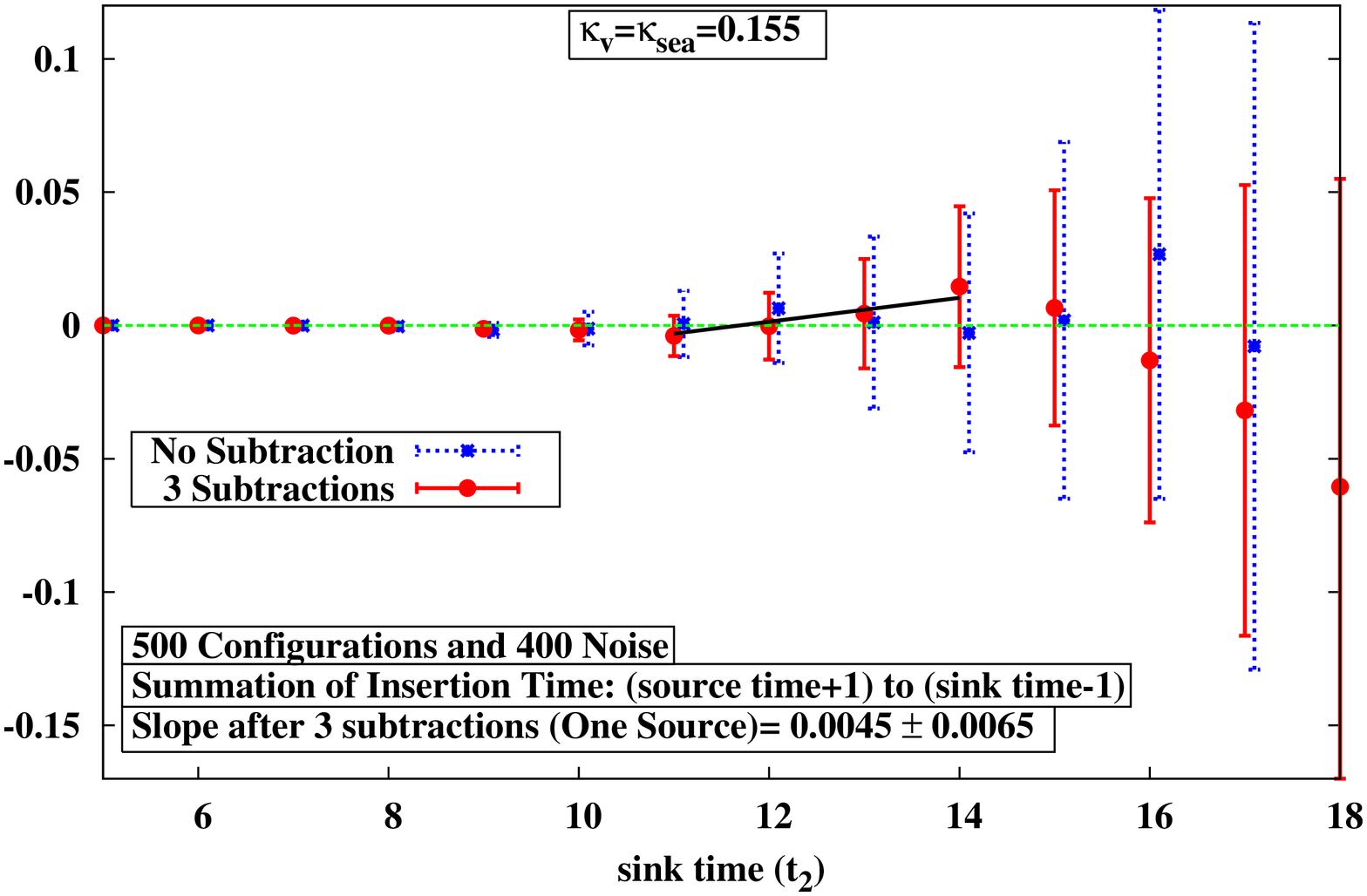}}
\label{dixsq_ud_4ii155}}
\subfigure[]
{\rotatebox{270}{\includegraphics[width=5.5cm, height=0.46\hsize]{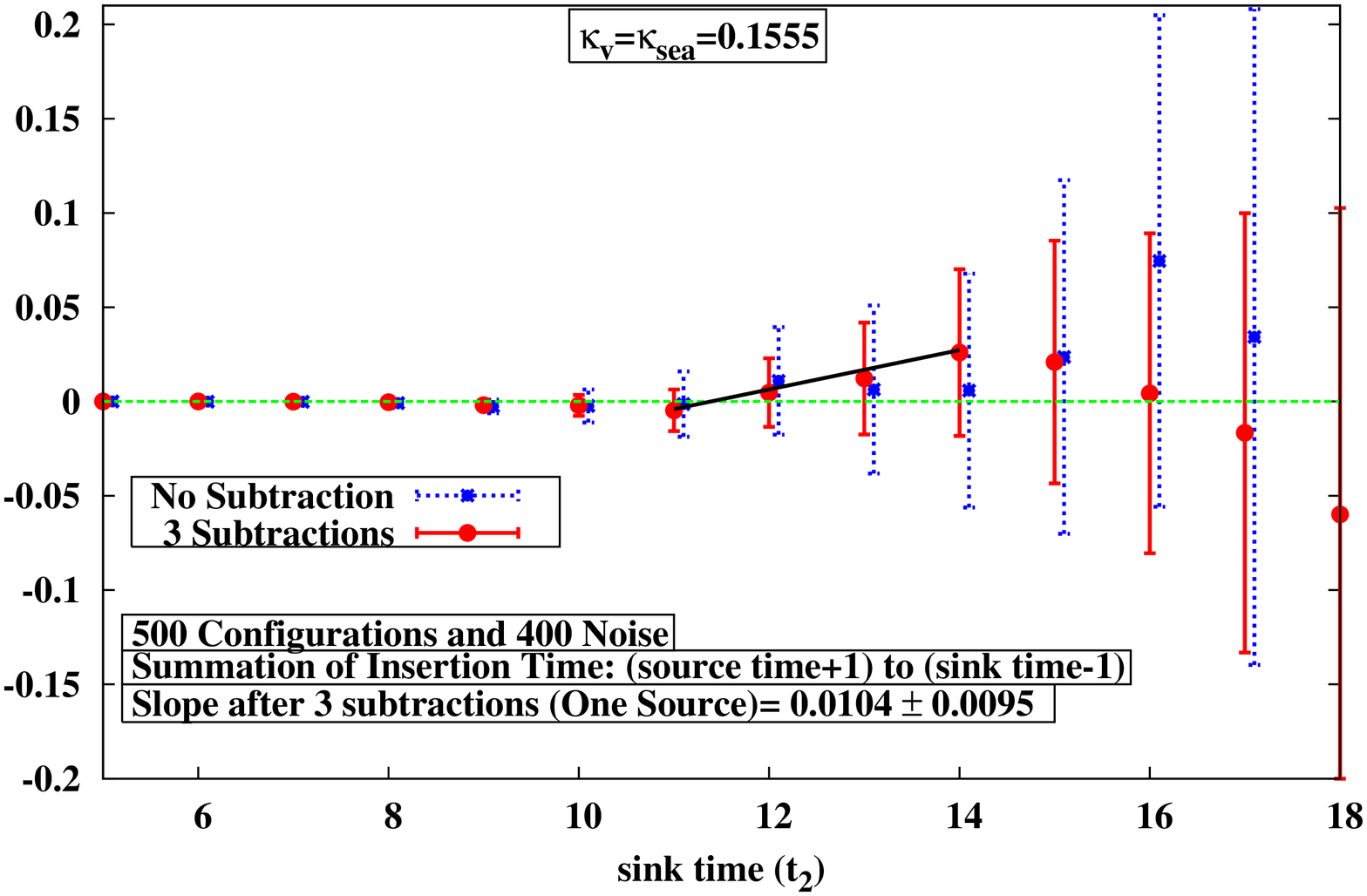}}
\label{dixsq_ud_4ii1555}}
        \hspace{0.5cm}
\subfigure[]
{\rotatebox{270}{\includegraphics[width=5.5cm, height=0.46\hsize]{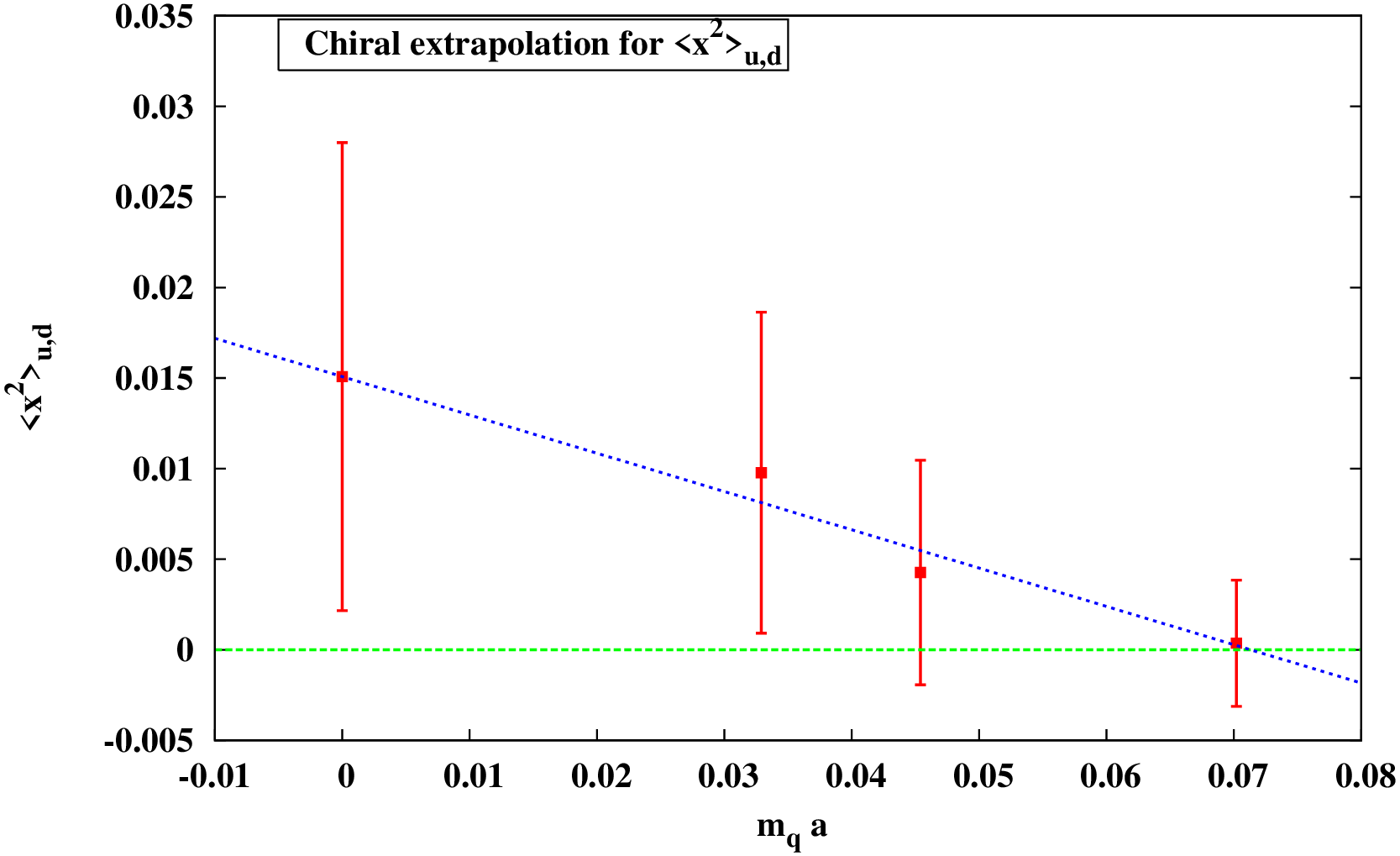}}
\label{up_chiralxsq4iiwsub}}
\caption{The ratio of the three-point to two-point functions (summed over insertion time) of the
${\cal O}_{4ii}-\frac{1}{2}\left({\cal O}_{4jj}+{\cal O}_{4kk}\right)$ operator,\ for up (down),\ is
plotted against the nucleon sink time $t_2$ at
(a) $\kappa_v = \kappa_s = 0.154$,
(b) $\kappa_v = \kappa_s = 0.155$, and
(c) $\kappa_v = \kappa_s = 0.1555$.
(d) is the linear extrapolation to the chiral limit for the second moment,\
$\langle x^2 \rangle_{u,d}$ (D.I.) plotted against $m_q a$.}
\label{fig:dis_xsq_ud}
\end{figure}

\begin{figure}[!hbtp]
\centering \subfigure[]
{\rotatebox{270}{\includegraphics[width=5.5cm, height=0.46\hsize]{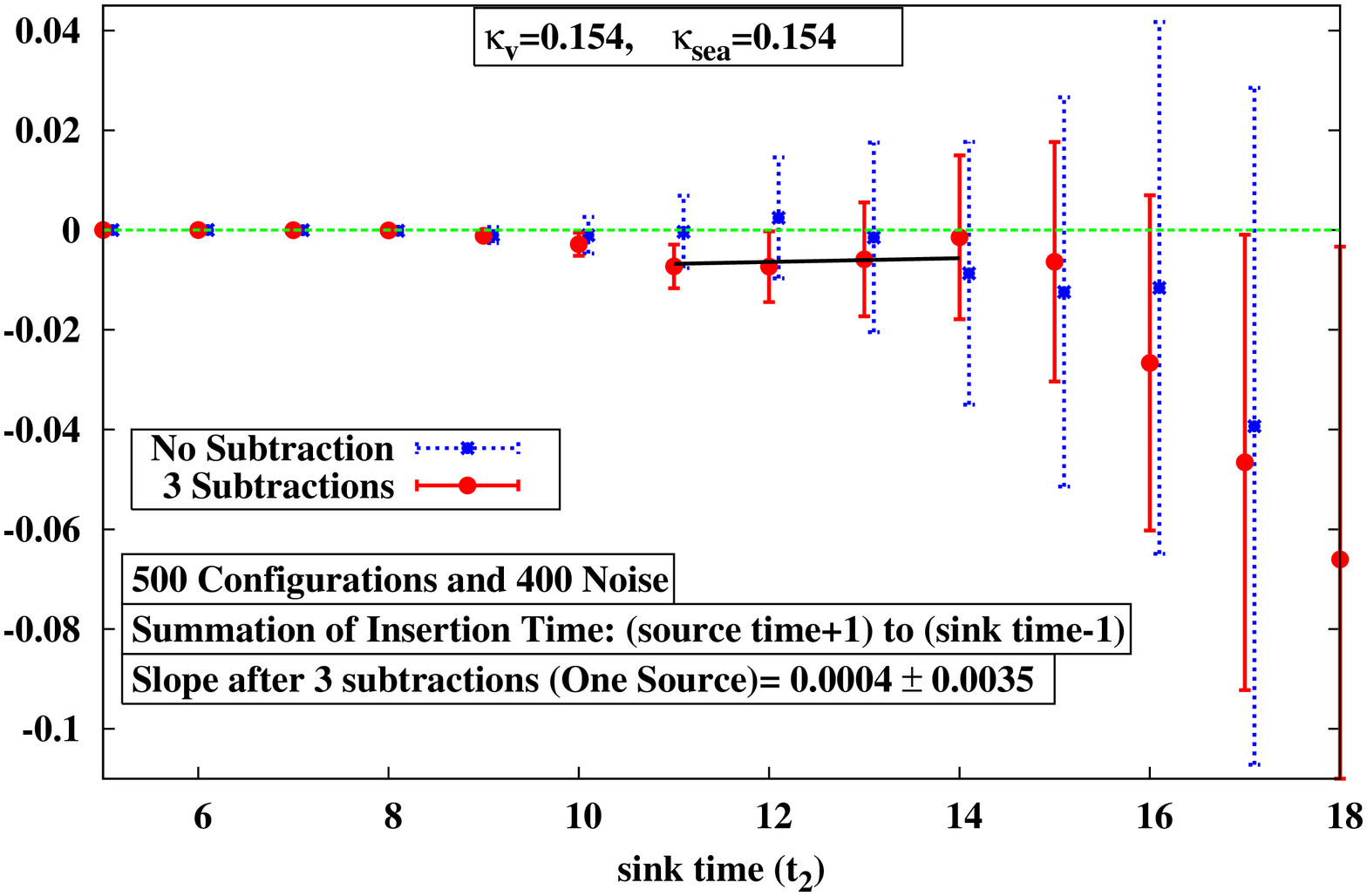}}
\label{dixsq_s_4ii154}}
        \hspace{0.5cm}
 \subfigure[]
{\rotatebox{270}{\includegraphics[width=5.5cm, height=0.46\hsize]{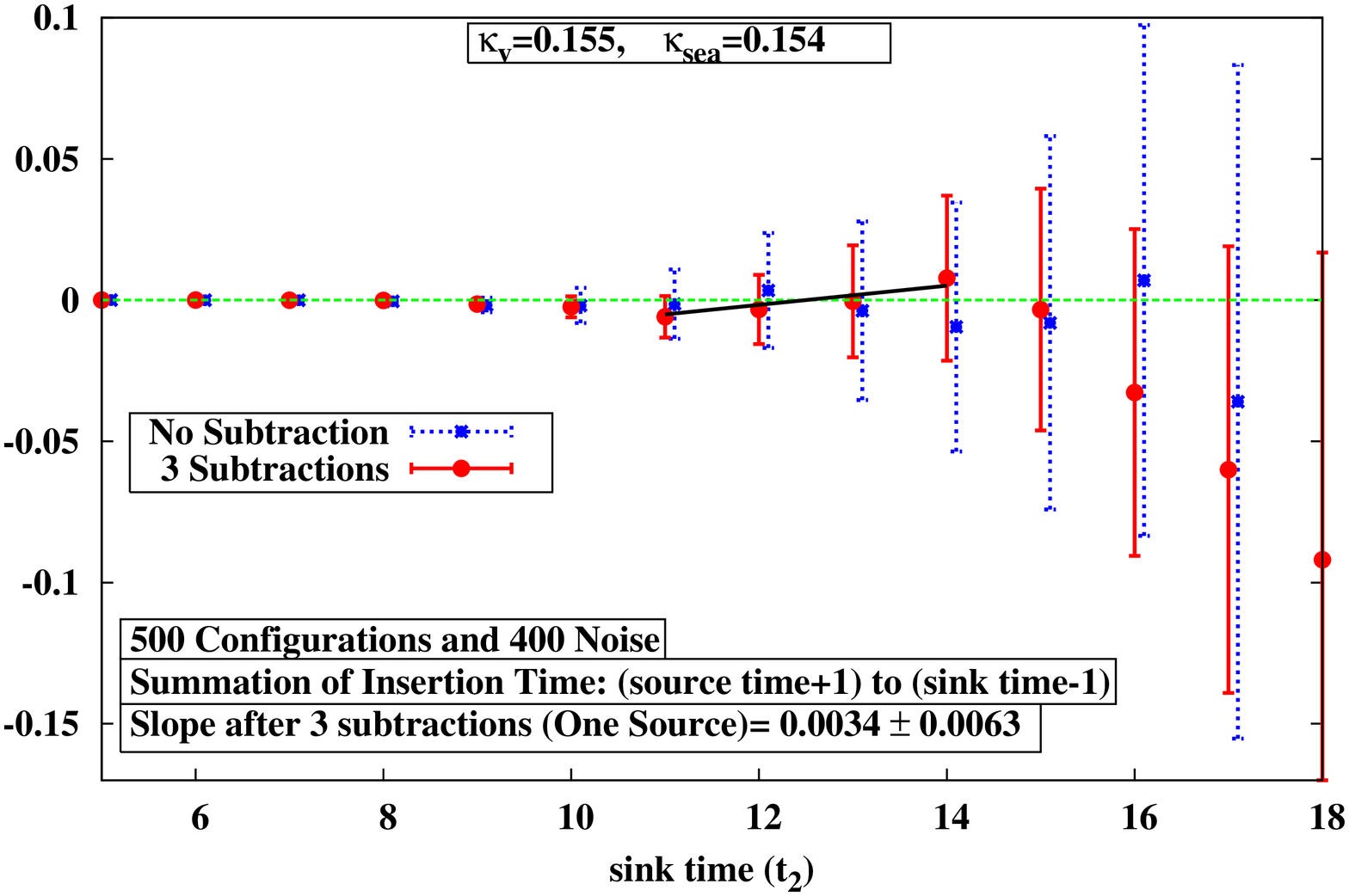}}
\label{dixsq_s_4ii155}}
\subfigure[]
{\rotatebox{270}{\includegraphics[width=5.5cm, height=0.46\hsize]{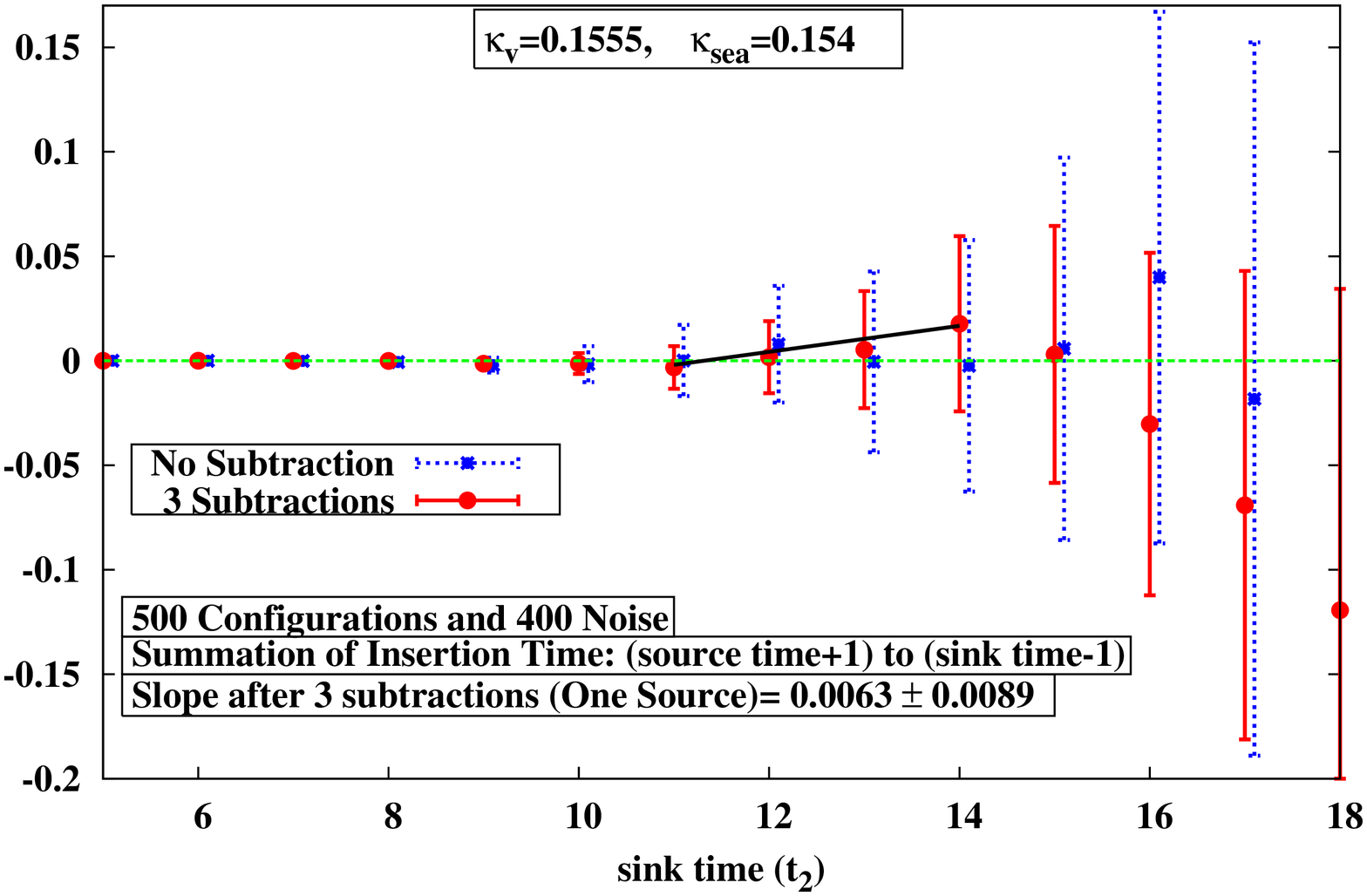}}
\label{dixsq_s_4ii1555}}
        \hspace{0.5cm}
\subfigure[]
{\rotatebox{270}{\includegraphics[width=5.5cm, height=0.46\hsize]{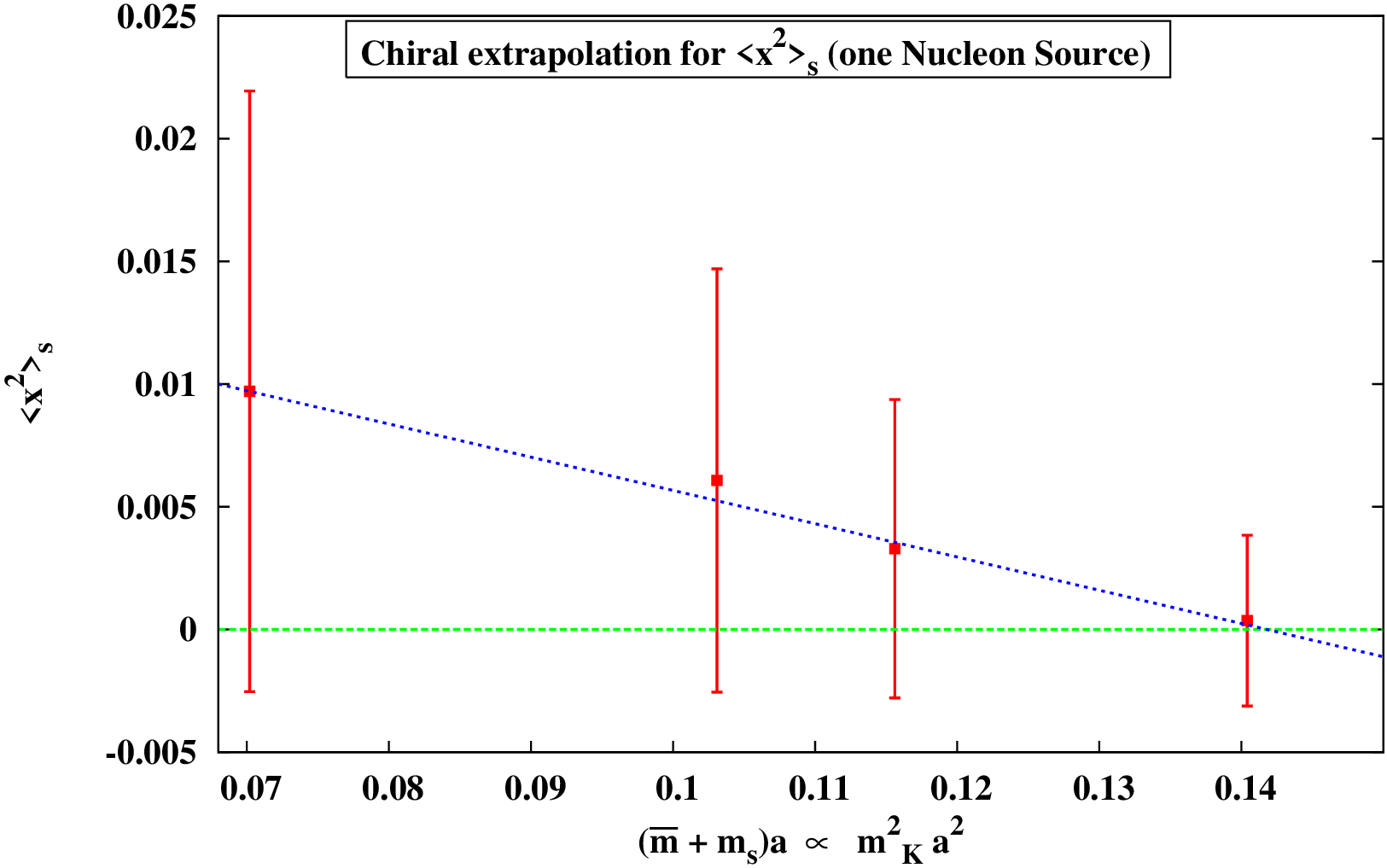}}
\label{strange_chiralxsq4iiwsub}}
\caption{The ratio of the
three-point to two-point functions (summed over insertion time) of the
${\cal O}_{4ii}-\frac{1}{2}\left({\cal O}_{4jj}+{\cal
O}_{4kk}\right)$ operator,\ for strange quarks,\ is plotted against
the nucleon sink time $t_2$ at
(a) $\kappa_v = 0.154$ and $\kappa_s = 0.154$,
(b) $\kappa_v = 0.155$ and $\kappa_s = 0.154$, and
(c) $\kappa_v = 0.1555$ and $\kappa_s = 0.154$.
(d) is the linear extrapolation to the chiral limit for the second moment,\
$\langle x^2 \rangle$ (D.I.),\ with fixed strange quark plotted
against $(\overline{m}+m_s) a$.}
\label{fig:dis_xsq_s}
\end{figure}

\begin{figure}[h]
\rotatebox{270}{\includegraphics[width=8cm,height=12cm]{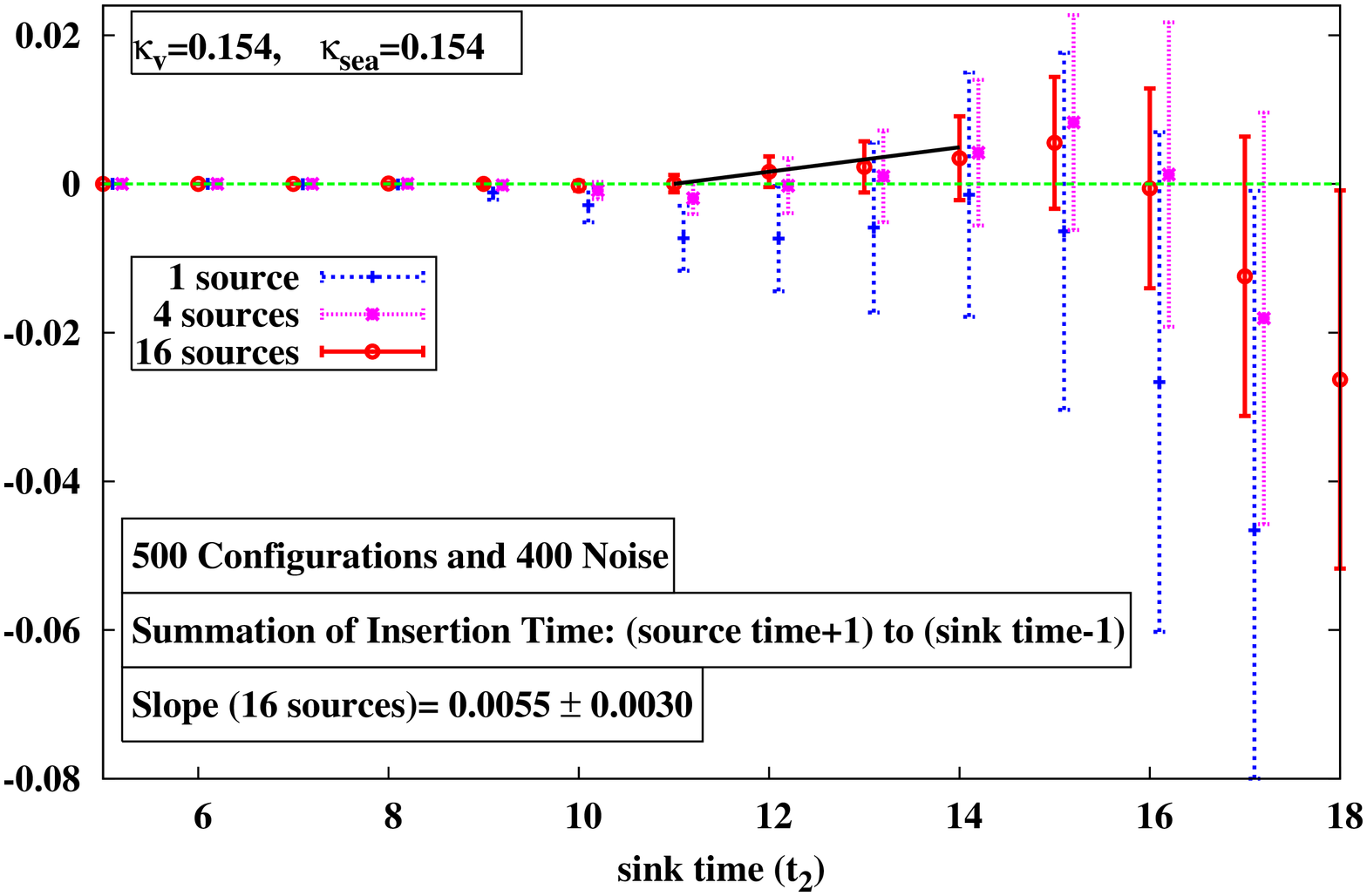}}
\caption{The ratio of the three-point to two-point functions for the
${\cal O}_{4ii}-\frac{1}{2}\left({\cal O}_{4jj}+{\cal O}_{4kk}\right)$
operator is plotted against the nucleon sink time $t_2$ at
$\kappa_v = 0.154$ and $\kappa_s = 0.154$ for 1, 4, and 16 sources after 3 subtractions.}
\label{fig:secmom_slope_all_source}
\end{figure}

\begin{figure}[!hbtp]
\centering \subfigure[]
{\rotatebox{270}{\includegraphics[width=5.5cm, height=0.46\hsize]{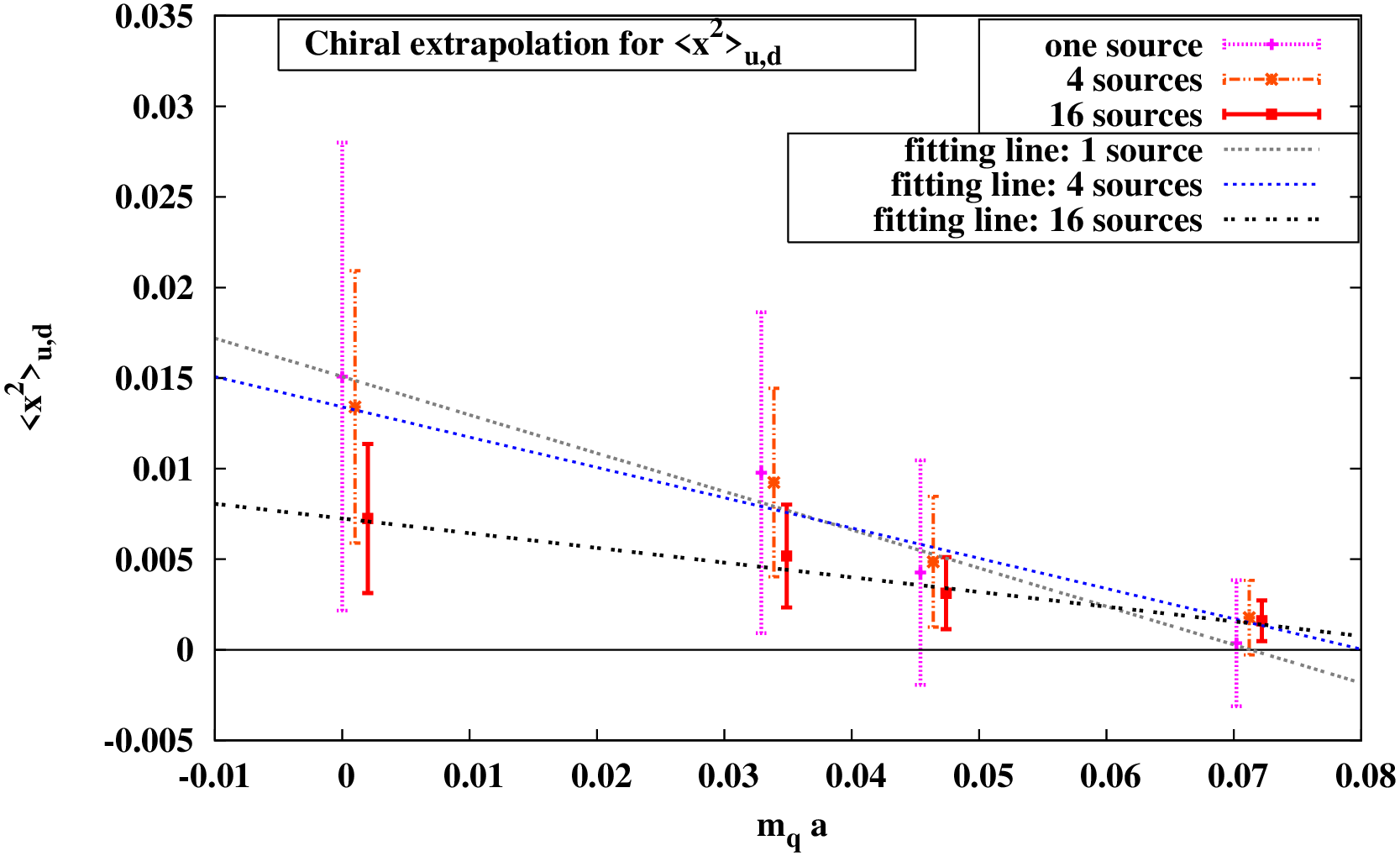}}
\label{dixsq_chiral_ud_all_source}}
        \hspace{0.5cm}
 \subfigure[]
{\rotatebox{270}{\includegraphics[width=5.5cm, height=0.46\hsize]{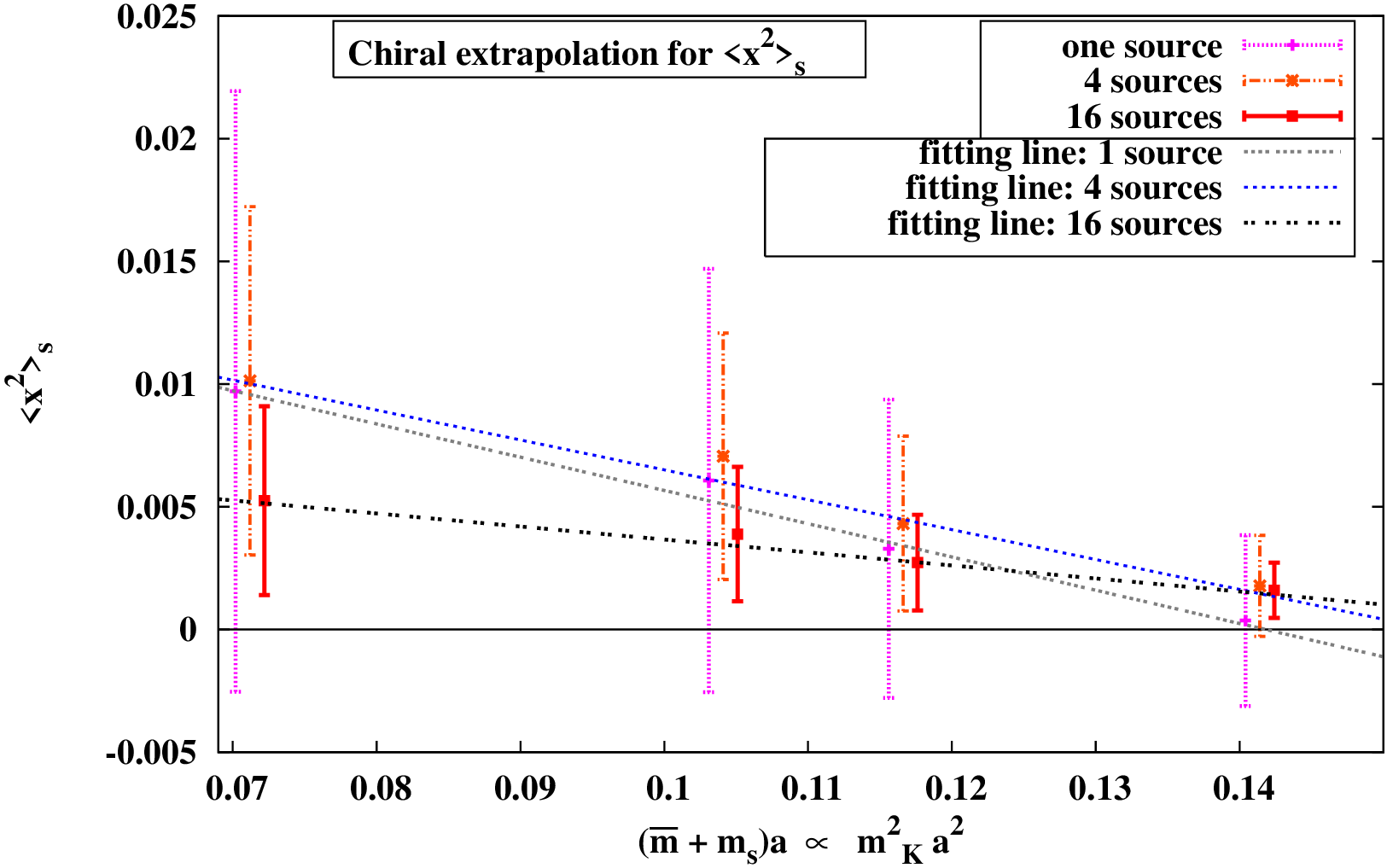}}
\label{dixsq_chiral_s_all_source}}
\caption{Linear extrapolation to the chiral limit for the second moment (a) for $\langle x^2
\rangle_{u,d}$, and (b) for $\langle x^2 \rangle_{s}$ with 1, 4, and 16 nucleon sources.}
\label{fig:dixsq_chiral_all_source}
\end{figure}

\begin{table}[h]	
\centering							
{\small					
\begin{tabular}{|p{1.1cm}|c|c|c|c|c|}												
\hline\hline
& & 1 source & 1 source & 4 sources & 16 sources\\	
& & (No Sub) & (4Sub)	& (4Sub) & (4Sub)\\								
\hline \hline													
\multirow{2}{*}{$\langle x^2 \rangle_{u,d}$} & $\kappa_v = \kappa_{\rm sea}=0.154$ & 0.0020 $\pm$ 0.0060 & 0.0004 $\pm$ 0.0035 & 0.0018 $\pm$  0.0021 & 0.015 $\pm$  0.0011 \\
\cline{2-6}														
\multirow{3}{*}{ (D.I.)} & $\kappa_v = \kappa_{\rm sea}=0.155$ & 0.0056 $\pm$ 0.0098 & 0.0043 $\pm$ 0.0620 & 0.0049$\pm$  0.0036 & 0.0031 $\pm$  0.0020 \\	
\cline{2-6}														
& $\kappa_v = \kappa_{\rm sea}=0.1555$ & 0.0122 $\pm$ 0.0133 & 0.0098 $\pm$ 0.0089 & 0.0092 $\pm$  0.0052 & 0.0052 $\pm$  0.0028  \\	
\cline{2-6}														
& Linear Extrapolation & 0.0195 $\pm$  0.0213 & 0.0168 $\pm$ 0.0144 & 0.0150 $\pm$  0.0084 & 0.0081 $\pm$  0.0046 \\					
\hline\hline														
\multirow{2}{*}{$\langle x^2 \rangle_{s-\bar s}$} & $\kappa_v = 0.154, \kappa_{\rm sea}= 0.154$ & 0.0020 $\pm$ 0.0060 & 0.0004 $\pm$ 0.0035 & 0.0018 $\pm$  0.0021 & 0.015 $\pm$  0.0011 \\	
\cline{2-6}														
\multirow{3}{*}{ (D.I.)} & $\kappa_v = 0.155,  \kappa_{\rm sea} = 0.154$ & 0.0048 $\pm$ 0.0099 & 0.0033 $\pm$  0.0061 & 0.0043 $\pm$ 0.0036 & 0.010 $\pm$ 0.0071 \\	
\cline{2-6}														
& $\kappa_v = 0.1555, \kappa_{\rm sea} = 0.154$ & 0.0086 $\pm$ 0.0134 & 0.0061 $\pm$  0.0086 & 0.0071  $\pm$ 0.0050 & 0.0039 $\pm$ 0.0027  \\	
\cline{2-6}														
& Linear Extrapolation & 0.0132 $\pm$ 0.0210 & 0.0108 $\pm$ 0.0137 & 0.0113 $\pm$ 0.0079 &  0.0059 $\pm$ 0.0043\\					
\hline\hline														
\end{tabular}	
}												
\caption{$\langle x^2 \rangle$ (D.I.) for up (down) and strange quarks at various					
$\kappa$'s and the linearly extrapolated results to the chiral limit with different
number of nucleon sources.}													
\label{table:dix4ii}													
\end{table}

We will now present the results for the second moments obtained by
using the current ${\cal O}_{4ii}-\frac{1}{2}\left({\cal
O}_{4jj}+{\cal O}_{4kk}\right) (i \neq j\neq k;\, i,j,k=1,2,3)$ for up
(down) and strange quarks (see Figs.~\ref{fig:dis_xsq_ud}, \ref{fig:dis_xsq_s},
\ref{fig:secmom_slope_all_source} and \ref{fig:dixsq_chiral_all_source}.\
We average over the results from the three operators:
${\cal O}_{411}-\frac{1}{2}\left({\cal O}_{422}+{\cal O}_{433}\right)$,\
${\cal O}_{422}-\frac{1}{2}\left({\cal O}_{433}+{\cal
O}_{411}\right)$ and ${\cal O}_{433}-\frac{1}{2}\left({\cal
O}_{411}+{\cal O}_{422}\right)$.\ So ${\cal
O}_{4ii}-\frac{1}{2}\left({\cal O}_{4jj}+{\cal O}_{4kk}\right)$ will
mean average over 1,\ 2 and 3 directions from now on.\ In addition, \ we have
used three subtraction terms ($\kappa D,\ \kappa^2 D^2$, and
$\kappa^3 D^3$). For this operator,\ we have performed the same
analysis as for the operators for the first moments,\ with the fitting
done from $t_2=11$ to 14. The values are presented in Table~\ref{table:dix4ii}.\
Unfortunately,\ we did not see any clear signal either for
$\langle x^2 \rangle_{u,d}$ $(= \langle x^2 \rangle_{u - \bar u} = \langle x^2 \rangle_{d - \bar d})$
or for $\langle x^2 \rangle_{s - \bar s}$, even with unbiased subtractions and multiples sources.
The statistical errors are large and the error bars overlap
either with zero or the signal is at best $1\sigma$ to $1.5\sigma$ away from zero for various nucleon
sources. For the best case (16 nucleon sources), the range for $\langle x^2 \rangle_{u,d}$ is
$[0.0035, 0.0127]$ and that for $\langle x^2 \rangle_{s - \bar s}$ is $[0.0016, 0.0102]$.\
We conclude that they are consistent with zero in the present calculation.


\subsection{Connected Insertions}
\label{CI}

In this section,\ we will present the results for the first and second
moments for the case of connected insertions.\ We will consider both
up and down quark currents. Unlike in D.I., they are different in the case of C.I.
\ As stated earlier,\ we consider nucleon momentum in the $x$-direction
only and we fix the nucleon sink time at $t_2=16$.

\subsubsection{First Moments}

\begin{figure}[!hbtp]
\centering \subfigure[]
{\rotatebox{270}{\includegraphics[width=5.5cm, height=0.46\hsize]{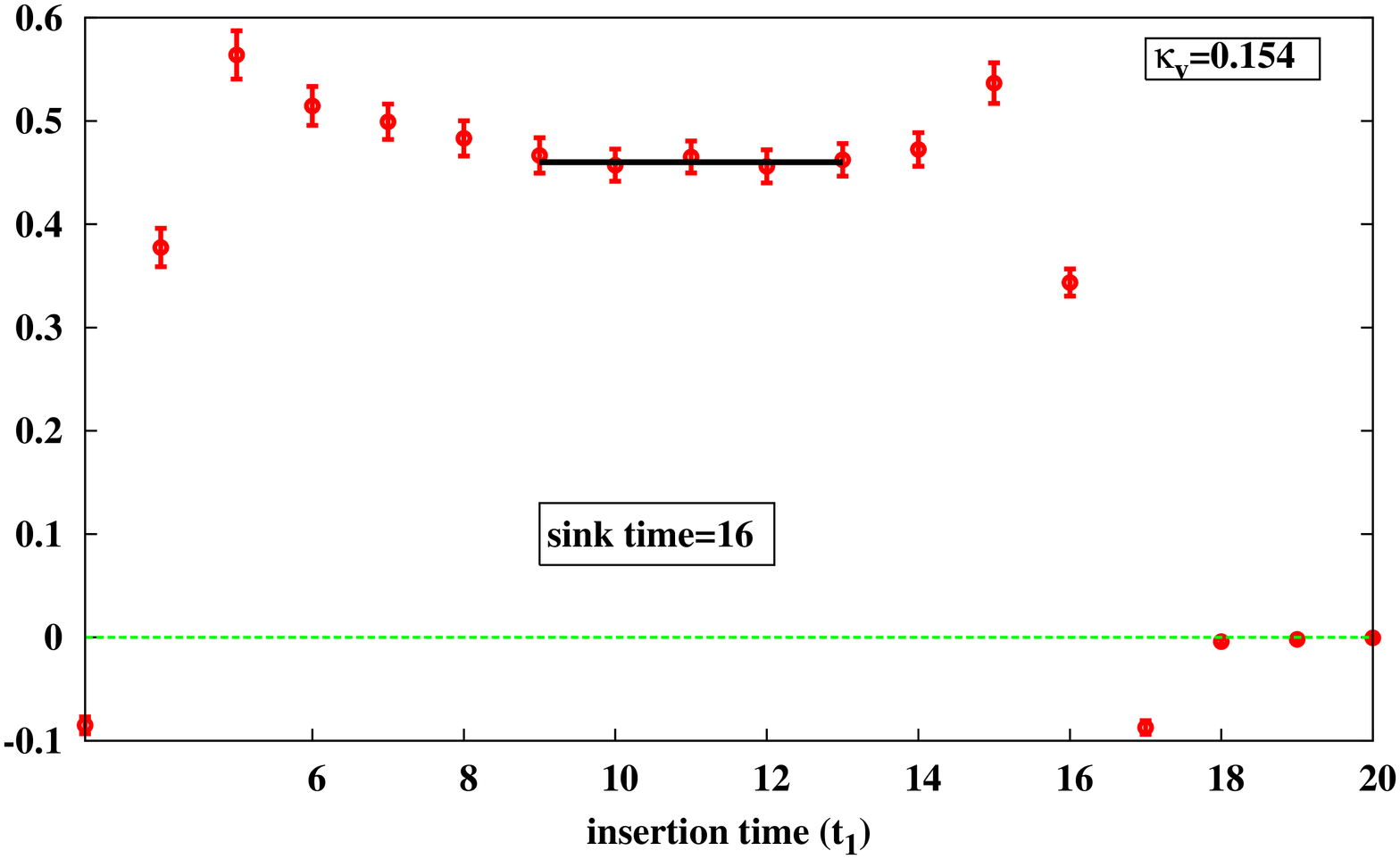}}
\label{ci154u41}}
        \hspace{0.5cm}
 \subfigure[]
{\rotatebox{270}{\includegraphics[width=5.5cm, height=0.46\hsize]{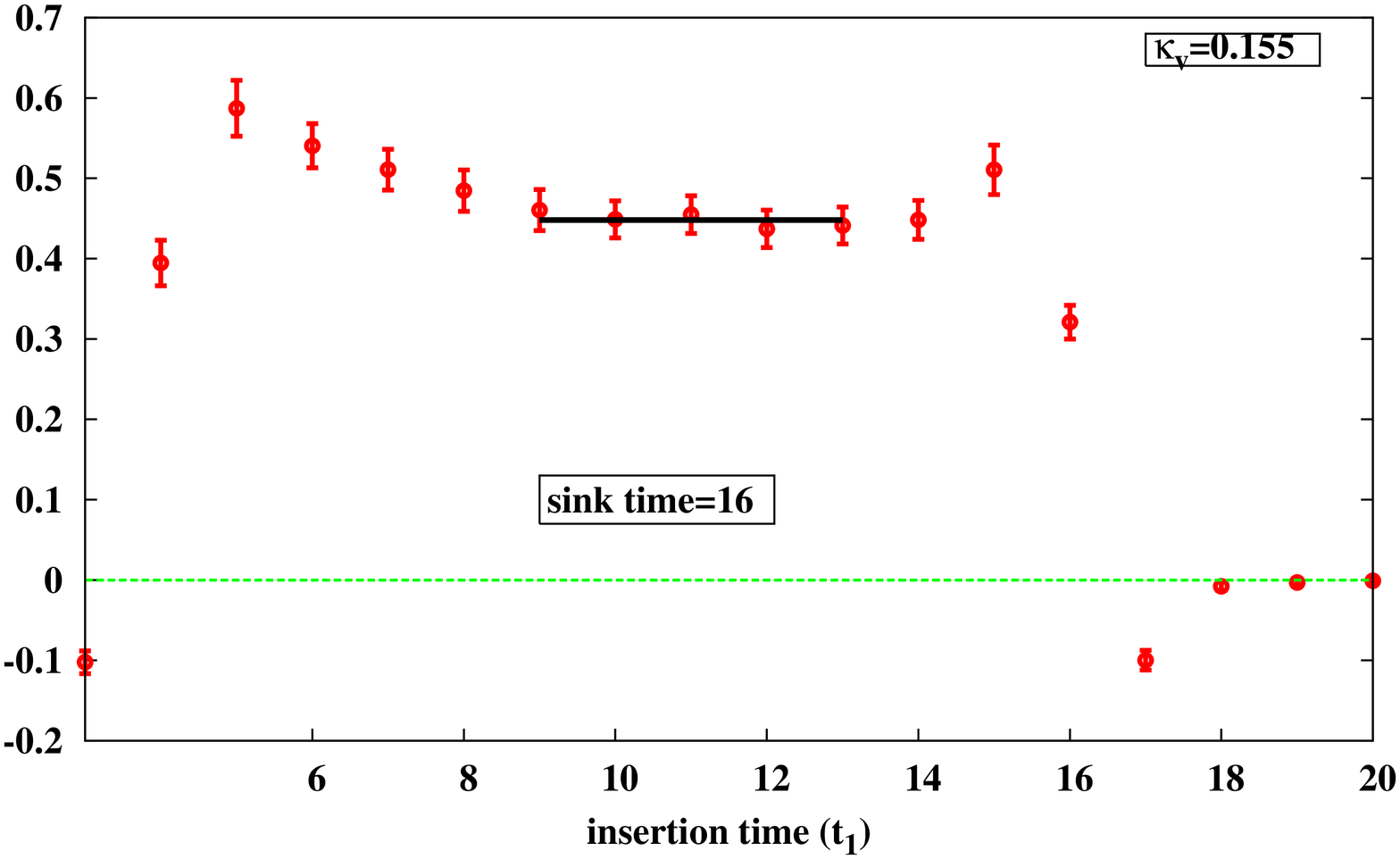}}
\label{ci155u41}}
\subfigure[]
{\rotatebox{270}{\includegraphics[width=5.5cm, height=0.46\hsize]{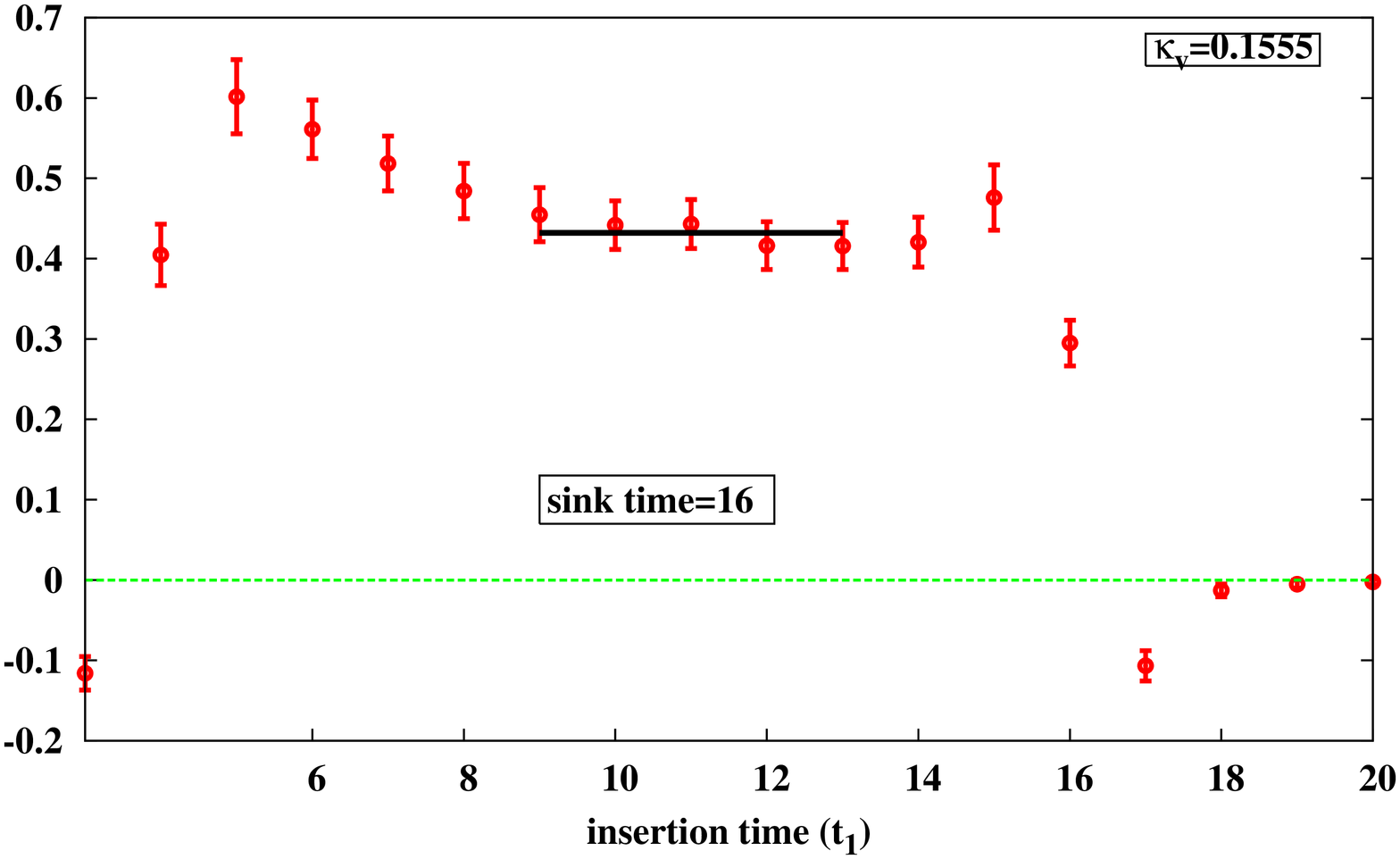}}
\label{ci1555u41}}
        \hspace{0.5cm}
\subfigure[]
{\rotatebox{270}{\includegraphics[width=5.5cm, height=0.46\hsize]{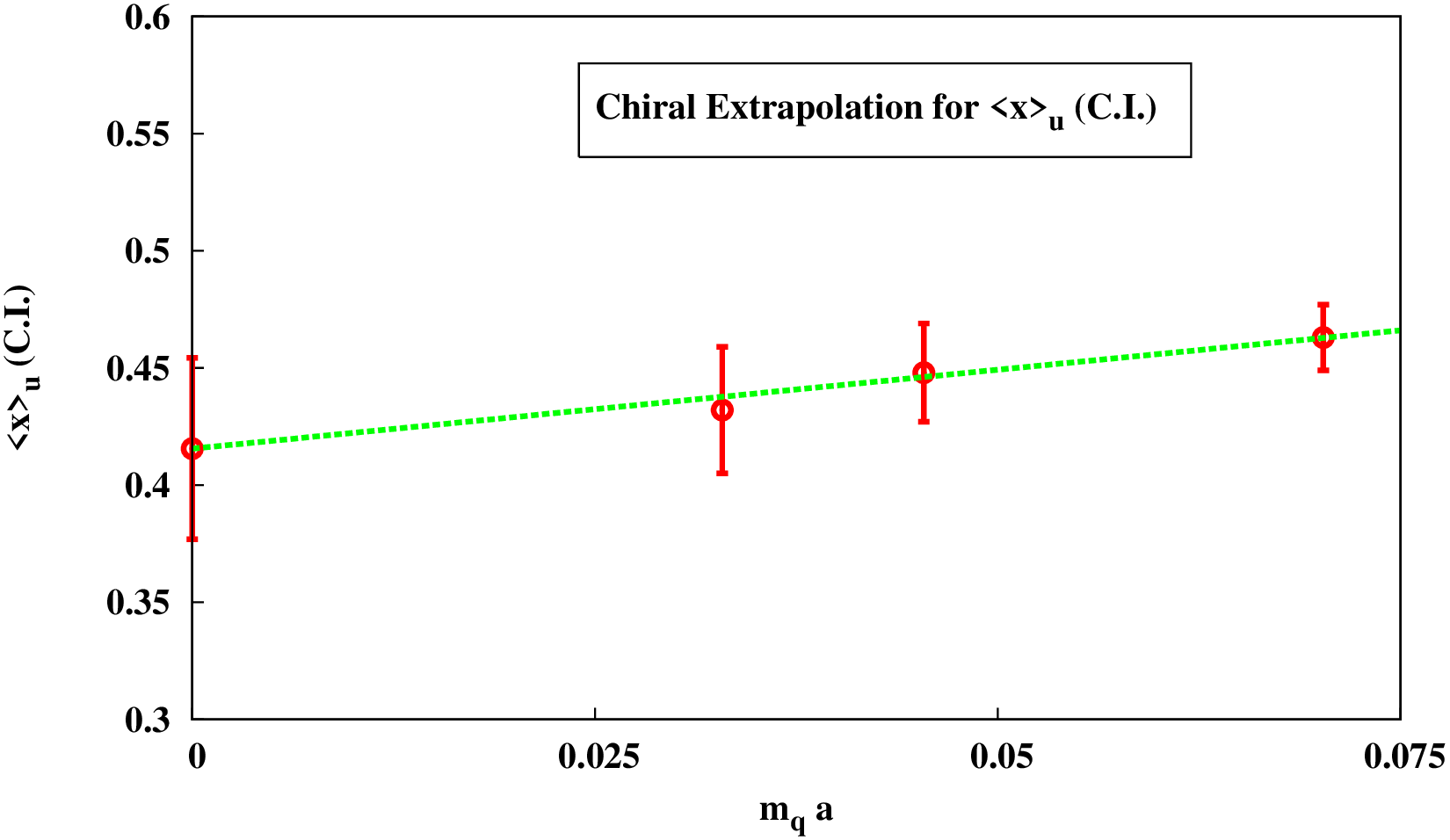}}
\label{contchiral41uci}}
 \caption{The ratio of the three-point to
two-point functions (C.I.),\ with fixed sink time,\ $t_2=16$,\ for
the ${\cal O}_{41}$ operator,\ for up quarks,\ is plotted against
the current insertion time ($t$) at
(a) $\kappa_v = 0.154$,
(b) $\kappa_v = 0.155$, and
(c) $\kappa_v = 0.1555$.
(d) is a linear extrapolation to the chiral limit plotted against $m_q a$.}
\end{figure}

First,\ we will discuss the results for the first moments by using
the current ${\cal O}_{41}$ for up  quarks.\ In
Figs.~\ref{ci154u41},~\ref{ci155u41} and~\ref{ci1555u41}, we
plot the ratio in Eq.~(\ref{ch:ciratio4i}) against the
current insertion time $t_1$ for $\kappa_v = 0.154, 0.155$, and 0.1555,
respectively.\ In these figures,\ we see that there is a plateau region
from the time slice 9 to 13.\ To extract the values of
$\langle x \rangle_{u + \bar u}$ (C.I.) at each $\kappa_v$,\ we fit a constant
between the time slice 9 to 13.\ These values and the corresponding errors
are listed in Table~\ref{table:cix}.

After obtaining $\langle x \rangle_{u + \bar u}$ (C.I.) at finite quark
mass,\ we then linearly extrapolate the valence quarks to the chiral
limit.\ Before extrapolation,\ we have converted the values of
$\langle x \rangle_{u + \bar u}$ (C.I.) to those of tadpole improved values
by using the factors in Eq.~(\ref{tadpole_fac}).\ As in the case
of $\langle x \rangle_{u,d}$ (D.I.),\ we have extrapolated with the
form $A + B m_q a$.\ In Table~\ref{table:cix},\ we have listed the
renormalized linearly extrapolated value of $\langle x \rangle_{u + \bar u}$ (C.I.).
As stated in Eq.~(\ref{renorm1.74}),\ the renormalization factor for this operator is 0.972.


Next,\ we consider the current for the down quark.\ Similar procedure
is followed as for the up quarks to obtain
the constants,\ which give the values of $\langle x \rangle_{d  + \bar d}$'s.\
Again,\ the fitting is performed from the insertion time 9 to 13.\ The
values of $\langle x \rangle_{d  + \bar d}$'s and their corresponding errors
are listed in Table~\ref{table:cix}.

\begin{figure}[!hbtp]
\centering \subfigure[]
{\rotatebox{270}{\includegraphics[width=5.5cm, height=0.46\hsize]{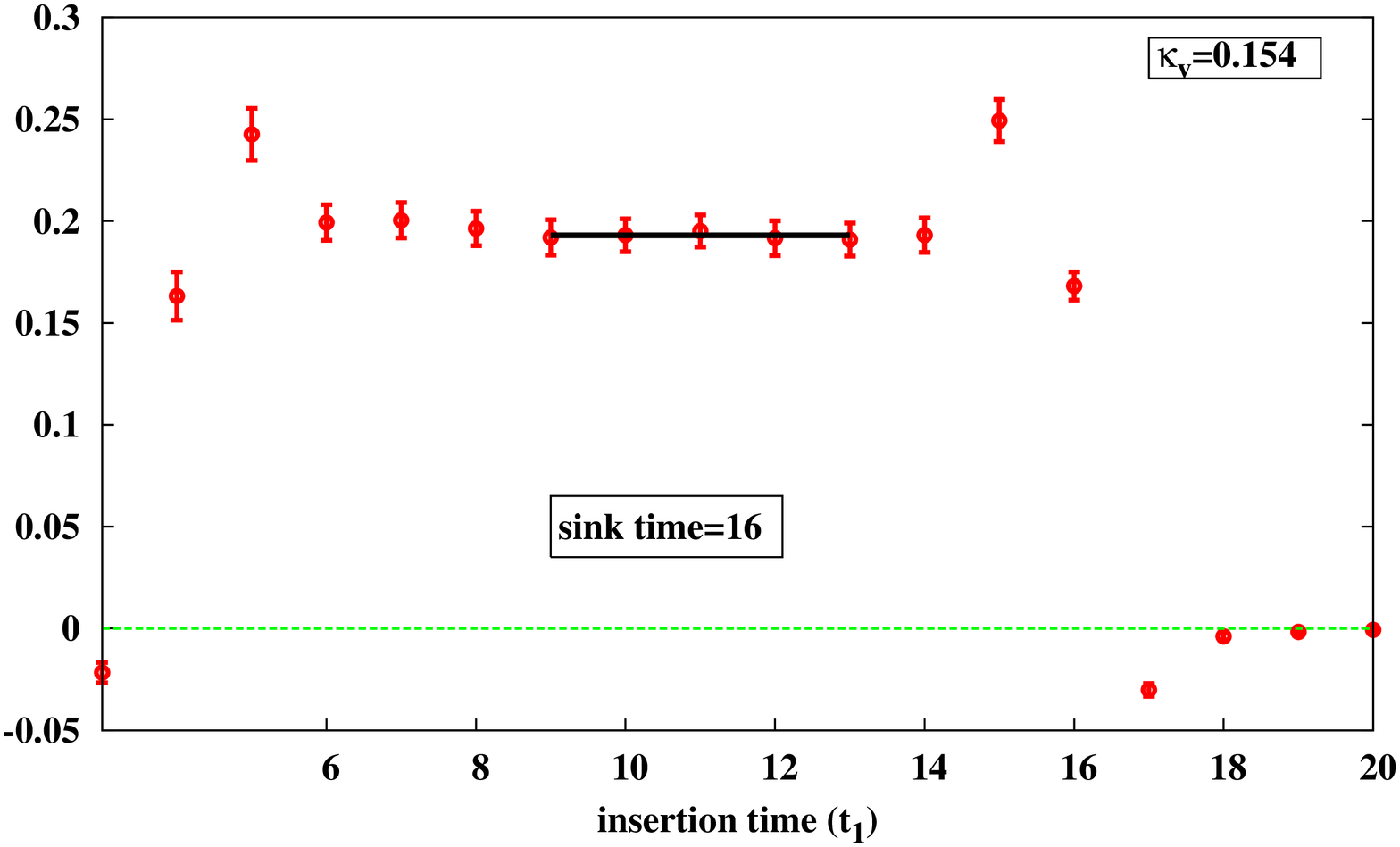}}
\label{ci154d41}}
        \hspace{0.5cm}
 \subfigure[]
{\rotatebox{270}{\includegraphics[width=5.5cm, height=0.46\hsize]{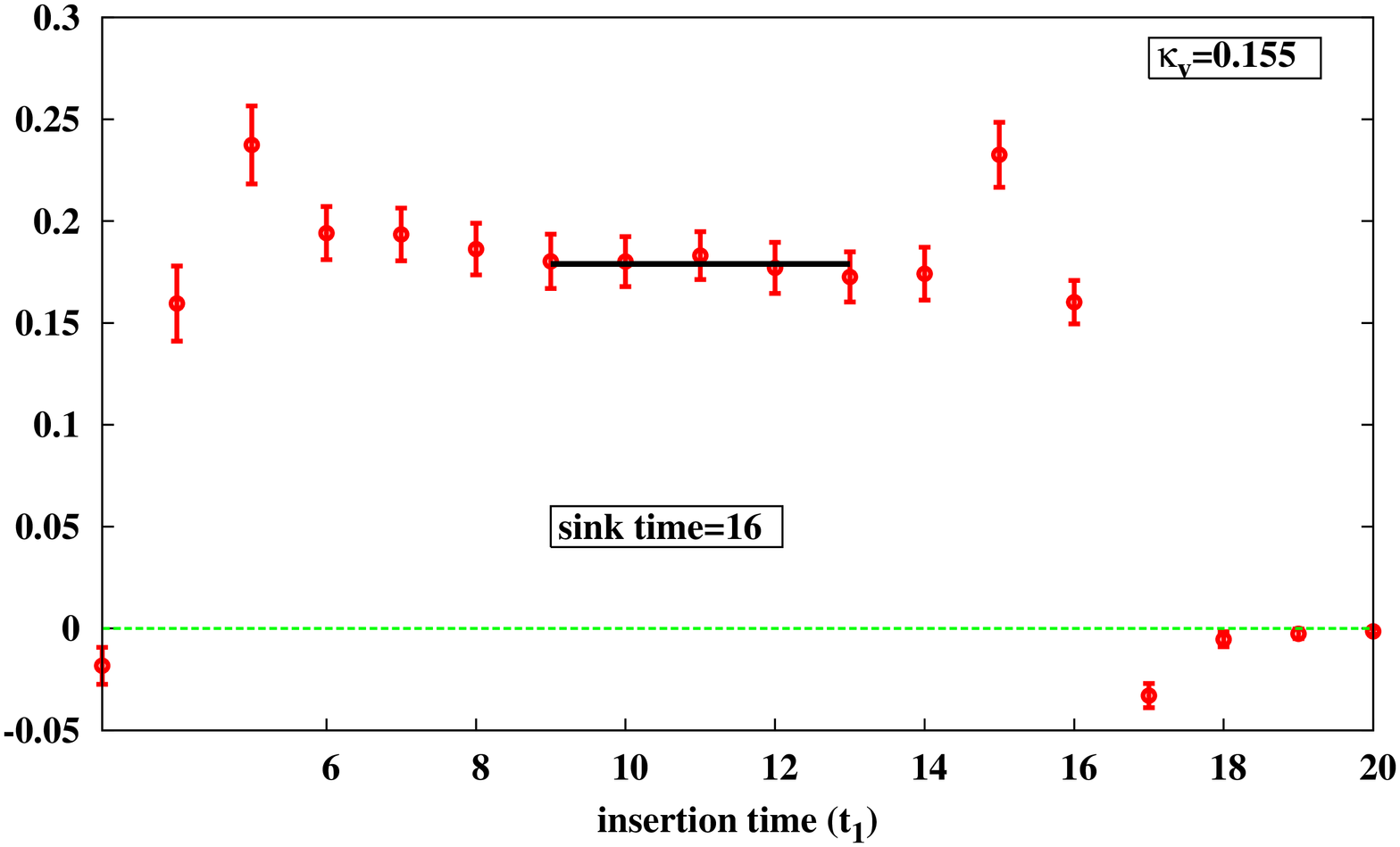}}
\label{ci155d41}}
\subfigure[]
{\rotatebox{270}{\includegraphics[width=5.5cm, height=0.46\hsize]{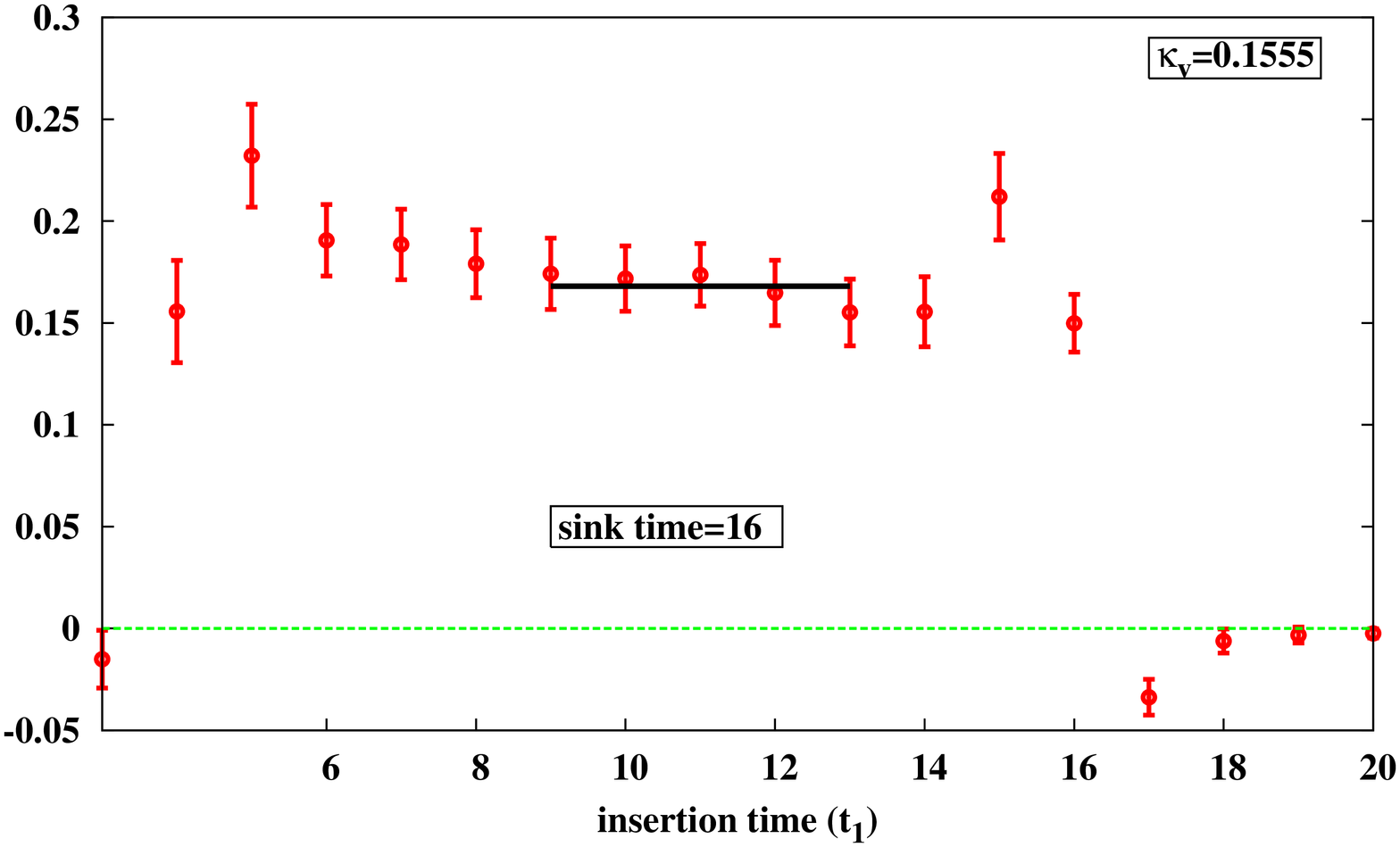}}
\label{ci1555d41}}
        \hspace{0.5cm}
\subfigure[]
{\rotatebox{270}{\includegraphics[width=5.5cm, height=0.46\hsize]{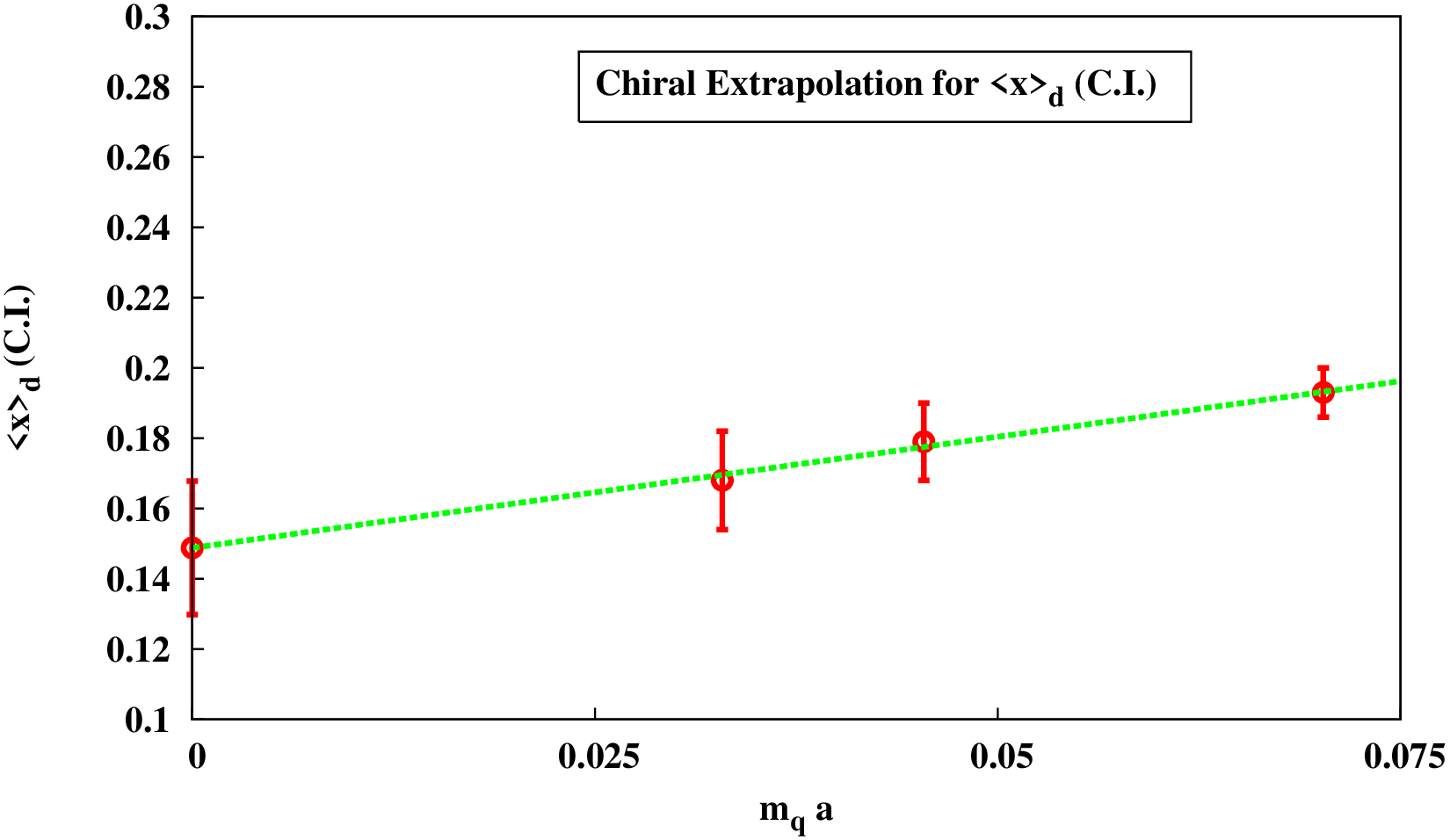}}
\label{contchiral41dci}}
\caption{The ratio of the three-point to
two-point functions (C.I.),\ with fixed sink time,\ $t_2=16$,\ for
the ${\cal O}_{41}$ operator,\ for down quarks,\ is plotted against
the current insertion time ($t$) at
(a) $\kappa_v = 0.154$,
(b) $\kappa_v = 0.155$, and
(c) $\kappa_v = 0.1555$.
(d) is a linear extrapolation to the chiral limit plotted against $m_q a$.}
\end{figure}




Now,\ we will consider the current ${\cal
O}_{44}-\frac{1}{3}\left({\cal O}_{11}+{\cal O}_{22}+{\cal
O}_{33}\right)$ for both up and down quarks.\ The fitting is performed
from the insertion time 9 to 11 for both up and down quark
currents.\ The values of the $\langle x \rangle_{u  + \bar u}$ and $\langle x
\rangle_{d + \bar d}$ (C.I.) for this operator,\ along with their errors,\
are listed in Table~\ref{table:cix}.\
The renormalization factor for this current is 0.953 as obtained
from Eq.~(\ref{renorm1.74}).


\begin{figure}[!hbtp]
\centering \subfigure[]
{\rotatebox{270}{\includegraphics[width=5.5cm, height=0.46\hsize]{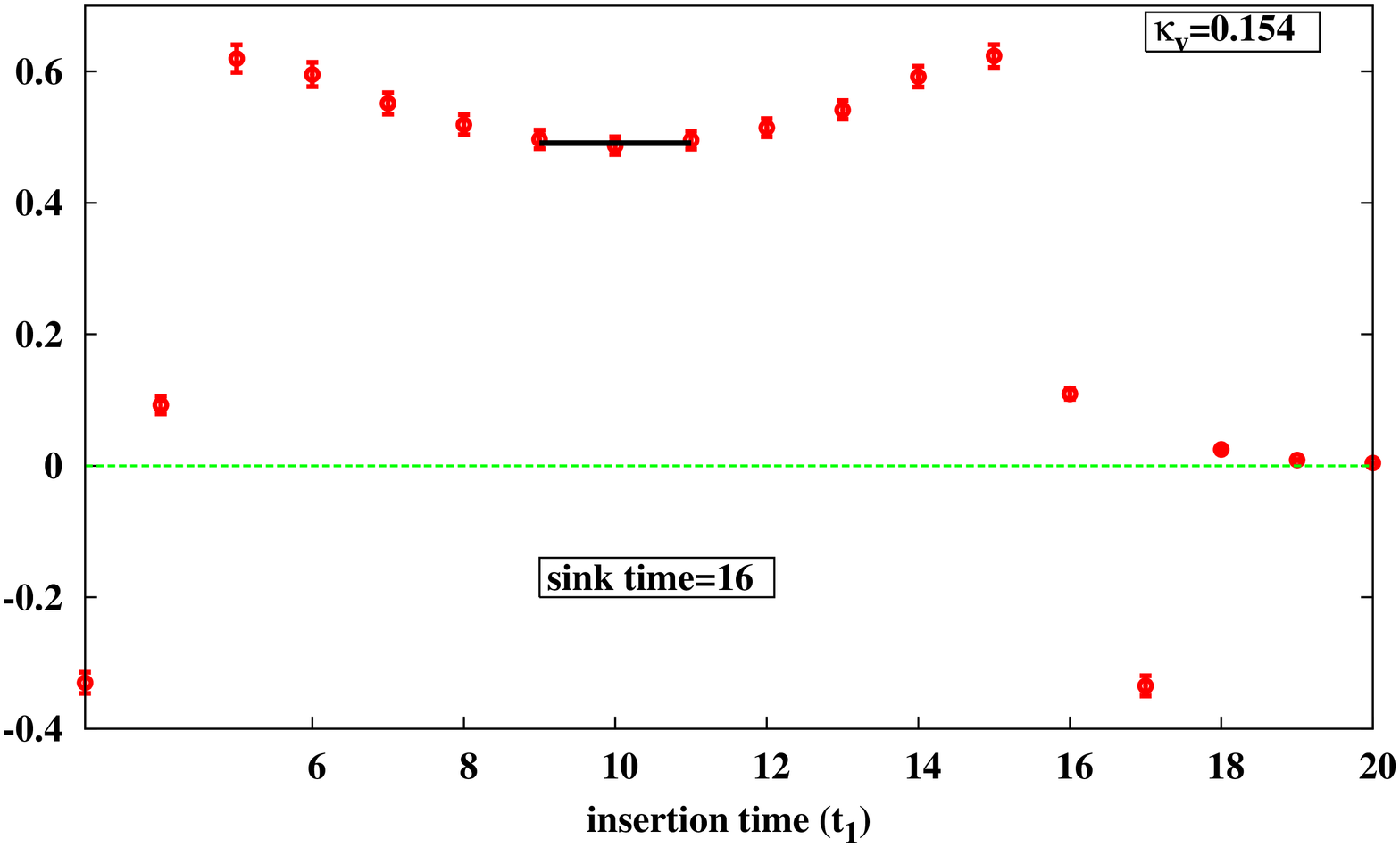}}
\label{ci154u44}}
        \hspace{0.5cm}
 \subfigure[]
{\rotatebox{270}{\includegraphics[width=5.5cm, height=0.46\hsize]{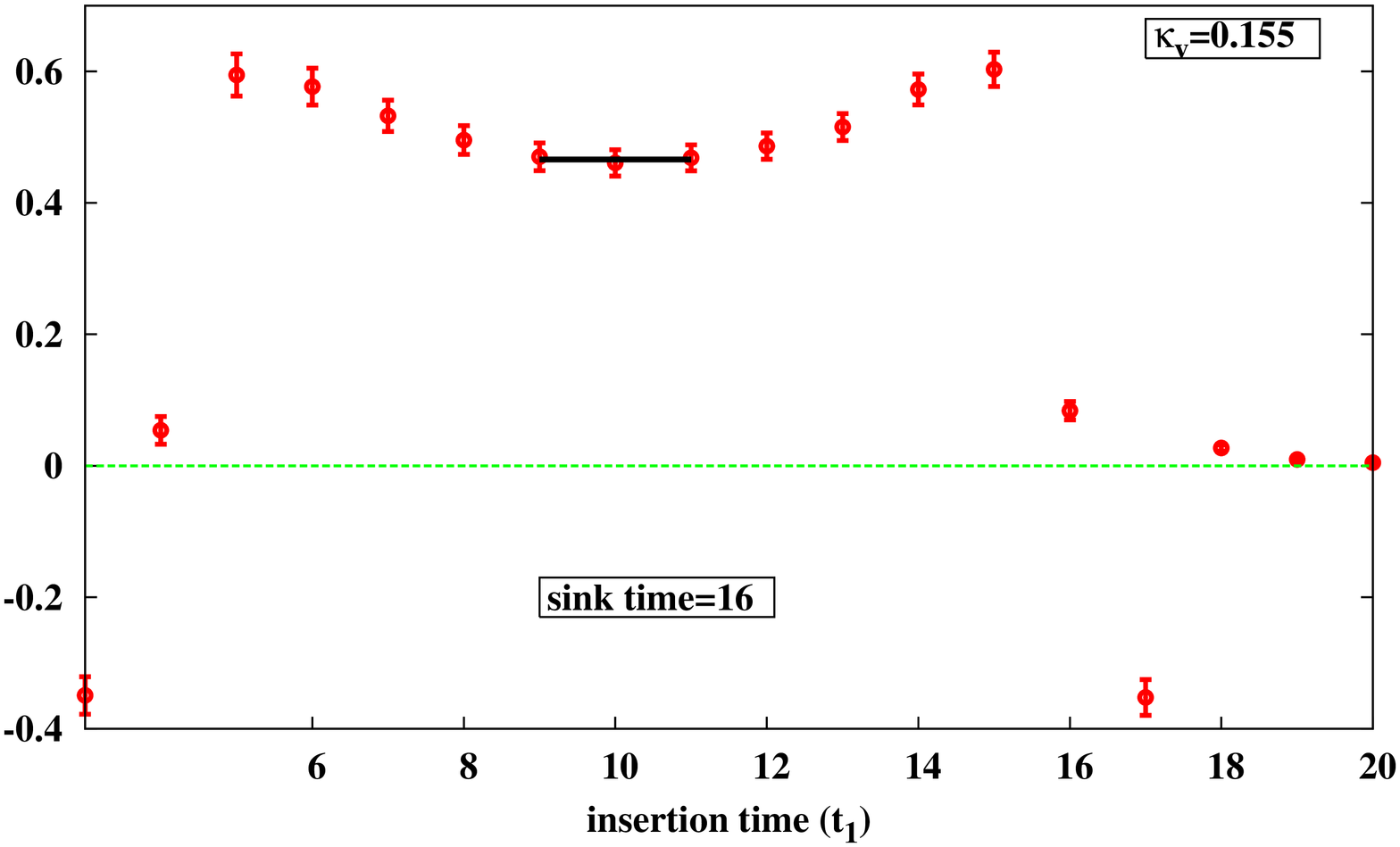}}
\label{ci155u44}}
\subfigure[]
{\rotatebox{270}{\includegraphics[width=5.5cm, height=0.46\hsize]{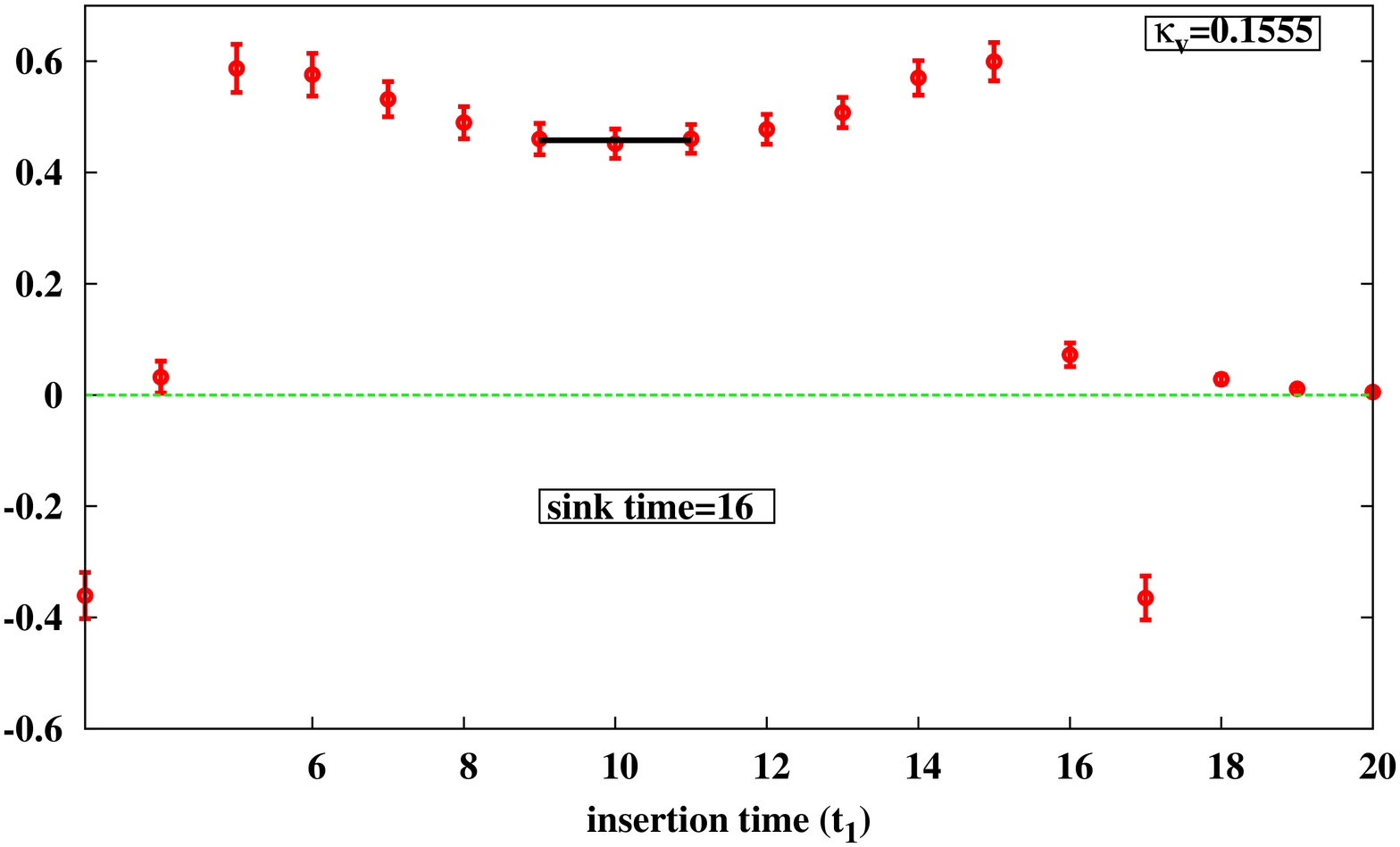}}
\label{ci1555u44}}
        \hspace{0.4cm}
\subfigure[]
{\rotatebox{270}{\includegraphics[width=5.5cm, height=0.46\hsize]{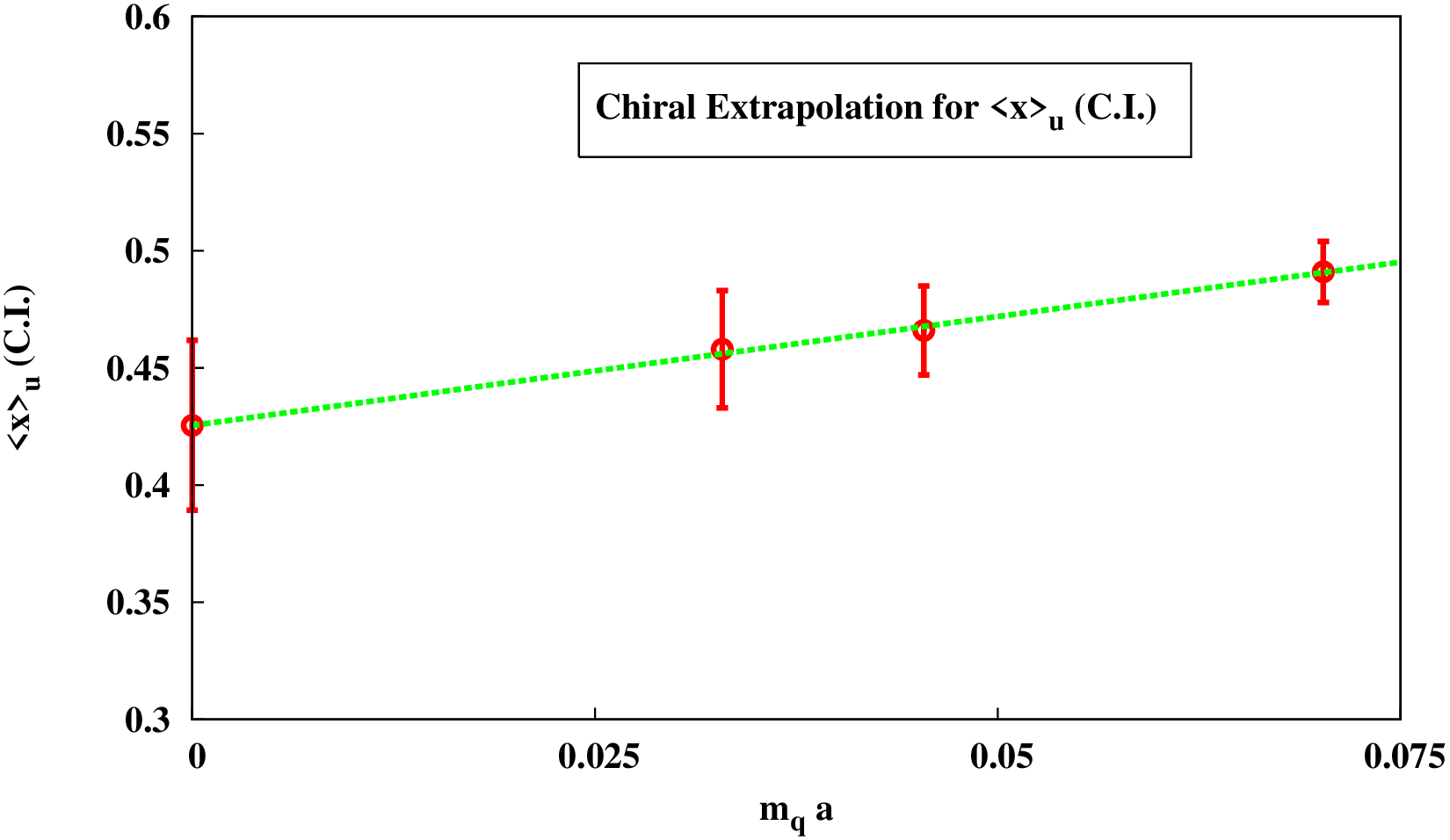}}
\label{contchiral44uci}}
\caption{The ratio of the three-point to two-point functions (C.I.),\ with fixed sink time,\ $t_2=16$,\ for
the ${\cal O}_{44}-\frac{1}{3}\left({\cal O}_{11}+{\cal O}_{22}+{\cal O}_{33}\right)$
operator,\ for up quarks,\ is plotted against the current insertion time $t_1$ at
(a) $\kappa_v = 0.154$,
(b) $\kappa_v = 0.155$, and
(c) $\kappa_v = 0.1555$.
(d) is a linear extrapolation to the chiral limit plotted against $m_q a$.}
\end{figure}


\begin{figure}[!hbtp]
\centering \subfigure[]
{\rotatebox{270}{\includegraphics[width=5.5cm, height=0.46\hsize]{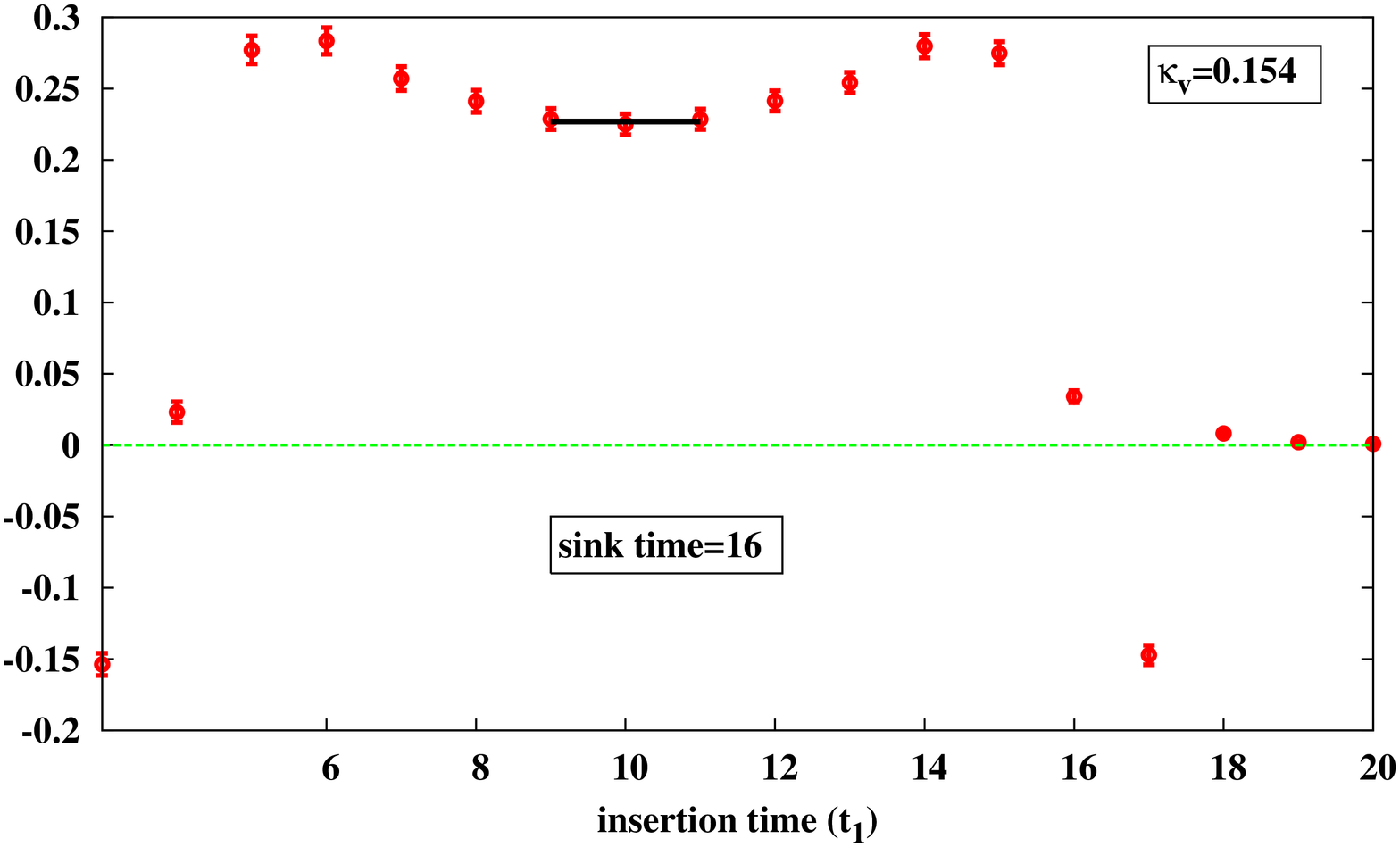}}
\label{ci154d44}}
        \hspace{0.5cm}
 \subfigure[]
{\rotatebox{270}{\includegraphics[width=5.5cm, height=0.46\hsize]{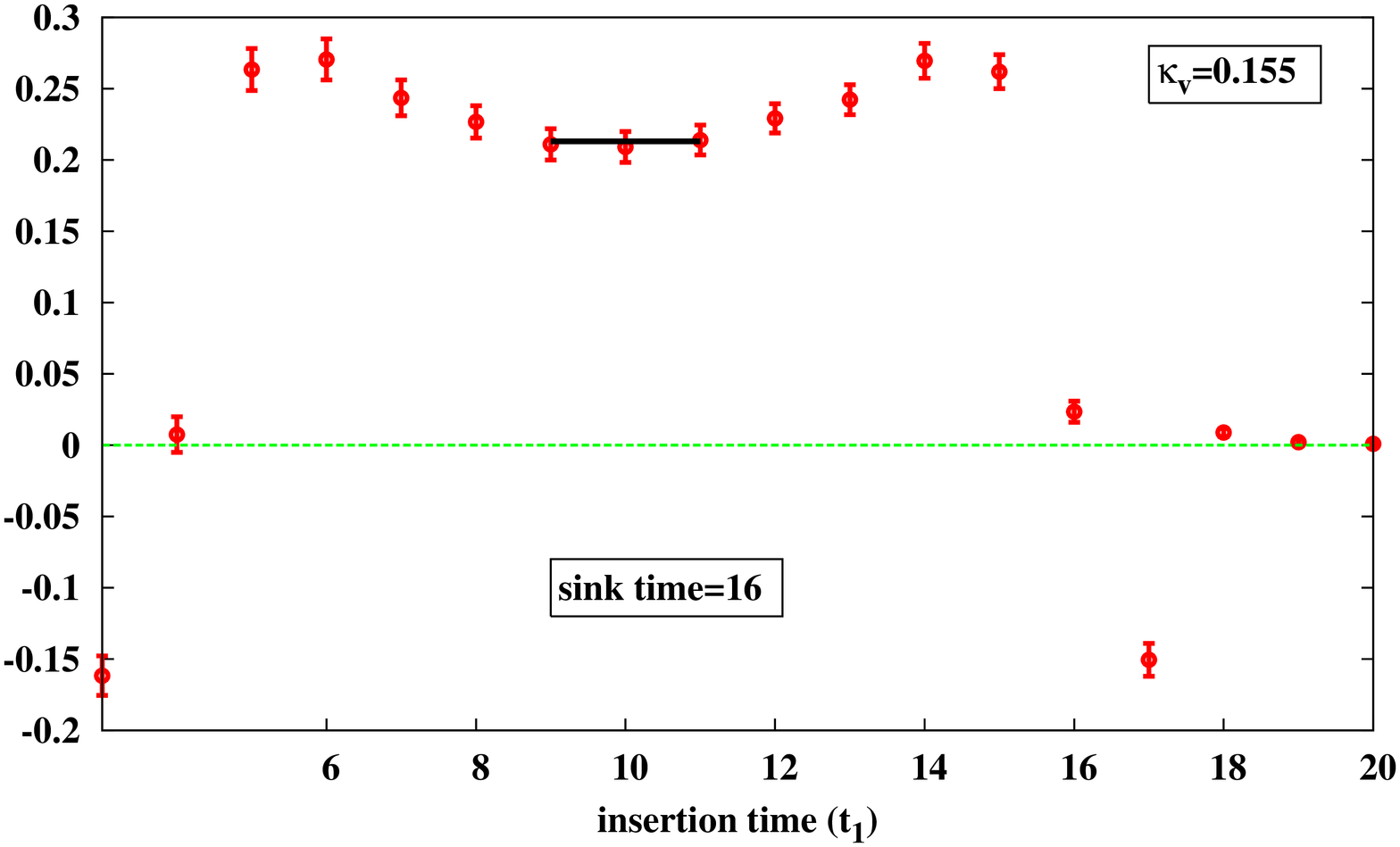}}
\label{ci155d44}}
\subfigure[]
{\rotatebox{270}{\includegraphics[width=5.5cm, height=0.46\hsize]{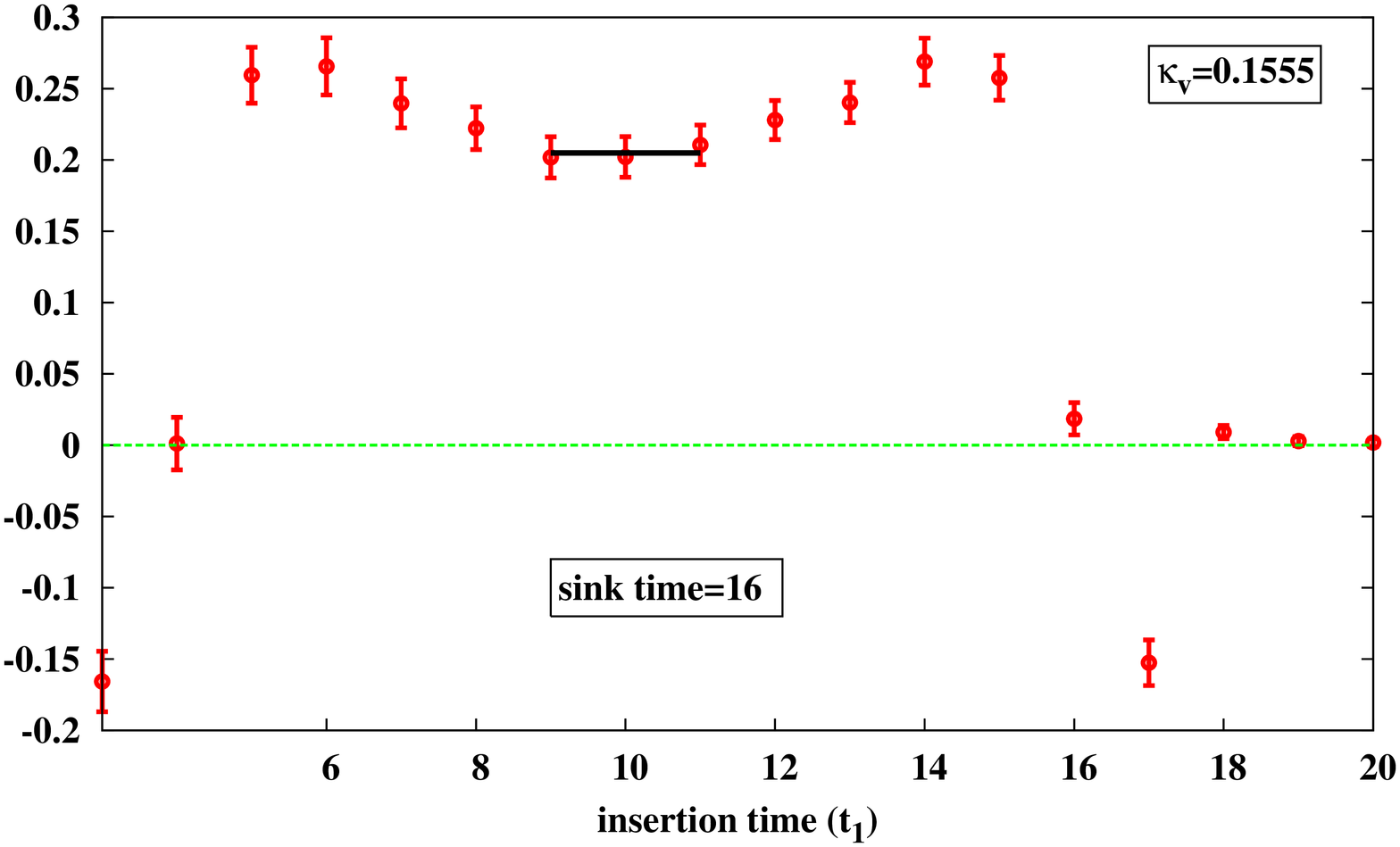}}
\label{ci1555d44}}
        \hspace{0.4cm}
\subfigure[]
{\rotatebox{270}{\includegraphics[width=5.5cm, height=0.46\hsize]{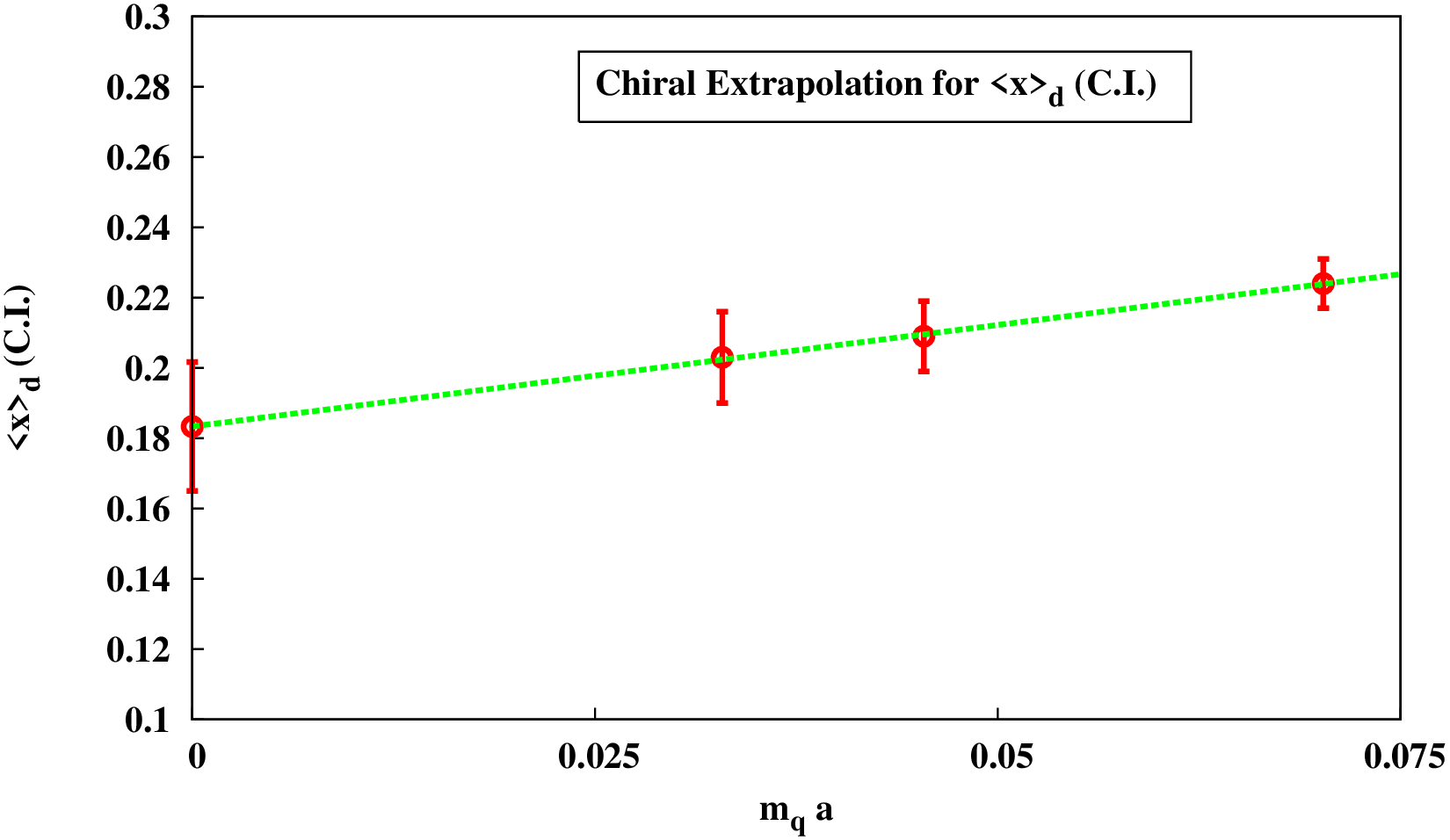}}
\label{contchiral44dci}}
\caption{The ratio of the three-point to two-point functions
(C.I.),\ with fixed sink time,\ $t_2 = 16$,\ for the ${\cal
O}_{44}-\frac{1}{3}\left({\cal O}_{11}+{\cal O}_{22}+{\cal
O}_{33}\right)$ operator,\ for down quarks,\ is plotted against the
current insertion time $t_1$ at
(a) $\kappa_v = 0.154$,
(b) $\kappa_v = 0.155$, and
(c) $\kappa_v = 0.1555$.
(d) is a linear extrapolation to the chiral limit plotted against $m_q a$.}
\end{figure}

\begin{table} [h]
\centering {
\begin{tabular}{|c|c|c|}
\hline\hline
\multirow{4}{*}{$\langle x \rangle^{41}_{u + \bar u}$ (C.I.)} & $\kappa_v = 0.154$  & 0.463 $\pm$ 0.014  \\
\cline{2-3}
& $\kappa_v = 0.155$  & 0.448 $\pm$ 0.021  \\
\cline{2-3}
& $\kappa_v = 0.1555$ & 0.432 $\pm$ 0.027  \\
\cline{2-3}
& Linear Extrapolation & 0.408 $\pm$ 0.038 \\
\hline\hline
\multirow{4}{*}{$\langle x \rangle^{41}_{d + \bar d}$ (C.I.)} & $\kappa_v = 0.154$  & 0.193 $\pm$ 0.007  \\
\cline{2-3}
& $\kappa_v = 0.155$  & 0.179 $\pm$ 0.011  \\
\cline{2-3}
& $\kappa_v = 0.1555$ & 0.168 $\pm$ 0.014  \\
\cline{2-3}
& Linear Extrapolation & 0.148 $\pm$ 0.019 \\
\hline\hline
\multirow{4}{*}{$\langle x \rangle^{44}_{u + \bar u}$ (C.I.)} & $\kappa_v = 0.154$  & 0.491 $\pm$ 0.013  \\
\cline{2-3}
& $\kappa_v = 0.155$  & 0.466 $\pm$ 0.019  \\
\cline{2-3}
& $\kappa_v = 0.1555$ & 0.458 $\pm$ 0.025  \\
\cline{2-3}
 & Linear Extrapolation & 0.420 $\pm$ 0.035 \\
\hline\hline
\multirow{4}{*}{$\langle x \rangle^{44}_{d + \bar d}$ (C.I.)} & $\kappa_v = 0.154$  & 0.224 $\pm$ 0.007  \\
\cline{2-3}
& $\kappa_v = 0.155$  & 0.209 $\pm$ 0.010  \\
\cline{2-3}
& $\kappa_v = 0.1555$ & 0.203 $\pm$ 0.013  \\
\cline{2-3}
& Linear Extrapolation & 0.181 $\pm$ 0.018 \\
\hline\hline
\end{tabular}}
\caption{Table for the values of $\langle x \rangle$ (C.I.) for
up and down quarks at various kappa values and after linear extrapolation
to the chiral limit for the ${\cal O}_{41}$ and
${\cal O}_{44}-\frac{1}{3}\left({\cal O}_{11}+{\cal O}_{22}+{\cal O}_{33}\right)$
operators.}
\label{table:cix}
\end{table}

\subsubsection{Second Moment}

For the second moment,\ we consider the operator ${\cal
O}_{411}-\frac{1}{2}\left({\cal O}_{422}+{\cal O}_{433}\right)$ for
both the up and down quarks.\ We did the similar analysis as in the case
of the first moments.\ The fitting is performed from the insertion time 9 to
11 for both up and down quark currents.\ The values of the $\langle
x^2 \rangle_{u - \bar u}$ and \mbox{$\langle x^2 \rangle_{d - \bar d}$ (C.I.)} for this
operator,\ along with their errors,\ are listed in Table~\ref{table:cixsq}.\
The renormalization factor for this current is 1.116 which we have obtained from
Eq.~(\ref{renorm1.74}).

\begin{figure}[!hbtp]
\centering \subfigure[]
{\rotatebox{270}{\includegraphics[width=5.5cm, height=0.46\hsize]{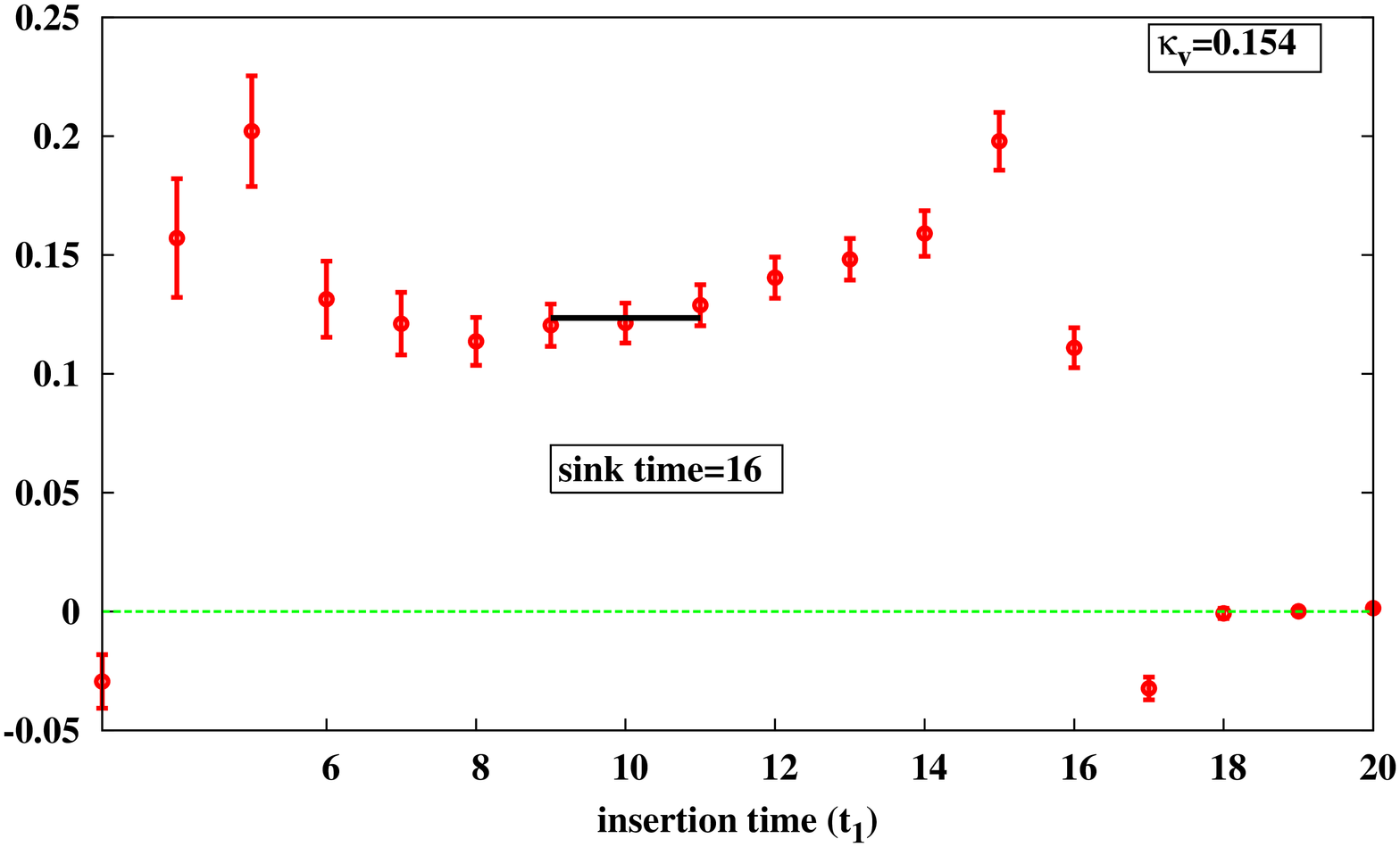}}
\label{ci154u411}}
        \hspace{0.5cm}
 \subfigure[]
{\rotatebox{270}{\includegraphics[width=5.5cm, height=0.46\hsize]{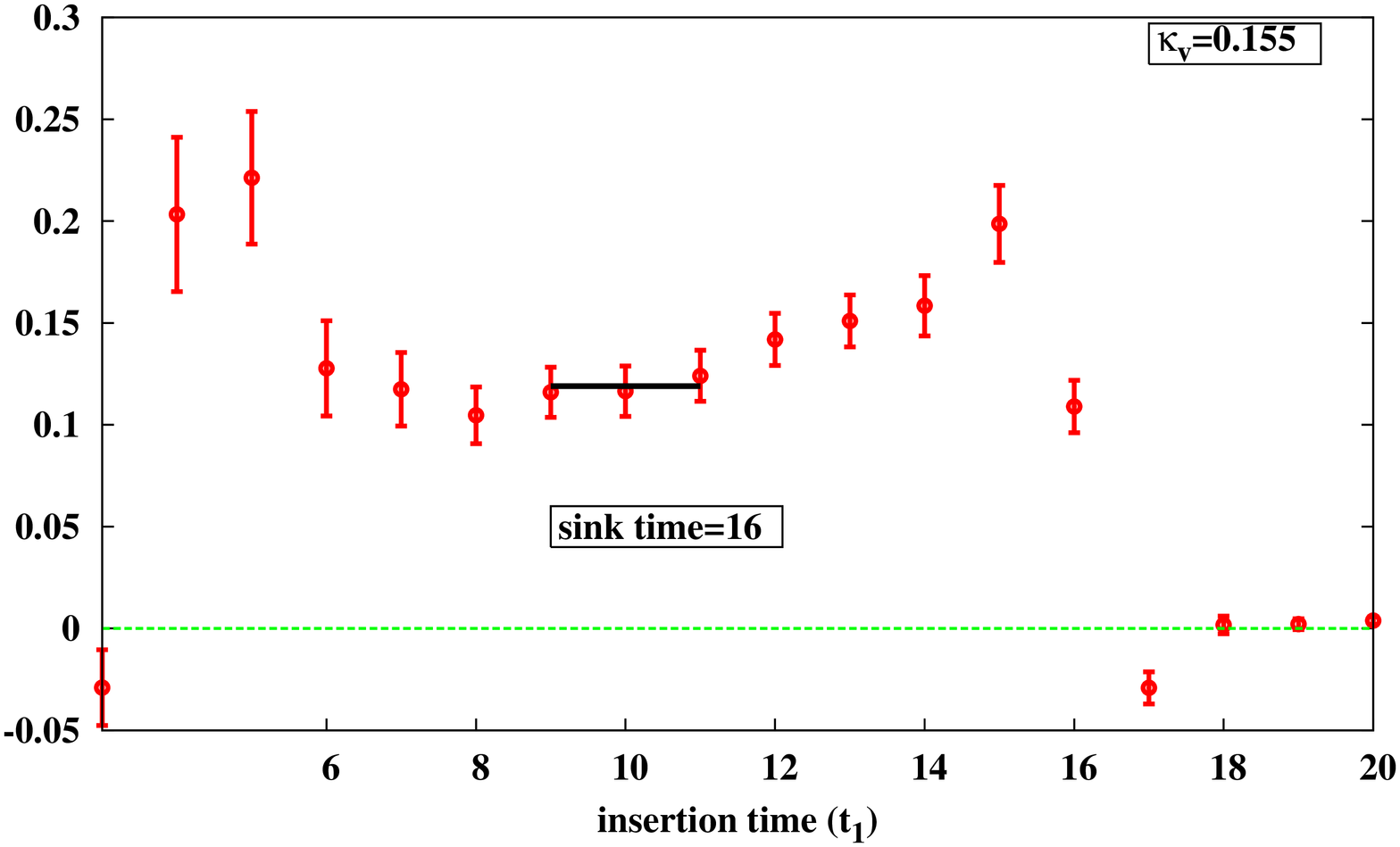}}
\label{ci155u411}}
\subfigure[]
{\rotatebox{270}{\includegraphics[width=5.5cm, height=0.46\hsize]{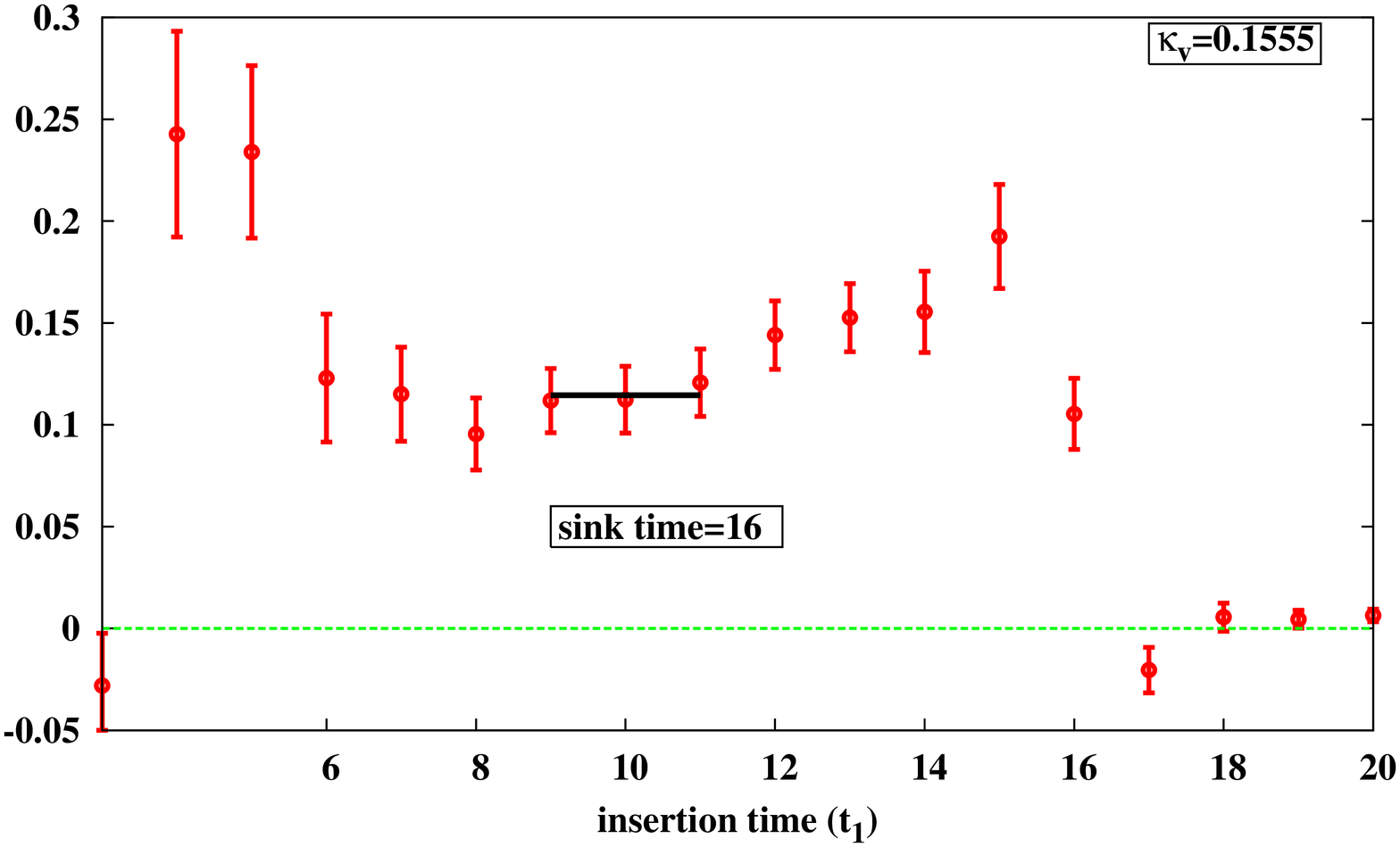}}
\label{ci1555u411}}
        \hspace{0.5cm}
\subfigure[]
{\rotatebox{270}{\includegraphics[width=5.5cm, height=0.46\hsize]{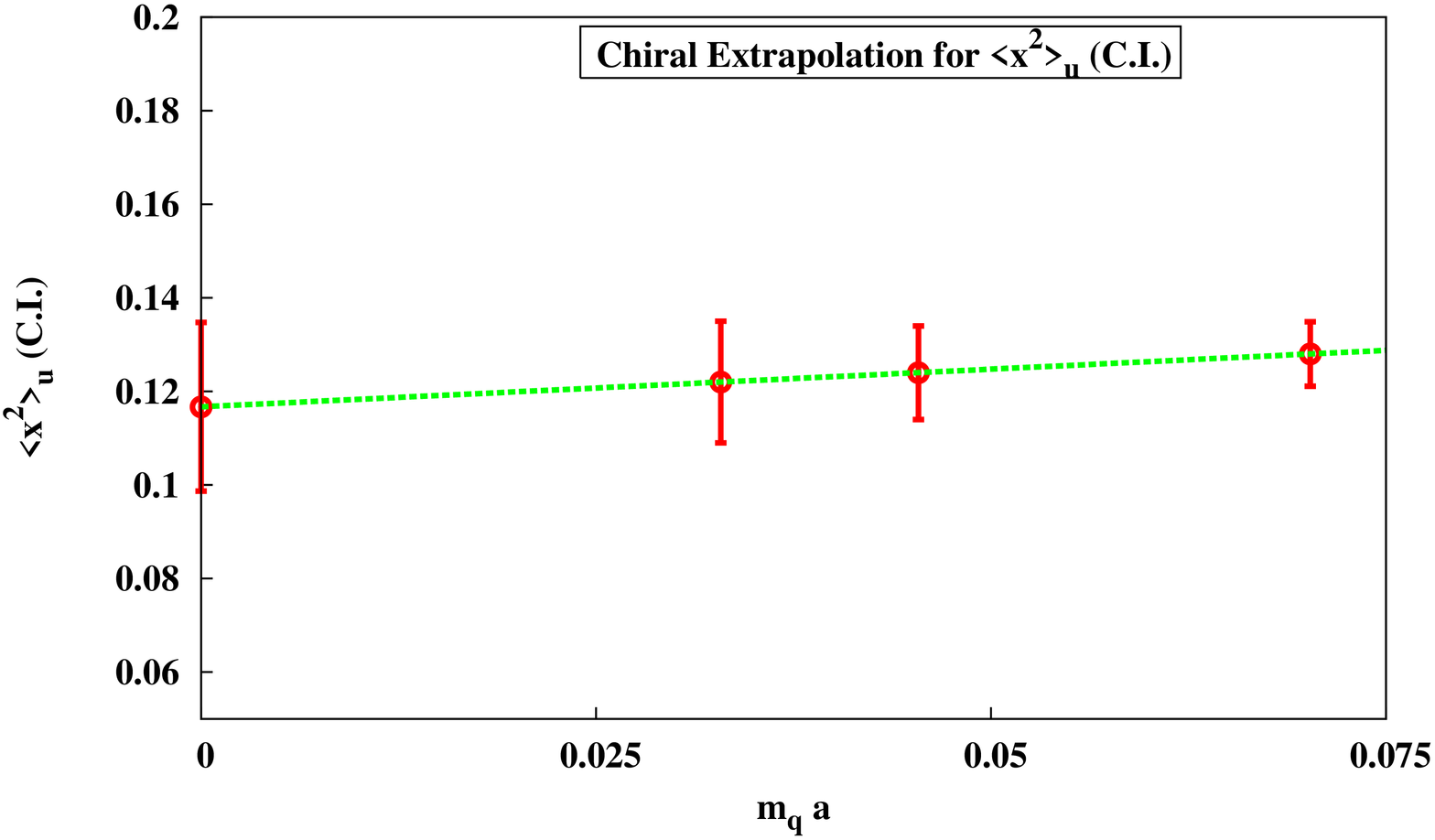}}
\label{contchiral411uci} }
\caption{The ratio of the three-point to two-point functions
(C.I.),\ with fixed sink time,\ $t_2 = 16$,\ for the
${\cal O}_{411}-\frac{1}{2}\left({\cal O}_{422}+{\cal O}_{433}\right)$
operator,\ for up quarks,\ is plotted against the current insertion
time ($t$) at
(a) $\kappa_v = 0.154$,
(b) $\kappa_v = 0.155$, and
(c) $\kappa_v = 0.1555$.
(d) is a linear extrapolation to the chiral limit plotted against $m_q a$.}
\end{figure}


\begin{figure}[!hbtp]
\centering \subfigure[]
{\rotatebox{270}{\includegraphics[width=5.5cm, height=0.46\hsize]{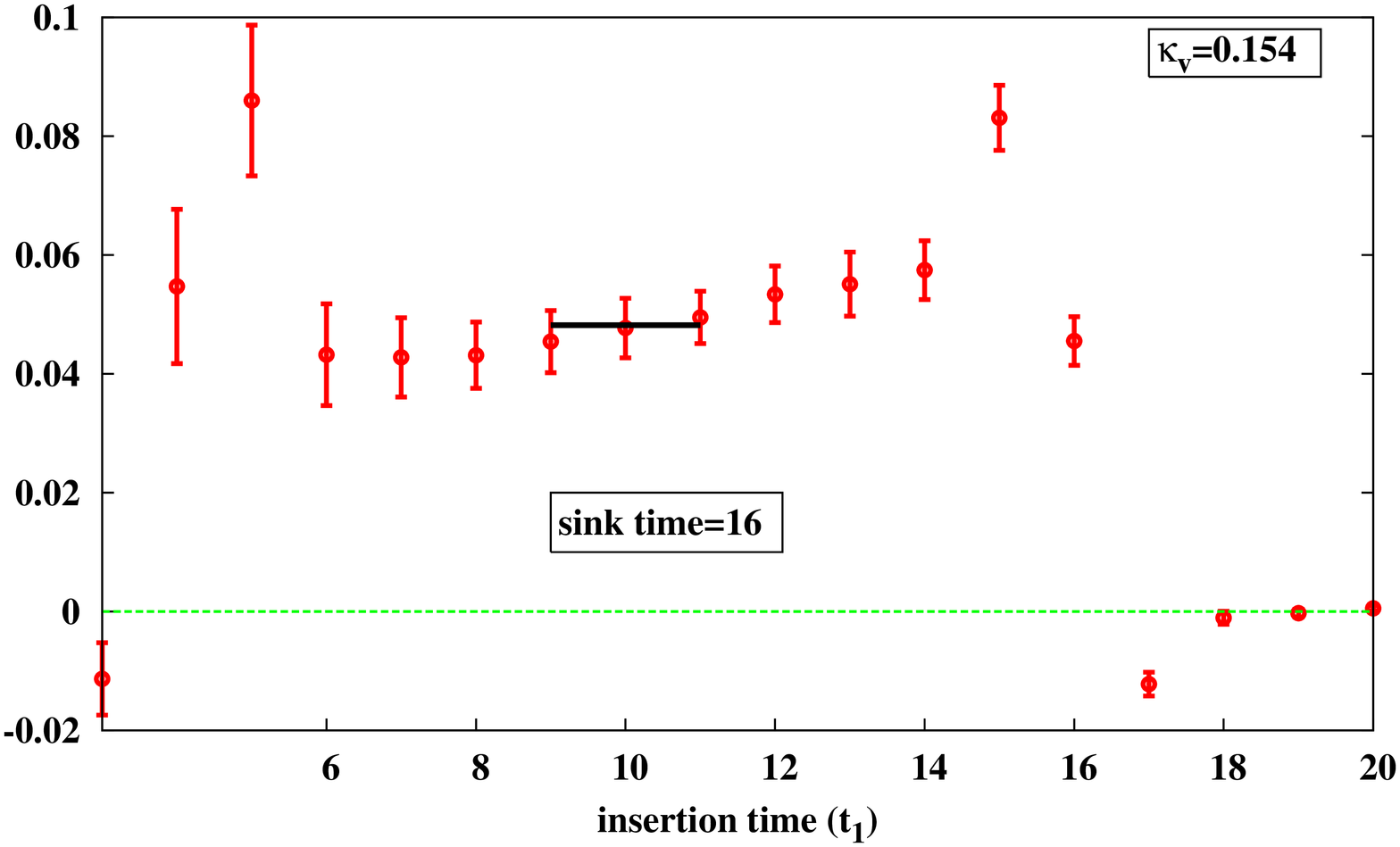}}
\label{ci154d411}}
        \hspace{0.5cm}
 \subfigure[]
{\rotatebox{270}{\includegraphics[width=5.5cm, height=0.46\hsize]{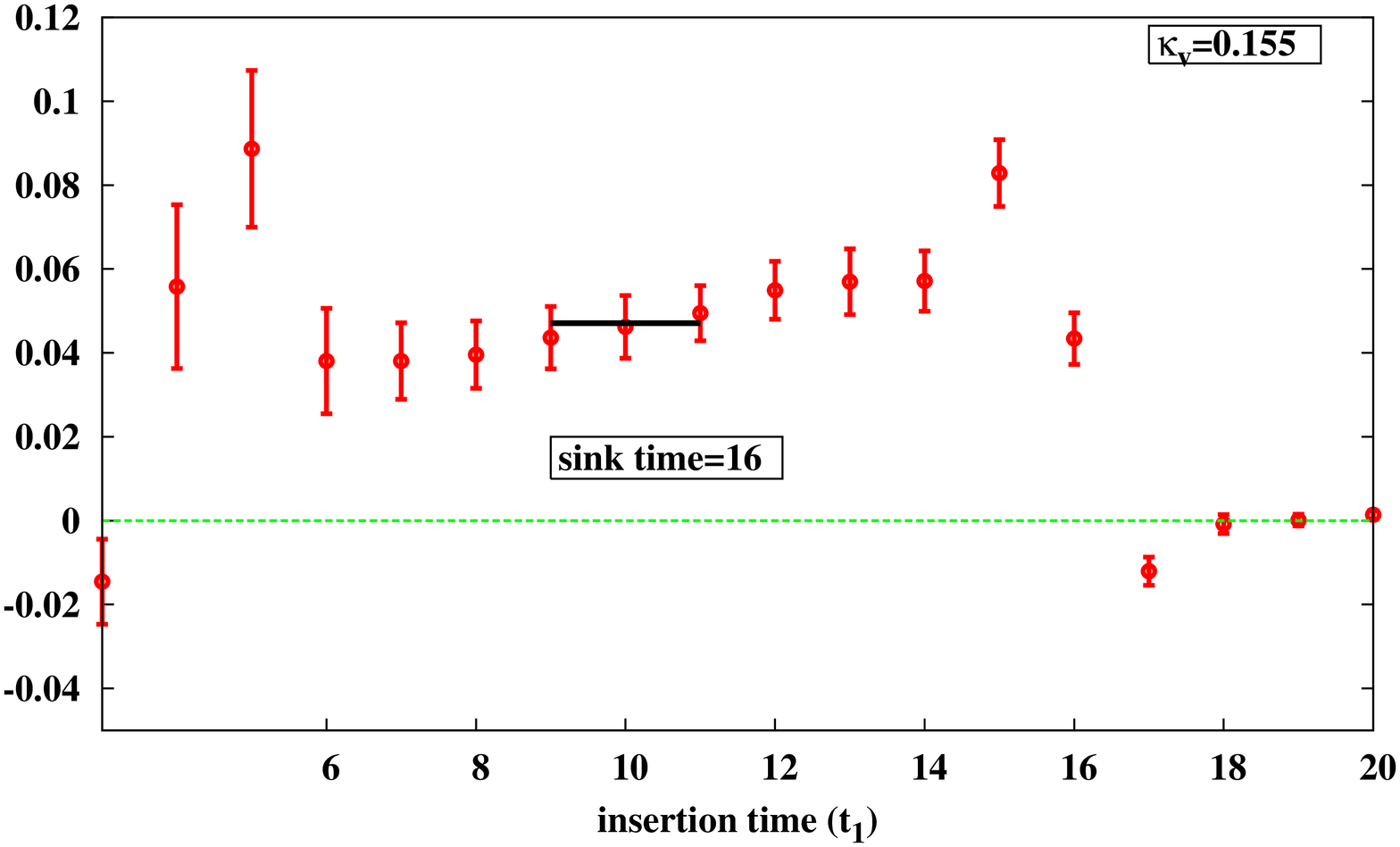}}
\label{ci155d411}}
\subfigure[]
{\rotatebox{270}{\includegraphics[width=5.5cm, height=0.46\hsize]{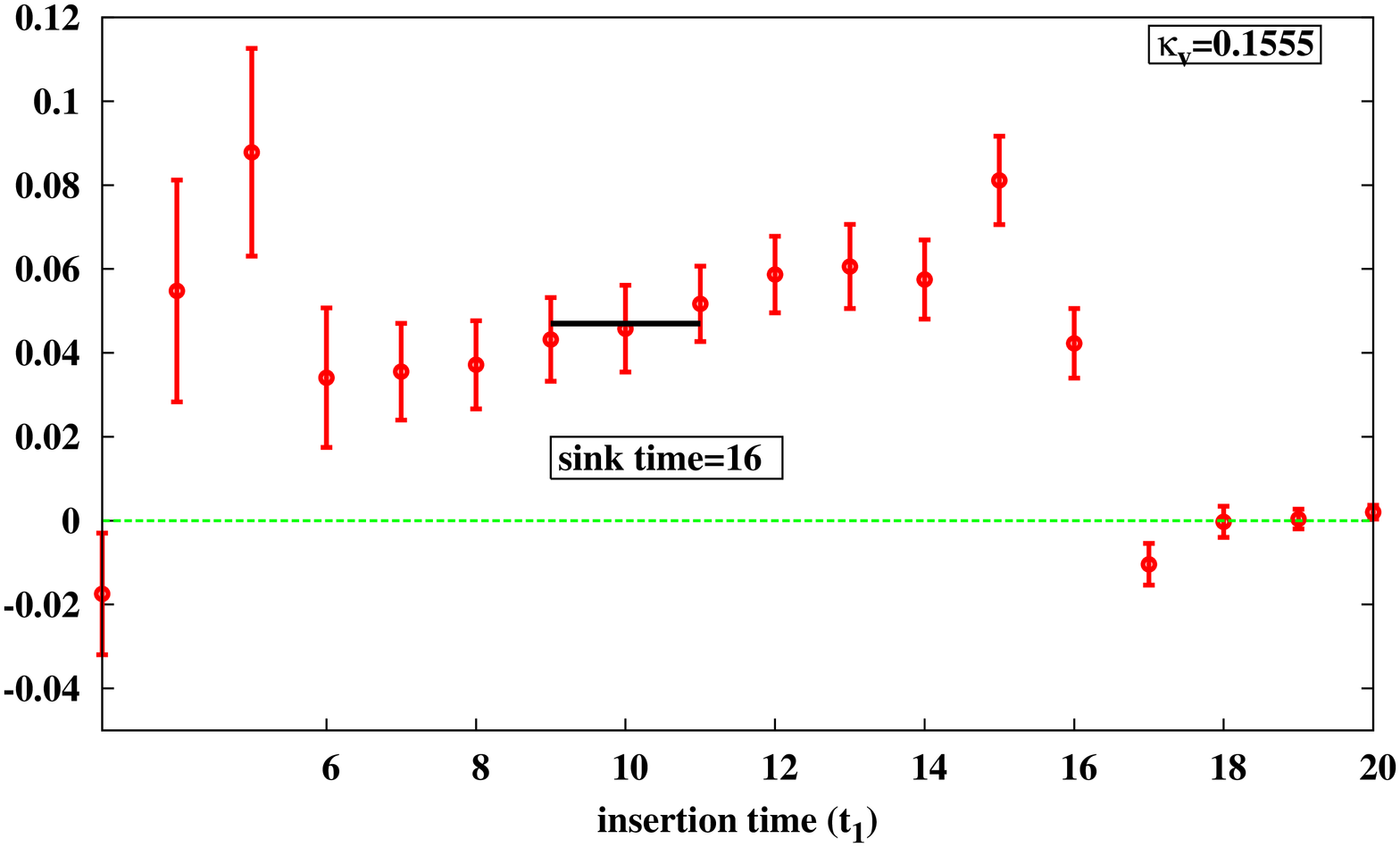}}
\label{ci1555d411}}
        \hspace{0.5cm}
\subfigure[]
{\rotatebox{270}{\includegraphics[width=5.5cm, height=0.46\hsize]{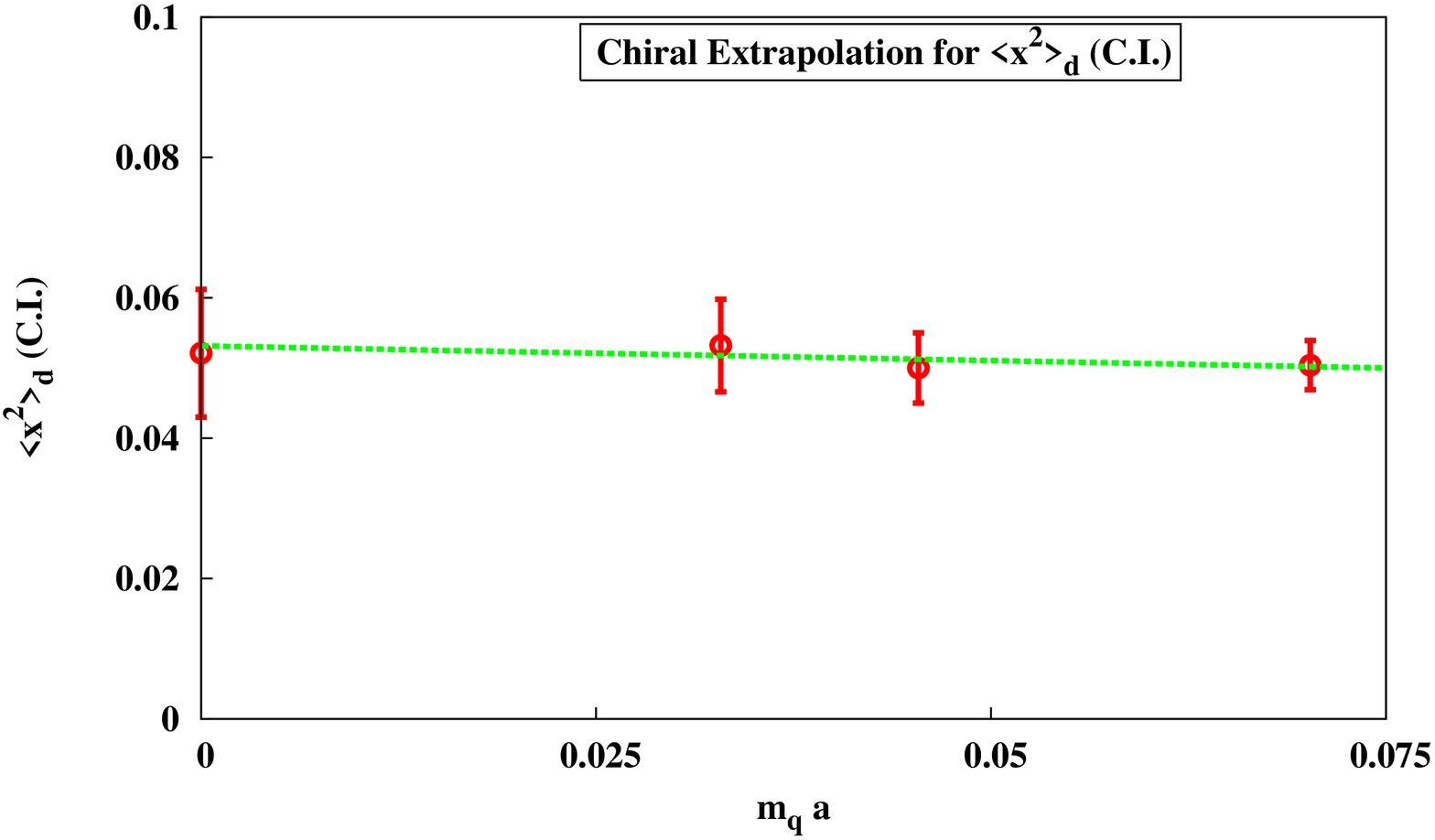}}
\label{contchiral411dci}}
\caption{The ratio of the three-point to two-point functions
(C.I.),\ with fixed sink time,\ $t_2 = 16$,\ for the
${\cal O}_{411}-\frac{1}{2}\left({\cal O}_{422}+{\cal O}_{433}\right)$
operator,\ for down quarks,\ is plotted against the current
insertion time $t_1$ at
(a) $\kappa_v = 0.154$,
(b) $\kappa_v = 0.155$, and
(c) $\kappa_v = 0.1555$.
(d) is a linear extrapolation to the chiral limit plotted against $m_q a$.}
\end{figure}

\begin{table} [h]
\centering {
\begin{tabular}{|c|c|c|}
\hline\hline
\multirow{4}{*}{$\langle x^2 \rangle^{411}_{u - \bar u}$ (C.I.)} & $\kappa_v = 0.154$  & 0.128 $\pm$ 0.007  \\
\cline{2-3}
& $\kappa_v = 0.155$  & 0.124 $\pm$ 0.010  \\
\cline{2-3}
& $\kappa_v = 0.1555$ & 0.122 $\pm$ 0.013  \\
\cline{2-3}
& Linear Extrapolation & 0.117 $\pm$ 0.018 \\
\hline\hline
\multirow{4}{*}{$\langle x^2 \rangle^{411}_{d - \bar d}$ (C.I.)} & $\kappa_v = 0.154$ & 0.0504 $\pm$ 0.0035  \\
\cline{2-3}
& $\kappa_v = 0.155$  & 0.0500 $\pm$ 0.0050  \\
\cline{2-3}
& $\kappa_v = 0.1555$ & 0.0532 $\pm$ 0.0066  \\
\cline{2-3}
& Linear Extrapolation  & 0.0521 $\pm$ 0.0091 \\
\hline\hline
\end{tabular}}
\caption{Table for the values of $\langle x^2 \rangle^{411}$(C.I.)
for up and down quarks at various kappa values and after linear
extrapolation to the chiral limit for the
${\cal O}_{411}-\frac{1}{2}\left({\cal O}_{422}+{\cal O}_{433}\right)$
operator.}
\label{table:cixsq}
\end{table}

\begin{turnpage}

\begin{table}[h]
\centering
\begin{tabular}{|c|c|c|c|c|c|c|}
\hline\hline
\multirow{3}{*}{\bf Moments }& Kentucky & QCDSF  & QCDSF  & LHPC  & LHPC  &  Experiment \\
  & (quenched)  & (quenched) & (quenched) & (quenched) & (full QCD) &  CTEQ3M \\
&($\mu^2= 4 \mbox{GeV}^2$) & ($\mu^2\simeq 5 \mbox{GeV}^2$) & ($\mu^2= 4 \mbox{GeV}^2$) & ($\mu^2= 4 \mbox{GeV}^2$) &  ($\mu^2= 4 \mbox{GeV}^2$) &($\mu^2= 4 \mbox{GeV}^2$) \\
\hline\hline
&&&&&&\\
$\langle x \rangle^{41}_{u + \bar u}$  & 0.408 (38)  & 0.410(34)   & 0.452 (26)  & 0.454 (29)   &  0.459 (29)   & 0.284  \\
&&&&&&\\
$\langle x \rangle^{44}_{u + \bar u}$  & 0.420 (35)  &    &    &      &      &        \\
\hline
&&&&&&\\
$\langle x \rangle^{41}_{d + \bar d}$  & 0.148  (19)  & 0.180 (16)  & 0.189 (12)  & 0.203 (14)   &  0.190 (17)   & 0.102   \\
&&&&&&\\
$\langle x \rangle^{44}_{d + \bar d}$  & 0.181 (18)  &    &    &     &      &         \\
\hline
&&&&&&\\
$\langle x^2 \rangle^{411}_{u - \bar u}$ & 0.117 (18)  & 0.108 (16)  & 0.104 (20)  & 0.119 (61)   &  0.176 (63)   & 0.083    \\
\hline
&&&&&&\\
$\langle x^2 \rangle^{411}_{d  - \bar d}$ & 0.052 (9)   & 0.036 (8)   & 0.037 (10)  & 0.029 (32)   &  0.031 (30)   & 0.025    \\
\hline\hline
\end{tabular}
\caption{Comparison of results for renormalized first and second
moments (connected insertion) in the chiral limit with other lattice
calculations and phenomenology in $\overline{\mbox{MS}}$ scheme.}
\label{compresults}
\end{table}
\end{turnpage}

We list our results in comparison with those from previous calculations in
Table~\ref{compresults}.\ Except for our calculations,\ other groups
have averaged the results of the first moments obtained from the two operators ${\cal
O}_{41}$ and ${\cal O}_{44}-\frac{1}{3}\left({\cal O}_{11}+{\cal
O}_{22}+{\cal O}_{33}\right)$. As we can see, all the results agree with each
other within errors.

For the QCDSF calculations at $\mu=5 \mbox{GeV}^2$ (second column),\ the
reported values of $\langle x \rangle$ are the averages of two different
procedures in~\cite{gockeler2, best2},\ whereas,\ the result of
$\langle x^2 \rangle$ is obtained from~\cite{gockeler2}.\ For both
procedures, two different lattices ($16^3\times32$ and $24^3\times32$
) are used for several $\kappa$ values.\ For the $16^3\times 32$ lattice,\
the number of gauge configurations involved is $O(1000)$ and for
the $24^3\times32$ lattice,\ the number of gauge configurations involved
is $O(100)$.\ All the calculations are performed for $\beta=6.0$.\
For the calculations at $\mu= 4\ \mbox{GeV}^2$~\cite{gockeler3} (third
column),\ three $\kappa$ values,\ 0.155,\ 0.153, and 0.1515,\ are
used on a $16^3\times 32$ lattice.\ The number of independent
gauge configurations involved are 100,\ 600, and 400 at these
$\kappa$ values respectively.\ All the calculations are performed for
$\beta=6.0$ by using standard Wilson action in quenched
approximation.

For the LHPC calculations at $\mu= 4 \mbox{ GeV}^2$ (fourth
column)~\cite{dolgov},\ a $16^3\times 32$ lattice is used for three
$\kappa$ values.\ The number of gauge configurations used are 200 at
$\beta=6.0$ for each of the three $\kappa$'s in quenched
approximation.\ LHPC calculations are also performed in full (unquenched)
QCD on a $16^3\times 32$ lattice.\ Four different $\kappa$'s are
used at $\beta=5.6$ for 200 SESAM configurations and three different
$\kappa$'s are used at $\beta=6.0$ for 100 SCRI configurations.

 From this comparison,\ we see that our results are comparable with
other lattice calculations.\ But all the lattice calculations for
$\langle x\rangle_{u-d}$ which involves only the C.I. seem
to be larger than the experimental result. \ This is a well known
problem and is presumably due to the fact that quark masses are still too heavy
compared to the physical ones~\cite{Detmold:2001jb,hagler2}.

\clearpage

\section{Conclusion and Discussion}

    We have calculated the first and second moments of the proton's parton distribution functions
for both the connected and disconnected insertions in lattice QCD. The lattice calculations are
carried out on quenched $16^3 \times 24$ lattices with $\beta = 6.0$ and quark masses which
correspond to pion masses of 650(3),\ 538(4) and 478(4) MeV,\ and nucleon
masses at 1291(9),\ 1159(11) and 1093(13) MeV respectively. The physical results are obtained from
linear extrapolation to the physical point (to the chiral limit). The connected insertion results turn out
to be consistent with the previous quenched and full QCD calculations.

    The calculation of the moments for the sea quark distribution in the D.I. is carried out for the first
time. With 400 $Z_2$ noise, 500 gauge configurations, and with unbiased subtractions and 16
nucleon sources, we are able to obtain results with $\sim\,5\sigma$ signals for the first moments.
The result of $\langle x\rangle_s = 0.027 \pm 0.006$ can be used to constrain $\langle x\rangle_s$
in the phenomenological fitting of parton distribution functions which is uncertain in the
range $ 0.018 < \langle x\rangle_s < 0.04$~\cite{CTEQ_heavy}. More interestingly, we find that
\begin{equation}
\label{s-u/d-lat}
\left.\frac{\langle x\rangle_{s+\bar{s}}}{\langle x\rangle_{u + \bar u}}\right|_{\mbox{D.I.}} =
\left.\frac{\displaystyle\int dx\, x\, (s(x) + \bar{s}(x))}{\displaystyle\int dx\, x\, (u(x)+\bar{u}(x))}
\right|_{\mbox{D.I.}}
= 0.88 \pm 0.07,
\end{equation}
which is about twice as large as the average phenomenological value from fitting the parton distribution functions
to experiments~\cite{CTEQ_heavy}
\begin{equation}
\label{s-u/d-exp}
\frac{\displaystyle\int dx\, x\, [s(x)+\bar{s}(x)]}{\displaystyle\int dx\, x\, [{\bar{u}(x)+\bar{d}(x)}]}
\sim 0.27 - 0.67.
\end{equation}
This difference is understandable and has been anticipated from the path-integral formulation of
parton degrees of freedom~\cite{liu1}. The ratio in the lattice
calculation involves $u/d$ quarks in the disconnected insertion (quark loops)~\footnote{Note that the
strange quark appears only in the disconnected insertion.}, while the phenomenological
ratio involves the $\bar{u}(x)+\bar{d}(x)$ in the connected insertion as well. The discrepancy
suggests that the momentum carried by the $\bar{u}/\bar{d}$ quarks are roughly equally shared in the connected
sea and the disconnected sea. The fact that there is anti-quark in the connected sea is demonstrated
by the large $\bar{u}(x) -\bar{d}(x)$ difference from the Gottfried sum rule violation~\cite{liu2} which
cannot be accommodated by the small $u$ and $d$ difference due to isospin breaking in the disconnected sea.
The combined Gottfried sum rule violation and the discrepancy of the ratios in Eqs. (\ref{s-u/d-lat}) and
(\ref{s-u/d-exp}) suggests the following form for the anti-parton distribution functions~\cite{liu1}:
\begin{eqnarray}
\bar{u}(x) &=& a_0 x^{a_1} (1-x)^{a_2} + b_u x^{b_1}(1-x)^{b_2}, \nonumber\\
\bar{d}(x) &=& a_0 x^{a_1} (1-x)^{a_2} + b_d x^{b_1}(1-x)^{b_2}, \nonumber\\
\bar{s}(x) &=&  a_s x^{a_1}(1-x)^{a_2},
\end{eqnarray}
where the first terms are for the disconnected seas with pomeron exchanges so that
$a_1 \sim -1$ and the second terms are from the connected sea with reggeon exchanges so
that $ b_1 \sim - \alpha_R$. The present lattice calculation suggests
that $\displaystyle\frac{a_s}{a_0}$ can be constrained to the ratio in Eq. (\ref{s-u/d-lat}). We should emphasize
that the current result is based on a quenched lattice calculation with linear chiral extrapolation from relatively
heavy quarks. We should take it with a sizable grain of salt.
As far as phenomenological fittings are concerned, the $\bar{u}(x) - \bar{d}(x)$ has been
taken into account which has a small $x$ behavior of $\sim x^{-1/2}$\cite{CTEQ5}. However,
the conventional ansatz $\bar{s}(x) \sim \bar{u}(x) + \bar{d}(x)$ used in recent fittings~\cite{CTEQ_heavy,MRST}
is obviously inadequate~\cite{liu1}. One needs to differentiate the different small $x$ behaviors in
the connected sea and disconnected sea and fit the $\bar{u}(x), \bar{d}(x)$, and $\bar{s}(x)$ accordingly.

    We have made an attempt to calculate $\langle x^2\rangle_s$ to see if it is $s(x)$ or $\bar{s}(x)$ which
is leading in large $x$. Our result has a tendency to be positive, similar to the experimental tendency of
$s_{-} = \displaystyle\int dx\, x\, (s(x) - \bar{s}(x))$ being positive, but it is consistent with zero within error.
We will see if the signal is stronger with the inclusion of 2+1 flavor dynamical fermions.

   So far, our results are obtained in the quenched approximation with relatively large quark masses
and small volume. They are subjected to large systematic errors. We will focus
our attention next on the dynamical fermion calculation with 2+1 flavor dynamical clover fermion
configurations~\cite{CP-PACS1}, and systematically move to smaller quark masses, larger volumes, and
the continuum limit~\cite{CP-PACS2}.

\acknowledgements

The work was partially supported by U.S. DOE Grant No.\ DE-FG05-84ER40154. The research of M. Deka was
supported in part by Graduate Student Research Assistantship from Thomas Jefferson National Accelerator
Facility under Jefferson Lab-university affiliation related Nuclear Physics research. The research of
N. Mathur is supported under the grant DST-SR/S2/RJN-19/2007, India. We would also like to thank
Andrei Alexandru, Devdatta Mankame, Ivan Horv\'{a}th and Anyi Li for their help and useful suggestions.
The numerical computations were performed on the supercomputer at the Center for Computational Sciences,
University of Kentucky.


\phantomsection
\addcontentsline{toc}{chapter}{Appendices}

\appendix

\section{Three-Point Correlation Functions}
\label{app:threepoint}

\subsection{General Considerations}
\label{appsubsec:threept_general_cons}

For the forward matrix elements,\ the three point-function for any
general operator ${\cal O}$ (color and spin indices suppressed) is
defined as
\eqarray{
 G^{\alpha\beta}_{N {\cal O} N}(t_2, t_1, \vec{p})&=&
       \displaystyle\sum_{\vec{x}_2, \vec{x}_1}\, e^{-i\vec{p}.(\vec{x}_2 -\vec{x}_0)} \,
       \langle\, 0\, | \,\mbox{T}\, ( \chi^\alpha (x_2)\,{\cal O}(x_1)\,
       \bar{\chi}^\beta (x_0) )\, |\, 0 \,\rangle,
}
where\ $t=t_2$\ is the nucleon sink time, $t=t_1$ is the current insertion time,\
$t=t_0$\ is the nucleon source time, and $\vec{p}_i$ is the momentum of the nucleon,\
respectively.\ We now consider the following situations:

\bigskip

$\bullet$  When $t_2 > t_1 > t_0$,\ we get
\eqarray{
          & & G^{\alpha\beta}_{N {\cal O} N}(t_2, t_1, \vec{p})\nonumber\\
          & &\nonumber\\
          &=&\displaystyle\sum_{\vec{x}_2, \vec{x}_1} e^{-i\vec{p}.(\vec{x}_2-\vec{x}_0)}
          \displaystyle\sum_{n_1,\vec{q}_1, s_1} \,\displaystyle\sum_{n_2, \vec{q}_2, s_2}
          \langle \, 0\, | \,\chi^\alpha (x_2)\, |n_2, \vec{q}_2, s_2\rangle\,
          \langle n_2, \vec{q}_2, s_2|\, {\cal O}(x_1)\, |n_1, \vec{q}_1, s_1\rangle\nonumber\\
          & &\nonumber\\
          & &\langle n_1, \vec{q}_1, s_1| \,\bar{\chi}^\beta (x_0)\, | 0\rangle\nonumber\\
          &=&N^2 \displaystyle\sum_{s}\, e^{-E^{0+}_{p}(t_2-t_0)}\,\langle \, 0\, | \,
          \chi^\alpha (x_0)\, |\,0^+, \vec{p}, s\,\rangle \langle \,0^+, \vec{p}, s\,|\,
          {\cal O}(x_0)\,  |\,0^+, \vec{p}, s \rangle\nonumber\\
          & &\nonumber\\
          & &\langle \,0^+, \vec{p}, s \,| \,\bar{\chi}^\beta (x_0)\, |\, 0 \,\rangle   \nonumber\\
          & &\nonumber\\
          &+&N^2 \displaystyle\sum_{s}\, e^{-E^{0-}_{p}(t_2-t_0)}\,\langle \, 0\, |
          \,\chi^\alpha (x_0)\, |\,0^-, \vec{p}, s\,\rangle \langle \,0^-, \vec{p}, s\,|\,
          {\cal O}(x_0)\,  |\,0^-, \vec{p}, s \rangle\nonumber\\
          & &\nonumber\\
          & &\langle \,0^-, \vec{p}, s \,| \,\bar{\chi}^\beta (x_0)\, |\, 0 \,\rangle +
          \displaystyle\sum_{\theta=+,-}\bigg{[} e^{-E^{0(\theta)}_{p}(t_2 - t_1)}
          e^{- E^{1(\theta)}_{p} (t_1-t_0)} C^{(\theta)(1)\alpha\beta}(\vec{p}) \nonumber\\
          & &\nonumber\\
          &+& e^{-E^{1(\theta)}_{p}(t_2 - t_1)} e^{- E^{0(\theta)}_{p} (t_1-t_0)}C^{(\theta)(2)}(\vec{p})\bigg{]}\nonumber\\
          & &\nonumber\\
          &+&\displaystyle\sum_{\theta, \theta'=+,-}\displaystyle\sum^\infty_{n^{(\theta)}_1, n^{(\theta')}_2=1}
         \bigg{[} e^{-E^{n^{(\theta')}_2}_{p}(t_2 - t_1)}e^{- E^{n^{(\theta)}_1}_p (t_1-t_0)}
         f^{(1)\alpha\beta}(n^{(\theta)}_1, n^{(\theta')}_2, \vec{p})\bigg{]},
\label{threept1}
}
where $N$ is the number of lattice sites,\ and the superscript $+$ ($-$) represents
positive (negative) parity state.
\eqarray{
C^{(\theta)(1)\alpha\beta}(\vec{p}) &=&
          N^2 \, \langle \, 0\, | \,\chi^\alpha (x_0)\, |\,0^{(\theta)}, \vec{p}, s\,\rangle
          \langle \,0^{(\theta)}, \vec{p}, s\,|\, {\cal O}(x_0)\, |\,1^{(\theta)}, \vec{p}, s\,\rangle\nonumber\\
          & &\langle \,1^{(\theta)}, \vec{p}, s\,| \,\bar{\chi}^\beta (x_0)\, |\, 0 \,\rangle, \nonumber\\
          & &\nonumber\\
C^{(\theta)(2)\alpha\beta}(\vec{p}) &=&
          N^2 \,\langle \, 0\, | \,\chi^\alpha (x_0)\, |\,1^{(\theta)}, \vec{p}, s\,\rangle
          \langle \,1^{(\theta)}, \vec{p}, s\,|\, {\cal O}(x_0)\,  |\,0^{(\theta)}, \vec{p}, s\,\rangle\nonumber\\
          & &\langle \,0^{(\theta)}, \vec{p}, s\,| \,\bar{\chi}^\beta (x_0)\, |\, 0 \,\rangle, \nonumber\\
          & &\nonumber\\
f^{(1)\alpha\beta}(n^{(\theta)}_1, n^{(\theta')}_2, \vec{p})&=&
          N^2 \displaystyle\sum_{s}\langle\, 0\,|\,\chi^\alpha (x_0)\,|\,n^{(\theta')}_2, \vec{p}, s\,\rangle\,
          \langle \,n^{(\theta')}_2, \vec{p}, s\,|\, {\cal O}(x_0)\,|\,n^{(\theta)}_1, \vec{p}, s\rangle\nonumber\\
          & &\langle \,n^{(\theta)}_1, \vec{p}, s\,| \,\bar{\chi}^\beta (x_0)\, |\, 0 \, \rangle.
\label{threept2}
}

$\bullet$  When $t_1 > t_2 > t_0$,\ we get
\eqarray{
G^{\alpha\beta}_{N {\cal O} N}(t_2, t_1, \vec{p})&=&  N^2 \displaystyle\sum_{n_1, n_2, s}\,e^{-E^{n_2}_{0}(t_1-t_2)}\,e^{-E^{n_1}_{p}(t_2-t_0)}\,
          \langle \, 0\, | \,{\cal O}(x_0)\, |\,n_2, \vec{0}, s\, \rangle\nonumber\\
          & &\nonumber\\
          & &\langle \,n_2,\vec{0}, s\,|\, \chi^\alpha (x_0) \,  |\,n_1, \vec{p}, s\,\rangle\, \langle \,n_1, \vec{p}, s\,| \,\bar{\chi}^\beta (x_0)\, |\, 0 \,\rangle\nonumber\\
          & &\nonumber\\
          &=&0. \hspace{10mm}\left[ \mbox{since,\ for a twist-two operator},\
          \langle \, 0\, | \,{\cal O}(x_0)\, |\,n_2, \vec{0}, s\, \rangle = 0\ \right]
\label{threept3}
}

$\bullet$  When $t_2 > t_0 > t_1$,\ we get
\eqarray{
G^{\alpha\beta}_{N {\cal O} N}(t_2, t_1, \vec{p})&=&N^2 \displaystyle\sum_{n_1, n_2, s}\,e^{-E^{n_2}_{p}(t_2-t_0)}\,e^{-E^{n_1}_{0}(t_0-t_1)}\,
          \langle \, 0\, | \,\chi^\alpha (x_0)\, |\,n_2, \vec{p}, s\, \rangle\nonumber\\
          & &\nonumber\\
          & &\langle \,n_2,\vec{p}, s\,|\, \bar{\chi}^\beta (x_0) \,  |\,n_1, \vec{0}, s\,\rangle\, \langle \,n_1, \vec{0}, s\,| \,{\cal O}(x_0)\, |\, 0 \,\rangle\nonumber\\
          & &\nonumber\\
          &=& 0. \hspace{10mm} \left[\ \mbox{since,\ for a twist-two operator},\
          \langle \,n_1, \vec{0}, s\,| \,{\cal O}(x_0)\, |\, 0 \,\rangle =0\ \right]
\label{threept4}
}

\newpage

\subsection{Ratios for Disconnected Insertion}
\label{appsubsec:threept_ratio_di}

We will show a sample calculation for extracting the disconnected first and second moments by considering
the ${\cal O}={\cal O}_{4i}$ operator,\ where\ $i=1, 2, 3$.\ The calculation for the other operators will be
similar. For this operator,\ we will use the nucleon with one unit of lattice momentum in the $i$-th direction.\
First consider the first and the second terms of \mbox{Eq.~(\ref{threept1})} only.\ Taking the trace with
parity projection operator,\ $\Gamma^{\beta\alpha}$ in  Eq.~(\ref{threept1}),\ we get
\eqarray{
         & & \Gamma^{\beta\alpha}\, \times
         \left[ \mbox{first term + second term of Eq.~(\ref{threept1})}\right]\nonumber\\
         & &\nonumber\\
         &=& N^2\, \displaystyle\sum_s e^{-E^{0+}_p (t_2-t_0)}\,\frac{a^3}{(2\kappa)^{\frac{3}{2}}}\,
         \bigg{(}\frac{m^+}{NE^{0+}_p}\bigg{)}^{\frac{1}{2}}\phi^+ \, \frac{a^3}{(2\kappa)^{\frac{3}{2}}}\,
         \bigg{(}\frac{m^+}{NE^{0+}_p}\bigg{)}^{\frac{1}{2}}\nonumber\\
         & &\nonumber\\
         & & \phi^{+*} \,\bar{u}^{+\beta}(\vec{p}, s) \, \Gamma^{\beta\alpha}\,u^{+\alpha}(\vec{p}, s)\,
         \frac{m^+}{\kappa N E^{0+}_p}\frac{\langle x\rangle(-i p_4) \, p_i }{2m^+} \nonumber\\
         & &\nonumber\\
         &+&N^2\, \displaystyle\sum_s e^{-E^{0-}_p (t_2-t_0)}\,\frac{a^3}{(2\kappa)^{\frac{3}{2}}}\,
         \bigg{(}\frac{m^-}{NE^{0-}_p}\bigg{)}^{\frac{1}{2}}\phi^- \, \frac{a^3}{(2\kappa)^{\frac{3}{2}}}\,
         \bigg{(}\frac{m^-}{NE^{0-}_p}\bigg{)}^{\frac{1}{2}}\phi^{-*} \nonumber\\
         & &\nonumber\\
         & & (\bar{u}^{-\beta}(\vec{p}, s)\,\gamma_5)\,\Gamma^{\beta\alpha}\,(\gamma_5\,u^{-\alpha}(\vec{p}, s))\,
         \frac{m^-}{\kappa N E^{0-}_p}\frac{\langle x\rangle^{-}(-i p_4) \, p_i }{2m^-} \nonumber\\
         & &\nonumber\\
         &=&\frac{a^6}{(2 \kappa)^3}\,\frac{m^{+}}{ E^{0+}_p}\,\,|\phi^+|^2 \,e^{-E^{0+}_p (t_2-t_0)}\,
         \left( 1 + \frac{m^-}{E^{0-}_p}\,\frac{ E^{0+}_p }{m^+}\right)\,\frac{ \langle x\rangle\, p_i}{2\kappa},
\label{3pt4i}
}
where $\langle x \rangle^+ = \langle x \rangle$ is the first moment, $\langle x \rangle^-$
is an unknown constant.

We will now sum $\mbox{Tr}\,\left[\Gamma \,G_{N{\cal O}N}(t_2, t_1, \vec{p})\right]$ over the
current insertion time,\ $t_1$,\ from an initial time,\ $t_i = t_0 + 1$,\ to a final time,\ $t_f=t_2-1$,\
so that,\ $t_f > t_i > t_0$,\ where the nucleon source is at $t_0$,\ and $t_2$ is the sink time.\
Then,\ by using \mbox{Eqs.\ (\ref{threept1}),\ (\ref{threept2}) and (\ref{3pt4i})},\ we get
\eqarray{
  & &\displaystyle\sum^{t_f}_{t_1=t_i}\mbox{Tr}\,\left[\Gamma\, G_{N{\cal O}_{4i}N}(t_2, t_1, p_i)\right] =
  \displaystyle\sum^{t_f}_{t_1} \Gamma^{\beta\alpha}\, G^{\alpha\beta}_{N{\cal O}_{4i}N}(t_2, t_1, \vec{p})\nonumber\\
  & &\nonumber\\
  &=&\displaystyle\sum^{t_f}_{t_1=t_i} \frac{a^6}{(2 \kappa)^3}\,\frac{m}{E^{0+}_p}\,\,|\phi|^2 \,
  e^{-E^{0}_p (t_2-t_0)}\, \left( 1 + \frac{m^-}{E^{0-}_p}\,\frac{ E^{0^+}_p }{m}\right)\,
  \frac{ \langle x\rangle\, p_i}{2\kappa} \nonumber\\
  & &\nonumber\\
  &+&\displaystyle\sum_{\theta=+,-} \bigg{[}e^{-E^{0(\theta)}_{p}(t_2 - t_1)}\,
  e^{- E^{1(\theta)}_{p} (t_1-t_0)}\, \tilde C^{(\theta)(1)}_{4i} ( \vec{p})   \nonumber\\
  & &\nonumber\\
  &+&\, e^{-E^{1(\theta)}_{p}(t_2 - t_1)}\,e^{- E^{0(\theta)}_{p} (t_1-t_0)}\tilde C^{((\theta)2)}_{4i} ( \vec{p})\bigg]\nonumber\\
  & &\nonumber\\
  &+& \displaystyle\sum_{\theta,\theta'=+,-}\displaystyle\sum^\infty_{n^{(\theta)}_1, n^{(\theta)}_2=1} \bigg[
  e^{-E^{n^{(\theta')}_2}_{p}(t_2 - t_1)}\,e^{- E^{n^{(\theta)}_1}_p (t_1-t_0)}
  \tilde f^{(1)}_{4i}(n^{(\theta)}_1, n^{(\theta')}_2, \vec{p})\bigg].
\label{threept5}
}

\newlength{\ab}
\settowidth{\ab}{$(t_2-t_0)\gg 1$}

Dividing by the two-point function, we get
\eqarray{
  & &\displaystyle\sum^{t_f}_{t_1=t_i} \frac{\mbox{Tr}\,\left[G_{N{\cal O}_{4i}N}(t_2, t_1, \vec{p})\right]}
  {\mbox{Tr}\,[\Gamma\,G_{NN}(t_2, \vec{p})]} \nonumber\\
  & &\nonumber\\
  &=& \left[\frac{\langle x\rangle p_i}{2 \kappa}\, t_2 - \frac{\langle x\rangle p_i}{2 \kappa}\, t_i
   + \mbox{k}_1\mbox{k}_2 - \mbox{k}_1 \mbox{k}_5 - \left[\mbox{k}_1 \mbox{k}_3- \mbox{k}_1 \mbox{k}_4\right]
   \displaystyle\sum_{\theta=+,-} e^{-(E^{1(\theta)}_{p}-E^{0(\theta)}_{p})(t_2-t_0)}\right. \nonumber\\
   &&\nonumber\\
  &+&\left. \mbox{k}_1 \displaystyle\sum_{\theta'=+,-}\displaystyle\sum^\infty_{n'^{(\theta')}=1}
  e^{-(E^{n'^{(\theta')}}_{p}-E^{0(\theta')}_p) (t_2-t_0)} \, g_3(n'^{(\theta')},\vec{p})\right]
    \nonumber\\
  & &\nonumber\\
  &+& \left[1 + \mbox{k}_1 \displaystyle\sum^\infty_{m=1}\left( -\displaystyle\sum_{\theta=+,-}
  \displaystyle\sum^\infty_{n^{(\theta)}=1} e^{-(E^{n^{(\theta)}}_p-E^0_p)(t_2-t_0)}\,
  \tilde{f}(n^{(\theta)},\vec{p})\right)^m \right],
\label{threept6}
}
where
\eqarray{
\tilde f (n^{+,-},\vec{p}) &=& N\, \Gamma^{\alpha\beta}\displaystyle\sum_s \langle\,0\,|\,\chi^\alpha(x_0)\,|\,n^{+,-}, \vec{p},s\,
  \rangle\langle\,n^{+,-}, \vec{p},s\,|\,\bar{\chi}^\beta (x_0)\,|\,0\,\rangle,\nonumber\\
\mbox{k}_1 &=& \frac{a^6}{(2\kappa)^3}\,\frac{m}{E^{0}_p}\,|\phi|^2\,
    \left(1 + \frac{m^-}{E^{0-}_p}\,\frac{E^{0}_p}{m}\,\right) (= \mbox{constant}),\nonumber\\
\mbox{k}_2 &=&\displaystyle\sum_{\theta=+,-}\frac{e^{-(E^{1(\theta)}_{p}- E^{0(\theta)}_{p})(t_i-t_0)}}
    {1- e^{-(E^{1(\theta)}_{p}- E^{0(\theta)}_{p})}}\, \tilde{C}^{(\theta)(1)}_{4i} (= \mbox{constant}), \nonumber\\
\mbox{k}_3 &=&\displaystyle\sum_{\theta=+,-}\frac{1}{1- e^{-(E^{1(\theta)}_{p}- E^{0(\theta)}_{p})}}\,
     \tilde{C}^{(\theta)(1)}_{4i} (= \mbox{constant}),\nonumber\\
\mbox{k}_4 &=&\displaystyle\sum_{\theta=+,-}\frac{e^{(E^{1(\theta)}_{p}- E^{0(\theta)}_{p})(t_i-t_0)}}
   {1- e^{(E^{1(\theta)}_{p}- E^{0(\theta)}_{p})}}\, \tilde{C}^{(\theta)(2)}_{4i} (= \mbox{constant}), \nonumber\\
\mbox{k}_5 &=&\displaystyle\sum_{\theta=+,-}\frac{1}{1- e^{(E^{1(\theta)}_{p}- E^{0(\theta)}_{p})}}\,
   \tilde{C}^{(\theta)(2)}_{4i} (= \mbox{constant}), \nonumber\\
g_3(n^{(\theta)},\vec{p}) &=& \displaystyle\sum_{\theta=+,-}\displaystyle\sum^\infty_{n^{(\theta')}_1=1}
     \left[e^{-(E^{n^{(\theta)}}_{p}-E^{n^{(\theta')}_1}_{p})(t_0 - t_i)}
     \,\frac{\tilde{f}^{(1)}_{4i}(n^{(\theta')}_1, n^{(\theta)},\vec{p})}
     {1- e^{(E^{n^{(\theta)}}_{p}- E^{n^{(\theta')}_1}_{p})}} \right.\nonumber\\
    &-& \left. \frac{\tilde{f}^{(1)}_{4i}(n^{(\theta)}, n^{(\theta')}_1,\vec{p})}
    {1- e^{(E^{n^{(\theta')}_1}_{p}- E^{n^{(\theta)}}_{p})}} \right].
}

\noindent
From Eq.~(\ref{threept6}), we get
\eq{
{\displaystyle\sum^{t_f}_{t_1=t_i} \frac{\mbox{Tr}\,\left[G_{N{\cal O}_{4i}N}(t_2, t_1, \vec{p})\right]}
  {\mbox{Tr}\,[\,\Gamma\,G_{NN}(t_2, \vec{p})]} \hspace{6mm} \overrightarrow{\hspace{\ab}}}
  \hspace{-\ab}\raisebox{2ex}{$(t_2-t_0)\gg 1$} \hspace{6mm} \frac{ \langle x\rangle p_i}{2 \kappa}\,t_2  +  \mbox{const.},
\label{ratio_disconnected}
}
where
\eqarray{
\mbox{const.} &=& -\frac{ \langle x\rangle p_i}{2 \kappa}\,t_i  +  \mbox{k}_1 \mbox{k}_2  -  \mbox{k}_1 \mbox{k}_5.
}

\section{Unbiased Subtraction}
\label{app:subtraction}

The set of traceless matrices is obtained from the hopping parameter expansion of the inverse
of the fermion matrix $M$ which is given by
\eqarray{
S &=&  M^{-1} = I + \kappa D + \kappa^2 D^2 + \kappa^3 D^3 + \cdots,
}
where
\eqarray{
D_{x, y} &=& \displaystyle\sum^4_{\mu=1}\, \bigg{[} (1-\gamma_\mu)_{\alpha\beta}\, U^{ab}_\mu (x)\,
\delta_{x, y - a_\mu}\, + \,   (1+\gamma_\mu)_{\alpha\beta}\, U^{\dag ab}_\mu (x - a_\mu)\,
\delta_{x, y + a_\mu} \bigg{]},
}
$x, y$ are space-time co-ordinates,\ $\alpha, \beta$ are spin indices, and $a, b$ are color indices.\
If for some currents,\ some powers of $D$ are not traceless, we make them traceless by
subtracting their traces from themselves.\ Now,\ we are going to calculate the traces for all the
three currents.\ We suppress the argument $U$ from each of the propagators for convenience.

\subsection{Two-Index Operators}
\label{appsubsec:two_index_sub}

The disconnected part of ${\cal O}_{\mu\nu}$ is given by
\eqarray{
         & &\mbox{Loop}\nonumber\\
         &=&\frac{\lambda}{8a} \displaystyle\sum_{\vec{x}_1}\bigg[\mbox{Tr}\,\big{[}S^{(f)mn}(x_1 + a_\nu, x_1)\,\gamma_\mu\,
         U^{n m}_\nu(x_1)\big{]}\, - \, \mbox{Tr}\,\big{[}S^{(f)mn}(x_1 - a_\nu, x_1)\,\gamma_\mu\,
         U^{\dag n m}_\nu(x_1 - a_\nu)\big{]} \nonumber\\
         & &\nonumber\\
         &+&\mbox{Tr}\,\big{[}S^{(f)mn}(x_1, x_1 - a_\nu)\,\gamma_\mu\,U^{n m}_\nu(x_1-a_\nu)\big{]} -
         \mbox{Tr}\,\big{[}S^{(f)mn}(x_1, x_1 + a_\nu)\,\gamma_\mu\,
         U^{\dag n m}_\nu(x_1)\big{]}\nonumber\\
         & &\nonumber\\
         &+& \mu \leftrightarrow \nu \bigg],
\label{app:dismunu1}
}
where $\lambda = -i$ for $\mu = 4,\ \nu = 1,2,3$;\ $\lambda = +1$ for $\mu = \nu = 1,2,3$ and
$\lambda = -1$ for $\mu = \nu = 4$.
Let us consider the first term of the loop
\eqarray{
          & &\frac{\lambda}{8a} \displaystyle\sum_{\vec{x}_1}\,\mbox{Tr}\,\left[S^{(f)mn}(x_1 + a_\nu, x_1)\,\gamma_\mu\,
          U^{n m}_\nu(x_1)\right]\nonumber\\
          & &\nonumber\\
          &\simeq&\frac{\lambda}{8a} \displaystyle\sum_{\vec{x}_1} \,\displaystyle\sum_{m, \xi}\, \langle \eta^\dag_{x_1, n, \tau}\,
          X_{x_1 + a_\nu, m, \xi} \rangle \, (\gamma_\mu)_{\tau\xi}\, U^{ nm}_\nu (x_1)
          \hspace{3mm} [\mbox{using noise}]\nonumber\\
          & &\nonumber\\
          &=& \frac{\lambda}{8a} \displaystyle\sum_{\vec{x}_1} \, \displaystyle\sum_{m, \xi, k, \sigma}\,\frac{1}{L}\,
          \displaystyle\sum^L_{l=1}\, \eta^{l\, \dag}_{x_1, n, \tau}\,
          M^{(-1)mk}_{\xi\sigma}(x_1 + a_\nu, z)\,\eta^l_{z, k, \sigma}\,(\gamma_\mu)_{\tau\xi}\,
          U^{ nm}_\nu (x_1).
\label{app:dismunu2}
}

In Eq.~(\ref{app:dismunu2}),\ we will substitute the hopping parameter expansion of $M^{-1}$.
Then,\ from \mbox{Eq.~(\ref{app:dismunu2})},\ the term becomes
\eqarray{
          \displaystyle\sum_{\vec{x}_1}\,\mbox{Tr}\,\big{[}S^{(f)mn}(x_1 + a_\nu, x_1)\,
          \gamma_\mu\,U^{n m}_\nu(x_1)\big{]}
          &\approx&\displaystyle\sum_{\vec{x}_1}\,\displaystyle\sum_{m, \xi}\, \frac{1}{L}\,
          \displaystyle\sum^L_{l=1}\, \eta^{l\, \dag}_{x_1, n, \tau}\,
          \eta^l_{x_1 + a_\nu, m, \xi}\,(\gamma_\mu)_{\tau\xi}\, U^{ nm}_\nu (x_1).\nonumber\\
\label{app:dismunu3}
}

Similarly,\ if we consider the fourth term of the Loop in Eq.~(\ref{app:dismunu1}),\ we get
\eqarray{
          & &\frac{\lambda}{8a} \displaystyle\sum_{\vec{x}_1}\mbox{Tr}\,\big{[}S^{(f)mn}(x_1 , x_1+ a_\nu)\,\gamma_\mu\,
          U^{\dag n m}_\nu(x_1)\big{]}\nonumber\\
          & &\nonumber\\
          &\approx& \frac{\lambda}{8a}  \displaystyle\sum_{\vec{x}_1}\displaystyle\sum_{m, \xi}\frac{1}{L}\,
          \displaystyle\sum^L_{l=1}\, \eta^{l\, \dag}_{x_1+ a_\nu, n, \tau}\,\eta^l_{x_1, m, \xi} \,
          (\gamma_\mu)_{\tau\xi}\, U^{\dag nm}_\nu (x_1).
\label{app:dismunu4}
}

Combining Eqs.~(\ref{app:dismunu3}) and (\ref{app:dismunu4}), we find that
\eqarray{ \label{app:dis7}
        & &\frac{\lambda}{8a} \left[\mbox{Tr}\,\big{[}S^{(f)mn}(x_1 + a_\nu, x_1)\,\gamma_\mu\,U^{n m}_\nu(x_1)\big{]} -
        \mbox{Tr}\,\big{[}S^{(f)mn}(x_1 , x_1+ a_\nu)\,\gamma_\mu\,U^{\dag n m}_\nu(x_1)\big{]}\right]\nonumber\\
        & &\nonumber\\
        &=& \frac{\lambda}{8a} \displaystyle\sum_{\vec{x}_1}\,\displaystyle\sum_{m, \xi}\, \frac{1}{L}\, \displaystyle\sum^L_{l=1}\,
        2 \,\mbox{Im}\left[  \eta^{l\, \dag}_{x_1, n, \tau}\,\eta^l_{x_1 + a_\nu, m, \xi}\,(\gamma_\mu)_{\tau\xi}\,
        U^{ nm}_\nu (x_1) \right].
}
By considering similar combinations of other terms,\ we can prove that for the ${\cal O}_{\mu\mu}$ operator,
the real part of the loop is zero for the first term,\ ${\bf I}$, and for the ${\cal O}_{4i}$ operator,
the imaginary part is zero.\ So,\ we cannot consider the first
term,\ ${\bf I}$.\ If we substitute $M^{-1}$ by $\kappa^2 D^2$ and $\kappa^4 D^4$, the trace is going to
be zero, since there will be no plaquette term. But if we substitute $M^{-1}$ by $\kappa D$ and
$\kappa^3 D^3$, there is a possibility of having a non-zero trace.\ Since the operators,\
${\cal O}_{4i}$ and ${\cal O}_{\mu\mu}$,\ can have different traces (or,\ no traces),\ we will
consider them separately.

\bigskip

$\bullet$ {\bf First term of ${\cal O}_{4i}$:}
\eqarray{
        & &\frac{-i}{8a} \displaystyle\sum_{\vec{x}_1}\,\mbox{Tr}\,\left[S^{(f)mn}(x_1 + a_i, x_1)\,\gamma_4\,
        U^{n m}_i(x_1)\right]\nonumber\\
        &\simeq&\frac{-i}{8a} \displaystyle\sum_{\vec{x}_1} \, \displaystyle\sum_{m, \xi, k, \sigma}\,\frac{1}{L}\,
        \displaystyle\sum^L_{l=1}\, \eta^{l\, \dag}_{x_1, n, \tau}\, M^{(-1)mk}_{\xi\sigma}(x_1 + a_i, z)\,
        \eta^l_{z, k, \sigma}\,(\gamma_4)_{\tau\xi}\, U^{ nm}_i (x_1).
\label{ubs4ifirst}
}

If we substitute $M^{-1}(x_1 + a_i, z)$ in Eq.~(\ref{ubs4ifirst}) by $\kappa D$,\ the only non-vanishing
term could be,
\[\displaystyle\sum^4_{\mu=1}(1+\gamma_\mu)_{\alpha\beta}\,U^{\dag ab}_\mu (x - a_\mu)\,
\delta_{x, y + a_\mu}.\]

So we get
\eqarray{
          & &\frac{-i}{8a} \displaystyle\sum_{\vec{x}_1}\,\mbox{Tr}\,\left[S^{(f)mn}(x_1 + a_i, x_1)\,\gamma_4\,
          U^{n m}_i(x_1)\right]\nonumber\\
          & &\nonumber\\
          &\simeq& \frac{-i}{8a} \kappa \displaystyle\sum_{\vec{x}_1} \, \displaystyle\sum_{m, \xi, k, \sigma}\,
          \frac{1}{L}\, \displaystyle\sum^L_{l=1}\, \eta^{l\, \dag}_{x_1, n, \tau}\,
          \left[\displaystyle\sum^4_{\mu=1}(1+\gamma_\mu)_{\xi\sigma}\,U^{\dag mk}_\mu (x_1+a_i - a_\mu)\,
          \delta_{x_1+a_i, z+ a_\mu} \right] \nonumber\\
          & &\nonumber\\
          & &\eta^l_{z, k, \sigma}\,(\gamma_4)_{\tau\xi}\, U^{ nm}_i (x_1)  \nonumber\\
          & &\nonumber\\
          &=&\frac{-i}{8a} \kappa \displaystyle\sum_{\vec{x}_1} \, \displaystyle\sum_{m, \xi, k, \sigma}\,
          \delta_{(n, k)(\tau, \sigma)}\, (\gamma_4)_{\tau\xi}\,(1+\gamma_i)_{\xi\sigma}\,
          U^{\dag mk}_i (x_1)\, U^{ nm}_i (x_1) \hspace{10mm}\mbox{( putting $\mu=i$ )}\nonumber\\
          & &\nonumber\\
          &=&0.
}
We now replace $M^{(-1)}$ by $\kappa^3 D^3$, the non-vanishing terms
could be
\eqarray{
    & & \displaystyle\sum_{\mu^\prime, \nu^\prime, \rho} \bigg{[}\{(1-\gamma_{\mu^\prime})
    (1+\gamma_{\nu^\prime})(1+\gamma_\rho)\}_{\alpha\beta}
    \{U_{\mu^\prime} (x)U^\dag_{\nu^\prime} (x+a_{\mu^\prime}-a_{\nu^\prime})\nonumber\\
    & &\nonumber\\
    & &U^\dag_\rho (x + a_{\mu^\prime} - a_{\nu^\prime} - a_\rho)\}^{ab}\delta_{x + a_{\mu^\prime} -
    a_{\nu^\prime} ,y + a_\rho}\nonumber\\
    & &\nonumber\\
    &+&\{(1+\gamma_{\mu^\prime})(1-\gamma_{\nu^\prime})(1+\gamma_\rho)\}_{\alpha\beta}
    \{U^\dag_{\mu^\prime} (x -a_{\mu^\prime})U_{\nu^\prime} (x- a_{\mu^\prime})
    U^\dag_\rho (x - a_{\mu^\prime} + a_{\nu^\prime} - a_\rho)\}^{ab}\nonumber\\
    & &\nonumber\\
    & &\delta_{x - a_{\mu^\prime} + a_{\nu^\prime},y + a_\rho}\nonumber\\
    & &\nonumber\\
    &+& \{(1+\gamma_{\mu^\prime})(1+\gamma_{\nu^\prime})(1-\gamma_\rho)\}_{\alpha\beta}
    \{U^\dag_{\mu^\prime} (x - a_{\mu^\prime})U^\dag_{\nu^\prime} (x - a_{\mu^\prime} - a_{\nu^\prime})
    U_\rho (x - a_{\mu^\prime} - a_{\nu^\prime} )\}^{ab}\nonumber\\
    & &\nonumber\\
    & &\delta_{x - a_{\mu^\prime} - a_{\nu^\prime} ,y - a_\rho}\bigg{]}.
\label{app:k3d34i1}
}
Let us consider the first term of Eq.~(\ref{app:k3d34i1}).
\eqarray{
    & &\frac{-i}{8a} \displaystyle\sum_{\vec{x}_1}\,\mbox{Tr}\,\left[S^{(f)mn}(x_1 + a_i, x_1)\,\gamma_4\,
    U^{n m}_i(x_1)\right]\nonumber\\
    & &\nonumber\\
    &=&\frac{-i}{8a} \kappa^3 \displaystyle\sum_{\vec{x}_1} \, \displaystyle\sum_{m, \xi, k, \sigma}\,\frac{1}{L}\,
    \displaystyle\sum^L_{l=1}\, \eta^{l\, \dag}_{x_1, n, \tau}\,
    \displaystyle\sum_{\mu^\prime, \nu^\prime, \rho} \bigg{[}\{(1-\gamma_{\mu^\prime})
    (1+\gamma_{\nu^\prime})(1+\gamma_\rho)\}_{\xi\sigma}\{U_{\mu^\prime} (x_1+a_i)\nonumber\\
    & &\nonumber\\
    & & U^\dag_{\nu^\prime} (x_1+a_i+a_{\mu^\prime}-a_{\nu^\prime})\,
    U^\dag_\rho (x_1+a_i + a_{\mu^\prime} - a_{\nu^\prime} - a_\rho)\}^{mk}\,
    \delta_{x_1+a_i + a_{\mu^\prime} - a_{\nu^\prime} ,z + a_\rho}\bigg{]}\nonumber\\
    & &\nonumber\\
    & &\eta^l_{z, k, \sigma}\,(\gamma_4)_{\tau\xi}\, U^{ nm}_i (x_1).
}
In order to have non-vanishing delta function,\ we must have either $\mu^\prime=\rho,\ \nu^\prime=i$; \
or\  $\mu^\prime=\nu^\prime,\ \rho=i$; \ or \ $\mu^\prime=\rho=\nu^\prime=i$.\ Let's take the case
$\mu^\prime=\rho, \nu^\prime=i$.\ Then the first term of  Eq.~(\ref{app:k3d34i1}) gives
\eqarray{
          & &\frac{-i}{8a} \kappa^3 \displaystyle\sum_{\vec{x}_1} \, \displaystyle\sum_{m, \xi, k, \sigma}\,
          \displaystyle\sum_{\mu^\prime}\delta_{(n,k)(\tau, \sigma)}\,(\gamma_4)_{\tau\xi}\,
          (1-\gamma_{\mu^\prime})(1+\gamma_{i})(1+\gamma_{\mu^\prime})_{\xi\sigma}
          \{U_{\mu^\prime} (x_1+a_i)\nonumber\\
          & &\nonumber\\
          & &U^\dag_{i} (x_1+a_{\mu^\prime})\, U^\dag_{\mu^\prime}(x_1)\}^{mk}\,
          U^{ nm}_i (x_1)\nonumber\\
          & &\nonumber\\
          &=&\frac{-i}{8a} \kappa^3 \displaystyle\sum_{\vec{x}_1} \, \displaystyle\sum_{m, \xi}\,
          \displaystyle\sum_{\mu^\prime}\{(\gamma_4-\gamma_4\gamma_{\mu^\prime})(1+\gamma_{i})
          (1+\gamma_{\mu^\prime})_{\xi\xi}\{U_{\mu^\prime}(x_1+a_i)U^\dag_{i}(x_1+a_{\mu^\prime})
          U^\dag_{\mu^\prime}(x_1)\nonumber\\
          & &\nonumber\\
          & &U_i (x_1)\}^{mm}\nonumber\\
          & &\nonumber\\
          &=&0.
}
For other two possibilities $\mu^\prime=\nu^\prime,\ \rho=i$ \ and \ $\mu^\prime=\rho=\nu^\prime=i$,\
the trace for the spin part is zero due to the multiplication of $\gamma$ matrices.

It can be shown that second and third terms of Eq.~(\ref{app:k3d34i1}) also have zero traces.\
In a similar manner,\ we can show that other terms of the operator ${\cal O}_{4i}$  also have
zero trace.

\bigskip

Now,\ we will consider the operator,\ ${\cal O}_{\mu\mu}$.

\bigskip

$\bullet$ {\bf First term of ${\cal O}_{\mu\mu}$:}
\eqarray{
          & & \frac{\lambda}{8a} \displaystyle\sum_{\vec{x}_1}\,\mbox{Tr}\,\left[S^{(f)mn}(x_1 + a_\mu, x_1)\,
          \gamma_\mu\, U^{n m}_\mu(x_1)\right]\nonumber\\
          & &\nonumber\\
          &\simeq& \frac{\lambda}{8a} \displaystyle\sum_{\vec{x}_1} \, \displaystyle\sum_{m, \xi, k, \sigma}\,\frac{1}{L}\,
          \displaystyle\sum^L_{l=1}\, \eta^{l\, \dag}_{x_1, n, \tau}\,
          M^{(-1)mk}_{\xi\sigma}(x_1 + a_\mu, z)\, \eta^l_{z, k, \sigma}\,(\gamma_\mu)_{\tau\xi}\,
          U^{ nm}_\mu (x_1),
\label{app:mumu1}
}
where  $\lambda = +1$ for $\mu = \nu = 1,2,3$ and $\lambda = -1$ for $\mu = \nu = 4$.

\noindent
If we substitute $M^{-1}(x_1 + a_\mu, z)$ in Eq.~(\ref{app:mumu1}) by $\kappa D$, the only non-vanishing term
 could be
\[\displaystyle\sum^4_{\nu=1}(1+\gamma_\nu)_{\alpha\beta}\,U^{\dag ab}_\nu (x - a_\nu)\,
\delta_{x, y + a_\nu}.\]
So we get
\eqarray{
        & &\frac{\lambda}{8a}\displaystyle\sum_{\vec{x}_1}\,\mbox{Tr}\,\left[S^{(f)mn}(x_1 + a_\mu, x_1)\,\gamma_\mu\,
        U^{n m}_\mu(x_1)\right]\nonumber\\
        & &\nonumber\\
        &\simeq&\frac{\lambda}{8a}\kappa \displaystyle\sum_{\vec{x}_1} \, \displaystyle\sum_{m, \xi, k, \sigma}\,\frac{1}{L}\,
        \displaystyle\sum^L_{l=1}\, \eta^{l\, \dag}_{x_1, n, \tau}\, \left[\displaystyle\sum^4_{\nu=1}
        (1+\gamma_\nu)_{\xi\sigma}\,U^{\dag mk}_\nu (x_1+a_\mu - a_\nu)\, \delta_{x_1+a_\mu, z+ a_\nu}\right] \nonumber\\
        & &\nonumber\\
        & &\eta^l_{z, k, \sigma}\,(\gamma_\mu)_{\tau\xi}\, U^{ nm}_\mu (x_1) \nonumber\\
        & &\nonumber\\
        &=& \frac{\lambda}{8a}\kappa \displaystyle\sum_{\vec{x}_1} \, \displaystyle\sum_{m, \xi, k, \sigma}\,
        \delta_{(n,k)(\tau\sigma)}\,(\gamma_\mu)_{\tau\xi}\,(1+\gamma_\mu)_{\xi\sigma}\,
        U^{\dag mk}_\mu (x_1)\, U^{ nm}_ \mu(x_1)\hspace{5mm}\mbox{( putting $\nu=\mu$ )}\nonumber\\
        & &\nonumber\\
        &=& \frac{\lambda}{8a}12 \kappa V^3.
}
And,\ if we substitute $M^{-1}(x_1 + a_\mu, z)$ by $\kappa^3 D^3$,\ the  non-vanishing terms could be
\eqarray{
    & &\kappa^3 \displaystyle\sum_{\mu^\prime, \nu^\prime, \rho} \bigg{[}\{(1-\gamma_{\mu^\prime})
    (1+\gamma_{\nu^\prime})(1+\gamma_\rho)\}_{\alpha\beta}\{U_{\mu^\prime}(x)
    U^\dag_{\nu^\prime}(x+a_{\mu^\prime}-a_{\nu^\prime}) \nonumber\\
    & &U^\dag_\rho (x + a_{\mu^\prime} - a_{\nu^\prime} - a_\rho)\}^{ab}\delta_{x + a_{\mu^\prime} -
    a_{\nu^\prime}, y + a_\rho}\nonumber\\
    & &\nonumber\\
    &+&\{(1+\gamma_{\mu^\prime})(1-\gamma_{\nu^\prime})(1+\gamma_\rho)\}_{\alpha\beta}
    \{U^\dag_{\mu^\prime}(x -a_{\mu^\prime})U_{\nu^\prime} (x- a_{\mu^\prime})
    U^\dag_\rho(x - a_{\mu^\prime} + a_{\nu^\prime} - a_\rho)\}^{ab} \nonumber\\
    & &\nonumber\\
    & &\delta_{x - a_{\mu^\prime} + a_{\nu^\prime},y + a_\rho}\nonumber\\
    & &\nonumber\\
    &+& \{(1+\gamma_{\mu^\prime})(1+\gamma_{\nu^\prime})(1-\gamma_\rho)\}_{\alpha\beta}
    \{U^\dag_{\mu^\prime} (x - a_{\mu^\prime})U^\dag_{\nu^\prime} (x - a_{\mu^\prime} -
    a_{\nu^\prime})U_\rho (x - a_{\mu^\prime} - a_{\nu^\prime} )\}^{ab} \nonumber\\
    & &\nonumber\\
    & &\delta_{x - a_{\mu^\prime} - a_{\nu^\prime}, y - a_\rho}\bigg{]}.
\label{app:k3d3441}
}
Let us consider the first term of Eq.~(\ref{app:k3d3441})
\eqarray{
          & &\frac{\lambda}{8a}\displaystyle\sum_{\vec{x}_1}\,\mbox{Tr}\,\left[S^{(f)mn}(x_1 + a_\mu, x_1)\,
          \gamma_\mu\, U^{n m}_\mu(x_1)\right]\nonumber\\
          & &\nonumber\\
          &\simeq&\frac{\lambda}{8a}\kappa^3 \displaystyle\sum_{\vec{x}_1} \, \displaystyle\sum_{m, \xi, k, \sigma}\,
          \frac{1}{L}\, \displaystyle\sum^L_{l=1}\, \eta^{l\, \dag}_{x_1, n, \tau}\,
          \displaystyle\sum_{\mu^\prime, \nu^\prime, \rho} \bigg{[}\{(1-\gamma_{\mu^\prime})(1+\gamma_{\nu^\prime})
          (1+\gamma_\rho)\}_{\xi\sigma}\{U_{\mu^\prime} (x_1+a_\mu)\nonumber\\
          & &\nonumber\\
          & &U^\dag_{\nu^\prime} (x_1+a_\mu+a_{\mu^\prime}-a_{\nu^\prime}) \, U^\dag_\rho (x_1+a_\mu +
          a_{\mu^\prime} - a_{\nu^\prime} - a_\rho)\}^{mk}\delta_{x_1+a_\mu + a_{\mu^\prime} -
          a_{\nu^\prime}, z + a_\rho}\bigg{]}\nonumber\\
          & &\nonumber\\
          & &\eta^l_{z, k, \sigma}\,(\gamma_\mu)_{\tau\xi}\, U^{ nm}_\mu (x_1).
}
In order to have non-vanishing delta function,\ we must have either $\mu^\prime=\rho,\ \nu^\prime=\mu$, \
or\  $\mu^\prime=\nu^\prime,\ \rho=\mu$, \ or \ $\mu^\prime=\rho=\nu^\prime=\mu$.\ Let's take the case
$\mu^\prime=\rho, \nu^\prime=\mu$.\ Then the first term of Eq.~(\ref{app:k3d3441}) gives
\eqarray{
          & &\frac{\lambda}{8a}\kappa^3 \displaystyle\sum_{\vec{x}_1} \, \displaystyle\sum_{m, \xi, k, \sigma}\,
          \displaystyle\sum_{\mu^\prime}\delta_{(n,k)(\tau,\sigma)}\,(\gamma_\mu)_{\tau\xi}\,
          (1-\gamma_{\mu^\prime})(1+\gamma_\mu)(1+\gamma_{\mu^\prime})_{\xi\sigma}
          \{U_{\mu^\prime}(x_1+a_\mu)\nonumber\\
          & &\nonumber\\
          & &U^\dag_\mu(x_1+a_{\mu^\prime})\,U^\dag_{\mu^\prime}(x_1)\}^{mk}\, U^{ nm}_\mu (x_1)\nonumber\\
          & &\nonumber\\
          &=& 8 \frac{\lambda}{8a}\kappa^3 \displaystyle\sum_{\vec{x}_1} \, \displaystyle\sum_{m}\,
          \displaystyle\sum_{\mu^\prime\neq\mu}\{U_{\mu^\prime}(x_1+a_\mu)
          U^\dag_\mu (x_1+a_{\mu^\prime})U^\dag_{\mu^\prime}(x_1)U_ \mu(x_1)\}^{mm}.
}
For other two possibilities $\mu^\prime=\nu^\prime,\ \rho=\mu$ \ and \
$\mu^\prime=\rho=\nu^\prime=\mu$,\ the trace for the spin part is zero due to the
multiplication of $\gamma$ matrices.\ Let now us consider the second term of
Eq.~(\ref{app:k3d3441}).\ Then
\eqarray{
          & & \frac{\lambda}{8a}\displaystyle\sum_{\vec{x}_1}\,\mbox{Tr}\,\left[S^{(f)mn}(x_1 + a_\mu, x_1)\,
          \gamma_\mu\, U^{n m}_\mu(x_1)\right]\nonumber\\
          & &\nonumber\\
          &\simeq& \frac{\lambda}{8a}\kappa^3 \displaystyle\sum_{\vec{x}_1} \, \displaystyle\sum_{m, \xi, k, \sigma}\,
          \frac{1}{L}\, \displaystyle\sum^L_{l=1}\, \eta^{l\, \dag}_{x_1, n, \tau}\,
          \displaystyle\sum_{\mu^\prime, \nu^\prime, \rho} \bigg{[}\{(1+\gamma_{\mu^\prime})
          (1-\gamma_{\nu^\prime})(1+\gamma_\rho)\}_{\xi\sigma}
          \{U^{\dag}_{\mu^\prime} (x_1+a_\mu-a_{\mu^\prime})\nonumber\\
          & &\nonumber\\
          & &U_{\nu^\prime} (x_1+a_\mu-a_{\mu^\prime})\, U^\dag_\rho (x_1+a_\mu - a_{\mu^\prime} +
          a_{\nu^\prime} - a_\rho)\}^{mk}\delta_{x_1+a_\mu - a_{\mu^\prime} +
          a_{\nu^\prime},z + a_\rho}\bigg{]}\nonumber\\
          & &\nonumber\\
          & &\eta^l_{z, k, \sigma}\,(\gamma_\mu)_{\tau\xi}\, U^{ nm}_\mu (x_1).
}
In order to have non-vanishing delta function,\ we must have either $\mu^\prime=\mu,\
\rho=\nu^\prime$,\ or\  $\mu^\prime=\nu^\prime,\ \rho=\mu$, \ or \
$\mu^\prime=\rho=\nu^\prime=\mu$.\ Let's take the case $\mu^\prime=\mu,\ \nu^\prime=\rho$.\
Then the second term of Eq.~(\ref{app:k3d3441}) gives
 \eqarray{
          & &\frac{\lambda}{8a}\kappa^3 \displaystyle\sum_{\vec{x}_1} \, \displaystyle\sum_{m, \xi, k, \sigma}\,
          \displaystyle\sum_{\mu^\prime}\delta_{(n,k)(\tau,\sigma)}\,(\gamma_\mu)_{\tau\xi}\,
          \{(1+\gamma_{\mu})(1-\gamma_{\nu^\prime})(1+\gamma_{\nu^\prime})\}_{\xi\sigma}
          \{U^{\dag}_{\mu}(x_1) \,U_{\nu^\prime} (x_1) \nonumber\\
          & &\nonumber\\
          & & U^\dag_{\nu^\prime} (x_1)\}^{mk}\, U^{ nm}_\mu (x_1)\nonumber\\
          & &\nonumber\\
          &=&0.
}
For other two possibilities $\mu^\prime=\nu^\prime,\ \rho=\mu$, \ and \ $\mu^\prime=\rho=\nu^\prime=\mu$,\
the traces for the spin part are zero due to the multiplication of $\gamma$ matrices.\ Let us consider the
third term of Eq.~(\ref{app:k3d3441}).\ Then
\eqarray{
          & &\frac{\lambda}{8a}\displaystyle\sum_{\vec{x}_1}\,\mbox{Tr}\,\left[S^{(f)mn}(x_1 + a_\mu, x_1)\,\gamma_\mu\,
          U^{n m}_\mu(x_1)\right]\nonumber\\
          &=&\frac{\lambda}{8a}\kappa^3 \displaystyle\sum_{\vec{x}_1} \, \displaystyle\sum_{m, \xi, k, \sigma}\,
          \frac{1}{L}\, \displaystyle\sum^L_{l=1}\, \eta^{l\, \dag}_{x_1, n, \tau}\,
          \displaystyle\sum_{\mu^\prime, \nu^\prime, \rho} \bigg{[}\{(1+\gamma_{\mu^\prime})
          (1+\gamma_{\nu^\prime})(1-\gamma_\rho)\}_{\xi\sigma}
          \{U^\dag_{\mu^\prime} (x_1+a_\mu-a_{\mu^\prime})\nonumber\\
          & &\nonumber\\
          & &U^\dag_{\nu^\prime} (x_1+a_\mu-a_{\mu^\prime}-a_{\nu^\prime})\, U_\rho (x_1+a_\mu -
          a_{\mu^\prime} - a_{\nu^\prime})\}^{mk}
          \delta_{x_1+a_\mu - a_{\mu^\prime} - a_{\nu^\prime},z - a_\rho}\bigg{]}\nonumber\\
          & &\nonumber\\
          & &\eta^l_{z, k, \sigma}\,(\gamma_\mu)_{\tau\xi}\, U^{ nm}_\mu (x_1).
}
In order to have non-vanishing delta function,\ we must have either $\mu^\prime=\rho,\
\nu^\prime=\mu$, \ or \  $\mu^\prime=\mu,\ \rho=\nu^\prime$, \ or \
$\mu^\prime=\rho=\nu^\prime=\mu$.\ Let's take the case
$\mu^\prime=\rho,\ \nu^\prime=\mu$.\ Then the third term of Eq.~(\ref{app:k3d3441}) gives
\eqarray{
          & &\frac{\lambda}{8a}\kappa^3 \displaystyle\sum_{\vec{x}_1} \, \displaystyle\sum_{m, \xi, k, \sigma}\,
          \displaystyle\sum_{\mu^\prime}\delta_{(n,k)(\tau,\sigma)}\,(\gamma_\mu)_{\tau\xi}\,
          (1+\gamma_{\mu^\prime})(1+\gamma_\mu)(1-\gamma_{\mu^\prime})_{\xi\sigma}
          \{U^\dag_{\mu^\prime} (x_1+a_\mu-a_{\mu^\prime})\nonumber\\
          & &\nonumber\\
          & &U^\dag_{\mu} (x_1-a_{\mu^\prime})\,U_{\mu^\prime} (x_1- a_{\mu^\prime})\}^{mk}\,
          U^{ nm}_\mu (x_1)\nonumber\\
          & &\nonumber\\
          &=& \frac{\lambda}{8a}8\kappa^3 \displaystyle\sum_{\vec{x}_1} \, \displaystyle\sum_{m}\,
          \displaystyle\sum_{\mu^\prime\neq\mu}\{U^\dag_{\mu^\prime} (x_1+a_\mu-a_{\mu^\prime})
          U^\dag_{\mu} (x_1-a_{\mu^\prime})U_{\mu^\prime} (x_1- a_{\mu^\prime})U_\mu (x_1)\}^{mm}.
}
For other two possibilities $\mu^\prime=\mu,\ \rho=\nu^\prime$, \ and \ $\mu^\prime=\rho=\nu^\prime=\mu$,\
the traces for the spin part are zero due to the multiplication of $\gamma$ matrices.

Doing similar calculations for other terms of ${\cal O}_{\mu\mu}$, we get the trace for the $\kappa D$ term
\eqarray{
\mbox{Trace}&=& \frac{\lambda}{8a} 2\kappa\,\left[ 12 V^3 - ( - 12 V^3 ) + 12 V^3 - ( - 12 V^3 ) \right]
    = \frac{\lambda}{8a} 96  \kappa\, V^3,
\label{app:1pref44}
}
and for the $\kappa^3 D^3$ term
\eqarray{
          \mbox{Trace} &\simeq& \frac{\lambda}{8a} 16 \kappa^3 \displaystyle\sum_{\vec{x}_1} \, \displaystyle\sum_{m}\,
          \displaystyle\sum_{\mu^\prime\neq\mu}\Bigg{[}\bigg{[}\{U_{\mu^\prime} (x_1+a_\mu)
          U^\dag_\mu (x_1+a_{\mu^\prime})U^\dag_{\mu^\prime}(x_1)U_ \mu(x_1)\nonumber\\
          & &\nonumber\\
          &+&U^\dag_{\mu^\prime} (x_1+a_\mu - a_{\mu^\prime})\, U^\dag_{\mu} (x_1-a_{\mu^\prime})
          U_{\mu^\prime} (x_1- a_{\mu^\prime})U_\mu (x_1)\}^{mm}\bigg{]}\nonumber\\
          & &\nonumber\\
          &-& \bigg{[} - \{U_{\mu^\prime} (x_1-a_\mu)U_\mu(x_1-a_{\mu}+a_{\mu^\prime})
          U^\dag_{\mu^\prime}(x_1)U^{\dag}_\mu (x_1-a_\mu) + U^\dag_{\mu^\prime} (x_1-a_\mu-a_{\mu^\prime})
          \nonumber\\
          & &\nonumber\\
          & &U_{\mu} (x_1-a_\mu-a_{\mu^\prime})U_{\mu^\prime} (x_1 - a_{\mu^\prime})
          U^{\dag}_\mu (x_1-a_\mu)\}^{mm}\bigg{]}\nonumber\\
          & &\nonumber\\
          &+&\bigg{[}\{U_{\mu^\prime} (x_1-a_\mu)U_\mu(x_1-a_{\mu}+a_{\mu^\prime})U^\dag_{\mu^\prime}(x_1)
          U^{\dag}_\mu (x_1-a_\mu)\,  + \, U^\dag_{\mu^\prime} (x_1-a_\mu-a_{\mu^\prime})\nonumber\\
          & &\nonumber\\
          & &U_{\mu} (x_1-a_\mu-a_{\mu^\prime})U_{\mu^\prime} (x_1 - a_{\mu^\prime})
          U^{\dag}_\mu (x_1-a_\mu)\}^{mm}\bigg{]}^\dag  \nonumber\\
          & &\nonumber\\
          &-& \bigg{[} -\{U_{\mu^\prime} (x_1+a_\mu)U^\dag_\mu (x_1+a_{\mu^\prime})U^\dag_{\mu^\prime}(x_1)
          U_ \mu(x_1) + U^\dag_{\mu^\prime} (x_1+a_\mu-a_{\mu^\prime})\nonumber\\
          & &\nonumber\\
          & &U^\dag_{\mu} (x_1-a_{\mu^\prime})U_{\mu^\prime} (x_1- a_{\mu^\prime})
          U_\mu (x_1)\}^{mm}\bigg{]}^\dag \Bigg{]}\nonumber\\
          & &\nonumber\\
          &=& \frac{\lambda}{8a} 32\kappa^3 \displaystyle\sum_{\vec{x}_1} \, \displaystyle\sum_{m}\,
          \displaystyle\sum_{\mu^\prime\neq\mu}\mbox{Re}\bigg{[}\{U_{\mu^\prime} (x_1+a_\mu)
          U^\dag_\mu (x_1+a_{\mu^\prime})U^\dag_{\mu^\prime}(x_1)U_ \mu(x_1)  \nonumber\\
          & &\nonumber\\
          &+&U_\mu (x_1) U^\dag_{\mu^\prime} (x_1+a_\mu-a_{\mu^\prime})U^\dag_{\mu} (x_1-a_{\mu^\prime})
          U_{\mu^\prime} (x_1- a_{\mu^\prime})  \nonumber\\
          & &\nonumber\\
          &+&U_\mu(x_1-a_{\mu}+a_{\mu^\prime})U^\dag_{\mu^\prime}(x_1)U^{\dag}_\mu (x_1-a_\mu)
          U_{\mu^\prime} (x_1-a_\mu)\nonumber\\
          & &\nonumber\\
          &+& U_{\mu^\prime} (x_1 - a_{\mu^\prime})U^{\dag}_\mu (x_1-a_\mu)
          U^\dag_{\mu^\prime}(x_1-a_\mu-a_{\mu^\prime})U_{\mu} (x_1-a_\mu-a_{\mu^\prime})\}^{mm}\bigg{]}.
\label{app:3pref44}
}

\subsection{Three-Index Operators}
\label{appsubsec:three_index_sub}

As in the case of two-index operator,\ we can show that the first term $\bf I$ of the hopping expansion
for this operator is real.\ But the loop part is imaginary. So, we can not consider the first term.\ If
we substitute $M^{-1}$ by $\kappa D$ and $\kappa^3 D^3$, the trace is going to be zero, since there will
be no plaquette terms. But if we substitute $M^{-1}$ by $\kappa^2 D^2$,\ there is a possibility of
non-zero trace.\ One of the terms in the loop part of ${\cal O}_{4ii}$ is
\eqarray{
        & &\frac{-1}{24 a^2}\mbox{Tr}\,\bigg{[} S^{(f)mn}(x_1+a_4+a_i, x_1) \, (\gamma_i)\, \bigg{\{} U^{nm^\prime}_4(x_1)\,
        U^{m^\prime m}_i(x_1+a_4) \, + \, U^{nm^\prime}_i(x_1)\nonumber\\
        & &\nonumber\\
        & & U^{m^\prime m}_4(x_1+a_i) \bigg{\}}\bigg{]}\nonumber\\
        & &\nonumber\\
        &\simeq&\frac{-1}{24 a^2}\displaystyle\sum_{\vec{x}_1} \, \displaystyle\sum_{m, \xi, k, \sigma}\,\frac{1}{L}\,
        \displaystyle\sum^L_{l=1}\, \eta^{l\, \dag}_{x_1, n, \tau}\, M^{(-1)mk}_{\xi\sigma}(x_1 +a_4 + a_i, z)\,
        \eta^l_{z, k, \sigma}\,(\gamma_i)_{\tau\xi}\nonumber\\
        & &\nonumber\\
        & &(\,U^{nm^\prime}_4(x_1)U^{m^\prime m}_i(x_1+a_4)+U^{nm^\prime}_i(x_1)U^{m^\prime m}_4(x_1+a_i).
}
If we substitute $M^{(-1)mk}_{\xi\sigma}$ by $\kappa^2 D^2$ term,\ the only non-vanishing term could be
\[\displaystyle\sum^4_{\mu^\prime=1}\,\displaystyle\sum^4_{\nu^\prime=1}\,
\bigg{[}\{(1+\gamma_{\mu^\prime})(1+\gamma_{\nu^\prime})\}_{\alpha\beta}\,
\{U^\dag_{\mu^\prime} (x - a_{\mu^\prime})U^\dag_{\nu^\prime}(x-a_{\mu^\prime} - a_{\nu^\prime})\}^{a b}\,
\delta_{x-a_{\mu^\prime}, y + a_{\nu^\prime}}\bigg{]}.\]

\noindent
Therefore,\ we get
\eqarray{
          & &\frac{-1}{24 a^2}\displaystyle\sum_{\vec{x}_1}\,\mbox{Tr}\,\bigg[S^{(f)mn}( x_1+a_4+a_i, x_1)\,(\gamma_i)\,
          (\,U^{nm^\prime}_4(x_1)U^{m^\prime m}_i(x_1+a_4)\nonumber\\
          & &\nonumber\\
          &+&U^{nm^\prime}_i(x_1)U^{m^\prime m}_4(x_1+a_i)\,)\bigg]\nonumber\\
          & &\nonumber\\
          &\simeq&\frac{-1}{24 a^2}\kappa^2 \displaystyle\sum_{\vec{x}_1}\displaystyle\sum_{m, \xi, k, \sigma}\frac{1}{L}
          \displaystyle\sum^L_{l=1}\eta^{l\, \dag}_{x_1, n, \tau}\, \displaystyle\sum^4_{\mu^\prime=1,\nu^\prime=1}
          \,\bigg{[}\{(1+\gamma_{\mu^\prime})(1+\gamma_{\nu^\prime})\}_{\xi\sigma}\nonumber\\
          & &\nonumber\\
          & & \{U^\dag_{\mu^\prime}(x_1+a_4+a_i - a_{\mu^\prime})
          U^\dag_{\nu^\prime}(x_1+a_4+a_i-a_{\mu^\prime} - a_{\nu^\prime})\}^{mk}
          \delta_{x_1+a_4+a_i-a_{\mu^\prime}, z + a_{\nu^\prime}}\bigg{]}\nonumber\\
          & &\nonumber\\
          & &\eta^l_{z, k, \sigma} (\gamma_i)_{\tau\xi}(U^{nm^\prime}_4(x_1)U^{m^\prime m}_i(x_1+a_4) +
          U^{nm^\prime}_i(x_1)U^{m^\prime m}_4(x_1+a_i)\,)\nonumber\\
          & &\nonumber\\
          &=& \frac{-1}{24 a^2} \kappa^2 \displaystyle\sum_{\vec{x}_1}\displaystyle\sum_{m, \xi, k, \sigma}
          \delta_{(n, k)(\tau,\sigma)}\, (\gamma_i)_{\tau\xi}\, \{(1+\gamma_i)(1+\gamma_4)\}_{\xi\sigma}\,
          \{U^\dag_i(x_1+a_4)) U^\dag_4 (x_1)\}^{mk}\nonumber\\
          & &\nonumber\\
          & &\bigg{[} \{ U_4(x_1)U_i(x_1+a_4)\}^{nm} + \{U_i(x_1)U_4(x_1+a_i)\}^{nm} \bigg{]}\,
          (\mbox{putting $\mu^\prime=i$ and $\nu^\prime=4 $})\nonumber\\
          & &\nonumber\\
          &+& \kappa^2 \displaystyle\sum_{\vec{x}_1}
          \displaystyle\sum_{m, \xi, k, \sigma}\delta_{(n, k)(\tau, \sigma)} \,(\gamma_i)_{\tau\xi}\,
          \{(1+\gamma_4)(1+\gamma_i)\}_{\xi\sigma} \,\{U^\dag_4(x_1+a_i)) U^\dag_i (x_1)\}^{mk}\nonumber\\
          & &\nonumber\\
          & &\bigg{[}\{ U_4(x_1)U_i(x_1+a_4)\}^{nm} +\{U_i(x_1)U_4(x_1+a_i)\}^{nm} \bigg{]}\,
          (\mbox{putting $\mu^\prime=4$ and $\nu^\prime=i $})\nonumber\\
          & &\nonumber\\
          &=& 8 \frac{-1}{24 a^2}\kappa^2 \displaystyle\sum_{\vec{x}_1}\bigg{[}\displaystyle\sum_{m} \mbox {Re} \,
          \{\{U^\dag_i(x_1+a_4) U^\dag_4 (x_1) U_i(x_1)U_4(x_1+a_i)\}^{mm}\, + \, 3\,\bigg{]}.
}

\bigskip
\noindent
Calculating all the other terms in a similar manner, we can show that
\eqarray{
     \mbox{Trace} &=& \frac{-1}{24 a^2}8 \kappa^2 \displaystyle\sum_{\vec{x}_1}\bigg{[}\displaystyle\sum_{m}
                  \mbox {Re}\, \{\{U^\dag_i(x_1+a_4) U^\dag_4 (x_1) U_i(x_1)U_4(x_1+a_i)\}^{mm}\, + \, 3\,\bigg{]}\nonumber\\
                  & &\nonumber\\
                  &-& 8 \kappa^2 \displaystyle\sum_{\vec{x}_1}\bigg{[}\displaystyle\sum_{m}
                  \mbox {Im} \,\{U^\dag_4(x_1-a_i)U_i (x_1-a_i)U_4(x_1)U^{\dag}_i(x_1+a_4-a_i)\}^{mm}\,\bigg{]}\nonumber\\
                  & &\nonumber\\
                  &-& 8 \kappa^2 \displaystyle\sum_{\vec{x}_1}\bigg{[}\displaystyle\sum_{m}
                  \mbox {Im} \,\{U_4(x_1-a_4+a_i)U^\dag_i (x_1)U^{\dag}_4(x_1-a_4)U_i(x_1-a_4)\}^{mm}\,\bigg{]}\nonumber\\
                  & &\nonumber\\
                  &-& 8 \kappa^2 \displaystyle\sum_{\vec{x}_1}\bigg{[}\displaystyle\sum_{m}
                  \mbox {Re} \,\{U_i(x_1-a_4-a_i)U_4 (x_1-a_4)U^{\dag}_i(x_1-a_i)\nonumber\\
                  & &\nonumber\\
                  & &U^{\dag}_4(x_1-a_4-a_i)\}^{mm}\, + \, 3\,\bigg{]} \nonumber\\
                  & &\nonumber\\
                  &+& 8 \kappa^2 \displaystyle\sum_{\vec{x}_1}\bigg{[}\displaystyle\sum_{m}
                  \mbox {Re} \,\{U_i(x_1-a_4-a_i)U_4 (x_1-a_4)U^{\dag}_i(x_1-a_i)\nonumber\\
                  & &\nonumber\\
                  & &U^{\dag}_4(x_1-a_4-a_i)\}^{mm}\, + \, 3\,\bigg{]}\nonumber\\
                  & &\nonumber\\
                  &-& 8 \kappa^2 \displaystyle\sum_{\vec{x}_1}\bigg{[}\displaystyle\sum_{m}
                  \mbox {Im} \,\{U_4(x_1-a_4+a_i)U^\dag_i (x_1)U^{\dag}_4(x_1-a_4)U_i(x_1-a_4)\}^{mm}\,\bigg{]} \nonumber\\
                  & &\nonumber\\
                  &-& 8 \kappa^2 \displaystyle\sum_{\vec{x}_1}\bigg{[}\displaystyle\sum_{m}
                  \mbox {Im} \,\{U^\dag_4(x_1-a_i)U_i (x_1-a_i)U_4(x_1)U^{\dag}_i(x_1+a_4-a_i)\}^{mm}\,\bigg{]}\nonumber\\
                  & &\nonumber\\
                  &-& 8 \kappa^2 \displaystyle\sum_{\vec{x}_1}\bigg{[}\displaystyle\sum_{m}
                  \mbox {Re} \,\{\{U^\dag_i(x_1+a_4) U^\dag_4 (x_1) U_i(x_1)U_4(x_1+a_i)\}^{mm}\, + \, 3\,\bigg{]}\nonumber\\
                  & &\nonumber\\
                  &=& \frac{1}{24 a^2}16  \kappa^2 \displaystyle\sum_{\vec{x}_1}\bigg{[}\displaystyle\sum_{m}
                  \mbox {Im} \,\{U^\dag_4(x_1-a_i)U_i (x_1-a_i)U_4(x_1)U^{\dag}_i(x_1+a_4-a_i)\}^{mm} \nonumber\\
                  & &\nonumber\\
                  &+&\displaystyle\sum_{m} \mbox {Im} \,\{U_4(x_1-a_4+a_i)U^\dag_i (x_1)
                  U^{\dag}_4(x_1-a_4)U_i(x_1-a_4)\}^{mm}\,\bigg{]}.
}

\vskip 40pt

\phantomsection
\addcontentsline{toc}{chapter}{References}

\centerline{\bf  REFERENCES} 

\end{document}